\documentclass{aastex}          
\usepackage{spr-astr-addons}    

\usepackage[para]{threeparttable}
\usepackage{xcolor}
\usepackage{mhchem}
\usepackage{graphicx}
\usepackage{appendix}
\usepackage{sidecap}
\usepackage{url}
\usepackage{natbib}

\begin{document}
%
\title{A review of type Ia supernova spectra}

\shorttitle{Why SN~Ia Spectra Matter}
\shortauthors{Parrent, Friesen, \& Parthasarathy}

\author{J.\ Parrent\altaffilmark{1,2}, B.\ Friesen\altaffilmark{3}, and M.\ Parthasarathy\altaffilmark{4}}
\altaffiltext{1}{6127 Wilder Lab, Department of Physics \& Astronomy, Dartmouth College, Hanover, NH 03755, USA}
\altaffiltext{2}{Las Cumbres Observatory Global Telescope Network, Goleta, CA 93117, USA}
\altaffiltext{3}{Homer L. Dodge Department of Physics and Astronomy, University of Oklahoma, 440 W Brooks, Norman, OK 73019, USA}
\altaffiltext{4}{Inter-University Centre for Astronomy and Astrophysics (IUCAA), Post Bag 4, Ganeshkhind, Pune 411007, India}

\begin{abstract}
SN~2011fe was the nearest and best-observed type Ia supernova in a generation, and brought previous incomplete datasets into sharp contrast with the detailed new data. In retrospect, documenting spectroscopic behaviors of type Ia supernovae has been more often limited by sparse and incomplete temporal sampling than by consequences of signal-to-noise ratios, telluric features, or small sample sizes. As a result, type Ia supernovae have been primarily studied insofar as parameters discretized by relative epochs and incomplete temporal snapshots near maximum light. Here we discuss a necessary next step toward consistently modeling and directly measuring spectroscopic observables of type Ia supernova spectra. In addition, we analyze current spectroscopic data in the parameter space defined by empirical metrics, which will be relevant even after progenitors are observed and detailed models are refined.
\end{abstract}


%

\section{Introduction}\label{s:intro}

The transient nature of extragalactic type Ia supernovae (SN~Ia) prevent studies from conclusively singling out unobserved progenitor configurations \citep{Roelofs08,WLi11a,Kilic13}. It remains fairly certain that the progenitor system of SN~Ia comprises at least one compact C$+$O white dwarf \citep{Chandrasekhar57,Nugent11,Bloom12}. However, \emph{how} the state of this primary star reaches a critical point of disruption continues to elude astronomers. This is particularly so given that less than $\sim$ 15\% of locally observed white dwarfs have a mass a few 0.1M$_{\odot}$ greater than a solar mass; very few systems near the formal Chandrasekhar-mass limit\footnote[5]{By ``formal'' we are referring to the mass limit that omits stellar rotation (see Appendix of \citealt{Jeffery06}).}, M$_{Ch}$ $\approx$ 1.38 M$_{\odot}$ \citep{Vennes99,Liebert05,Napiwotzki05, Partha07,Napiwotzki07}. 

Thus far observational constraints of SN~Ia have been inconclusive in distinguishing between the following three separate theoretical considerations about possible progenitor scenarios. Along side perturbations in the critical mass limit or masses of the progenitors, e.g., from rotational support \citep{Muller85,Yoon05,Chen09,Hachisu12a,Tornambe13} or variances of white dwarf (WD) populations \citep{Kerkwijk10,Dan13}, the primary WD may reach the critical point by accretion of material from a low-mass, radially-confined secondary star \citep{Whelan73,Nomoto77,Hayden10SDSS,Bianco11,Bloom12,Hachisu12a,Wheeler12,Mazzali13,Chen13}, and/or through one of several white dwarf merger scenarios with a close binary companion \citep{Webbink84,Iben84,Paczynski85,Thompson11,BoWang13,Pakmor13}. In addition, the presence (or absence) of circumstellar material may not solely rule out particular progenitor systems as now both single- and double-degenerate systems are consistent with having polluted environments prior to the explosion \citep{Shen13,Phillips13}. 


Meanwhile, and within the context of a well-observed spectroscopically normal SN~2011fe, recent detailed models and spectrum synthesis along with SN Ia rates studies, a strong case for merging binaries as the progenitors of normal SN Ia has surfaced (c.f., \citealt{Kerkwijk10,WLi11b,Blondin12,Chomiuk13,Dan13,Moll13,Maoz13,Johansson14}). However, because no progenitor system has \emph{ever} been connected to any SN Ia, most observational constraints and trends are difficult to robustly impose on a standard model picture for even a single progenitor channel; the SN~Ia problem is yet to be confined for each SN Ia subtype. 

As for restricting SN Ia subtypes to candidate progenitor systems: (i) observed ``jumps'' between mean properties of SN Ia subtypes signify potential differences of progenitors and/or explosion mechanisms, (ii) the dispersions of individual subtypes are thought to arise from various abundance, density, metallicity, and/or temperature enhancements of the original progenitor system's post-explosion ejecta tomography, and (iii) ``transitional-type'' SN Ia complicate the already similar overlap of observed SN Ia properties \citep{Nugent95a,Lentz00,Benetti05,Branch09,Hoflich10,WangX12,WangX13,Dessart13models}. Moreover, our physical understanding of all observed SN~Ia subclasses remains based entirely on interpretations of idealized explosion models that are so far constrained and evaluated by ``goodness of fit'' comparisons to incomplete observations, particularly for SN~Ia spectra at all epochs. 

By default, spectra have been a limiting factor of supernova studies due to associated observational consequences, e.g., impromptu transient targets, variable intrinsic peak luminosities, a sparsity of complete datasets in wavelength and time, insufficient signal-to-noise ratios, and the ever-present obstacle of spectroscopic line blending \citep{Payne40}. Subsequently, two frequently relied upon empirical quantifiers of SN~Ia spectroscopic diversity have been the rate at which rest-frame 6100 \AA\ absorption minima shift redward vis-\`{a}-vis projected Doppler velocities of the absorbing Si-rich material \citep{Benetti05,WangX09Subtype} and absorption strength measurements (a.k.a. pseudo equivalent widths; pEWs) of several lines of interest (see \citealt{Branch06,Hachinger06,Silverman12maxlight,Blondin12}). Together these classification schemes more-or-less describe the same events by two interconnected parameter spaces (i.e. flux and expansion velocities, \citealt{Branch09,Foley11,Blondin12}) that are dependent on a multi-dimensional array of physical properties. Naturally, the necessary next step for supernova studies alike is the development of prescriptions for the physical diagnosis of spectroscopic behaviors (see \S2.2 and \citealt{Kerzendorf14}). 

For those supernova events that \emph{have} revealed the observed patterns of
SN~Ia properties, the majority are termed ``Branch-normal'' \citep{Branch93,WLi11b}, while others further away from the norm are historically said to be ``peculiar'' (e.g., SN~1991T, 1991bg; see \citealt{Flipper97} and references therein). Although, many non-standard events have since obscured the boundaries between both normal and peculiar varieties of SN~Ia, such as SN~1999aa \citep{Garavini04}, 2000cx \citep{Chornock00,WLi01,Rudy02}, 2001ay \citep{Krisciunas11}, 2002cx \citep{WLi03}, 2003fg \citep{Howell06,Jeffery06}, 2003hv \citep{Leloudas09,Mazzali11}, 2004dt \citep{Wang06,Altavilla07}, 2004eo \citep{Pastorello07a}, 2005gj \citep{Prieto07}, 2006bt \citep{Foley10b}, 2007ax \citep{Kasliwal08}, 2008ha \citep{Foley09,Foley10a}, 2009ig \citep{Foley12c,Marion13}, PTF10ops \citep{Maguire11}, PTF11kx \citep{Dilday12,Silverman13}, and 2012fr \citep{Maund13,Childress13}. 

The fact that certain subsets of normal SN~Ia constitute a near homogenous group of intrinsically bright events has led to their use as standardizable distance indicators \citep{Kowal68,Elias81,Branch92,Riess99,Perlmutter99,Schmidt04,Mandel11,Maeda11,Sullivan11SNLS,Hicken12}. However, this same attribute of homogeneity remains the greatest challenge in the individual study of SN~Ia given that the time-evolving spectrum of a supernova is unique unto itself from the earliest to the latest epochs.  

Because SN~Ia are invaluable tools for both cosmology and understanding progenitor populations,
a multitude of large scale surveys, searches, and observing campaigns\footnote[6]{e.g., The Automated Survey for SuperNovae (Assassin), The Backyard Observatory Supernova Search (BOSS), The Brazilian Supernova Search (BRASS), The Carnegie Supernova Project (CSP), The Catalina Real-Time Transient Survey (CRTS), The CHilean Automatic Supernovas sEarch (CHASE), The Dark Energy Survey (DES), The Equation of State: SupErNovae trace Cosmic Expansion (ESSENCE) Supernova Survey, The La Silla-QUEST Variability Survey (LSQ), Las Cumbres Observatory Global Telescope Network (LCOGT), The Lick Observatory Supernova Search (LOSS), The Mobile Astronomical System of the Telescope-Robots Supernova Search (MASTER), The Nearby Supernova Factory (SNfactory), The Optical Gravitational Lensing Experiment (OGLE-IV), The Palomar Transient Factory (PTF), The Panoramic Survey Telescope and Rapid Response System (Pan-STARRS), The Plaskett Spectroscopic Supernova Survey (PSSS), Public ESO Spectroscopic Survey of Transient Objects (PESSTO), The Puckett Observatory World Supernova Search, The ROTSE Supernova Verification Project (RSVP), The SDSS Supernova Survey, The Canada-France-Hawaii Telescope Legacy Survey Supernova Program (SNLS), The Southern inTermediate Redshift ESO Supernova Search (STRESS), The Texas Supernova Search (TSS); for more, see \url{http://www.rochesterastronomy.org/snimages/snlinks.html}.} are continually being carried out
with regularly improved precision. Subsequently, this build-up of competing resources has also resulted in an ever growing number of new and important discoveries, with less than complete information for each. In fact, with so many papers published each year on various aspects of SN~Ia, it can be difficult to keep track of new results and important developments, including the validity of past and present theoretical explosion simulations and their related observational interpretations (see \citealt{Maoz13} for the latest). 

Here we compile some of the discussions on spectroscopic properties of SN~Ia from the past decade of published works. In \S2 we overview the most common means for studying SN~Ia: light curves (\S2.1), spectra (\S2.2), and detailed explosion models (\S2.3). In particular, we overview how far the well-observed SN~2011fe has progressed the degree of confidence associated with reading highly blended SN~Ia spectra. Issues of SN~Ia diversity are discussed in \S3. Next, in \S4 we recall several SN~Ia that have made up the bulk of \emph{recent} advances in uncovering the extent of their properties and peculiarities (see also the Appendix for a guide of some recent events). Finally, in \S5 we summarize and conclude with some observational lessons of SN~2011fe.

\section{Common Subfields of Utility}\label{s:utility}

\subsection{Light curves}\label{ss:lightcurves}

The interaction between the radiation field and the ejecta can be interpreted to zeroth order with the bolometric light curve. For SN~Ia, the rise and fall of the light curve is said to be ``powered'' by $^{56}$Ni produced in the explosion \citep{Colgate69,Arnett82,Khokhlov93,Mazzali98,Pinto00b,Stritzinger05}. Additional sources \emph{are} expected to contribute to the overall luminosity behavior at various epochs\footnote[7]{Just a few examples include: C$+$O layer metallicity \citep{Lentz00,Timmes03,Meng11}, interaction with circumstellar material (CSM, see \citealt{Quimby06,Patat07,Simon07,Kasen10,Hayden10SDSS,Sternberg11,Foley12dust,Forster12,Shen13,Silverman13IaCSM,Raskin13}) or an enshrouding C$+$O envelope \citep{Scalzo12,Taubenberger13}, differences in total progenitor system masses \citep{Hachisu12a,Pakmor13,Chen13}, and directional dependent aspects of binary configurations (e.g., \citealt{Blondin11a,Moll13}).}. 

For example, \citet{Nomoto03} has suggested that the variation of the carbon mass fraction in the C+O WD (C/O), or the variation of the initial WD mass, causes the diversity of SN~Ia brightnesses (see \citealt{Hoflich10}). Similarly, \citet{Meng11} argue that C/O and progenitor metallicity, Z, are intimately related for a fixed WD mass, and particularly for high metallicities given that it results in lower 3$\alpha$ burning rates plus an increased reduction of carbon via $^{12}$C($\alpha$,$\gamma$)$^{16}$O. 
For Z $>$ Z$_{\odot}$ ($\sim$0.02), \citet{Meng11} find that both C/O and Z have an approximately equal influence on $^{56}$Ni production since, for a given WD mass, high progenitor metallicities (a greater abundance of species heavier than oxygen) and low C/O abundances (low carbon-rich fuel assuming a single-degenerate scenario) result in a low $^{56}$Ni yield and subsequently dimmer SN~Ia. For near solar metallicities or less, the carbon mass fraction plays a dominant role in $^{56}$Ni production \citep{Timmes03}. This then suggests that the average C/O ratio in the final state of the progenitor is an important \emph{physical} cause, in addition to metallicity, for the observed width-luminosity relationship (WLR\footnote[8]{A WLR is followed when a SN~Ia has a proportionately broader light curve for its intrinsic brightness at maximum light \citep{Phillips93}. \citet{Phillips99} later extended this correlation by incorporating measurement of the extinction via late time \emph{B} $-$ \emph{V} color measurements and \emph{B} $-$ \emph{V} and \emph{V} $-$ \emph{I} measurements at maximum light (see also \citealt{Germany04,Prieto06}). Because lights curves of faint SN Ia evolve promptly before 15 days post-maximum light, light curve shape measurements are better suited for evaluating the light curve ``stretch'' \citep{Conley08}.}) of normal SN~Ia light curves \citep{Umeda99b,Timmes03,Nomoto03,Bravo10,Meng11}. 

At the same time, the observed characteristics of SN~Ia light curves and spectra can be fairly matched by adopting radial and/or axial shifts in the distribution of $^{56}$Ni, possibly due to a delayed- and/or pulsational-detonation-like explosion mechanism (see \citealt{Khokhlov91PRD,Hoflich95,Baron08,Bravo09,Maedanature,Baron12,Dessart13models}) or a merger scenario (e.g., \citealt{Dan13,Moll13}). Central ignition densities are also expected to play a secondary role in the form of the WLR since they are dependent upon the accretion rate of H and/or He-rich material and cooling time \citep{Roepke05c,Hoflich10,Meng10,Krueger10,Sim13}, in addition to the spin-down timescales for differentially rotating WDs \citep{Hachisu12a,Tornambe13}. Generally, discerning which of these factors dominate the spectrophotometric variation from one SN~Ia to another remains a challenging task \citep{WangX12}. As a result, astronomers are still mapping a broad range of SN~Ia characteristics and trends (\S3). 

Meanwhile, cosmological parameters determined by SN~Ia light curves depend
on an accurate comparison of nearby and distant events\footnote[9]{Most SN~Ia distance determination methods rely on correlating a
distance dependent parameter and one or more distance independent
parameters. Subsequently, a number of methods have been developed to calibrate SN~Ia by multi-color light curve shapes (e.g., \citealt{Hamuy96,Nugent02,Knop03,Nobili05,Prieto06,Jha07,Conley08,Rodney09,Burns11}).}. For distant and therefore redshifted SN~Ia, a ``K-correction'' converts an observed magnitude to that which would be observed
in the rest frame in another bandpass filter, allowing for the comparison of SN~Ia brightnesses at various redshifts \citep{Hogg02}. Consequently, K-corrections require the spectral energy distribution (SED) of the SN~Ia and depend on SN~Ia broad-band colors and the diversity
of spectroscopic features \citep{Nugent02}. While some light curve fitters take a K-correction-less approach (e.g., \citealt{Guy05,Guy07,Conley08}), an SED is still required. A spectral template time series dataset is usually used since there exists remarkable
homogeneity in the observed optical spectra of ``normal'' SN~Ia (e.g., \citealt{Hsiao07}).

Unfortunately there do remain poorly understood differences regarding spectroscopic feature strengths and inferred expansion velocities for these and other types of thermonuclear supernovae (see \S2.2 and \S3). At best, the spectroscopic diversity of SN~Ia has been determined to be multidimensional \citep{Hatano00,Benetti05,Branch09,WangX09Subtype}. Verily, SN~Ia diversity studies require numerous large spectroscopic datasets in order to subvert many complex challenges faced when interpreting the data and extracting both projected Doppler velocities and ``feature strength'' measurements. However, studies that seek to primarily utilize SN~Ia broad band luminosities need only collect a handful of sporadically sampled spectra in order to type the supernova event as a bona fide SN~Ia. We note that interests in precision cosmology conflict at this point with the study of SN~Ia. This is primarily because obtaining \emph{UBVRI} photometry for hundreds of events is cheaper than collecting complete spectroscopy for a lesser number of SN~Ia at various redshifts.

Nevertheless, the brightness decline rate in the \emph{B}-band during the first 15 rest-frame days post-maximum light, $\Delta$m$_{15}$(\emph{B}), has proven useful for all SN~Ia surveys. \citet{Phillips93} noted that $\Delta$m$_{15}$(\emph{B}) is well correlated with the intrinsic luminosity, a.k.a. the width-luminosity relationship. Previously, \citet{Khokhlov93} did predict the existence of a WLR given that the light curve shape is sensitive to the time-dependent state of the ejected material. 

\citet{Kasen07} recently utilized multi-dimensional time-dependent Monte Carlo radiative transfer calculations of
Chandrasekhar-mass SN~Ia models to access the physical relationship between the luminosity and light curve decline rate. They found that the WLR is largely a consequence of the radiative transfer inherent to SN~Ia atmospheres, whereby the ionization evolution of iron redirects flux red ward and is hastened for dimmer and/or cooler SN~Ia. \citet{Woosley07} later explored the diversity of SN~Ia
light curves using a grid of 130 one-dimensional models. They concluded that a WLR is satisfied when SN~Ia burn $\sim$ 1.1 M$_{\odot}$ of material, with iron-group elements extending out to $\sim$ 8000 km~s$^{-1}$. 

Broadly speaking, the shape of the WLR is fundamentally influenced by the ionization evolution of iron group elements \citep{Kasen07}. However, since broad band luminosities are the sum of a supernova SED per wavelength interval, details of SN~Ia diversity risk being ``blurred out'' for large samples of SN Ia. Therefore, decoding the spectra of all SN~Ia subtypes, in addition to indirectly constraining detailed explosion models by the WLR, is of vital importance since variable signatures of iron-peak elements (IPEs) blend themselves within an SED typically populated by relatively strong features of overlapping signatures of intermediate-mass elements (IMEs).

\subsection{Spectra}\label{ss:spectra}

Supernova spectra detail information about the explosion and its local environment. To isolate and extract physical details (and determine their order of influence), several groups have invested greatly in advancing the computation of synthetic spectra for SN~Ia, particularly during the early phases of homologous expansion (e.g., \citealt{MazzaliLucy93,Hauschildt99,Kasen02,Thomas02,Hoflich02,Branch04b,Sauer06,Kasen06,Jeffery07Giant,Sim10RT,ThomasSYNAPPS,Hillier12,Hoffman13,Pauldrach13,Kerzendorf14}). Although, even the basic facets of the supernova radiation environment serve as obstacles for timely computations of physically accurate, statistically representative, and robustly certain synthetic spectra (e.g., consequences of expansion).

It is the time-dependent interaction of the radiation field with the expanding material that complicates drawing conclusions about the explosion physics from the observations\footnote[10]{There is general consensus that the observed spectroscopic diversity of most SN~Ia are influenced by: different configurations of $^{56}$Ni produced in the 
events \citep{Colgate69,Arnett82,Khokhlov93,Baron12}, their effective temperatures \citep{Nugent95a}, density profiles and the amount of IPEs present within the outermost layers of ejecta \citep{Hatano99,Baron06,Hachinger12}, global symmetries of Si-rich material \citep{Thomas02}, departures from spherical symmetry for Ca and Si-rich material at high velocities \citep{Wang07,Kasen09,Maeda10,Maund13,Moll13,Dessart13models}, efficiencies of flux redistribution \citep{Kasen06,Jack12}, the radial extent of stratified material resulting from a detonation phase \citep{Woosley07}, host galaxy dust \citep{Tripp99,Childress13Hosts}, and the metallicity of the progenitors \citep{Hoflich98,Lentz00,Timmes03,Howell09,Bravo10,Jackson10,WangX13}.}.  In a sense, there are two stages during which direct (and accessible) information about the progenitor system is driven away from being easily discernible within the post-explosion spectra: explosive nucleosynthesis and radiation transport\footnote[11]{Some relevant obstacles include: a high radiation
energy density in a low matter density environment, radiative versus local collisional processes (non-LTE conditions) and effects \citep{Baron96}, time-dependent effects and the dominance of line over continuum opacity \citep{Pinto00b,Pinto00a}, and relativistic flows as well as GR effects on line profiles \citep{Chen07,Knop09}. In addition, the entire
light emission is powered by decay-chain $\gamma$-rays, interactions with CSM, and is influenced by positrons, fast electrons, and Auger electrons in later phases \citep{Kozma92,Seitenzahl09}.}. That is to say, the ability to reproduce both the observed light curve and spectra, as well as the range of observed characteristics among SN~Ia, is essential towards validating and/or restricting any explosion model for a given subtype.

\begin{figure*}[Htbp]
\begin{center}
\includegraphics[width=3.3in, angle=90]{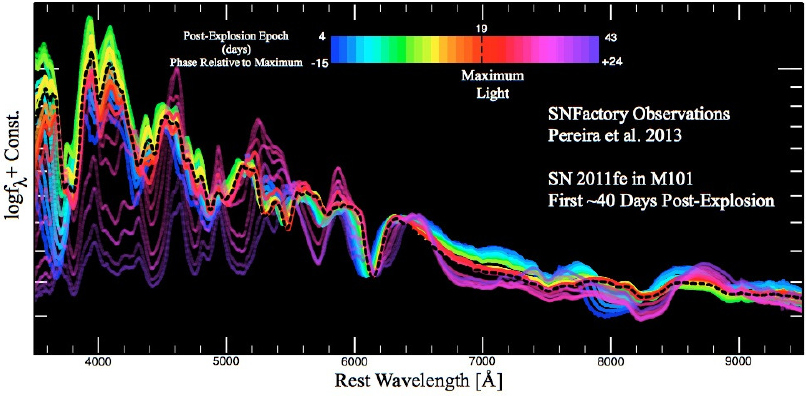}
\end{center}
\caption{Plotted is the SNFactory's early epoch dataset of SN~2011fe presented by \citet{Pereira13}. We have normalized and over-plotted each spectrum at the 6100 \AA\ P Cygni profile in order to show the relative locations of all ill-defined features as they evolve with the expansion of the ejecta. The quoted rise-time to maximum light (dashed black) is from \citet{Mazzali13}. }
\label{fig:pcygfiga} 
\end{figure*}

Moreover, this assumes the sources of observed spectroscopic signatures in all varieties of SN~Ia are known {\it a priori}, which is not necessarily the case given the immense volume of actively contributing atomic line transitions and continuum processes \citep{Baron95,Baron96,Kasen08,Bongard08,Sauer08}. In fact, several features throughout the spectra have been either \emph{tentatively} associated with a particular blend of atomic lines or identified with a multiple of conflicting suggestions (e.g., forbidden versus permitted lines at late or ``nebular'' transitional phases, see \citealt{Bowers97,Branch05,Friesen12,Dessart13}). Meanwhile others are simply misidentified or unresolved due to the inherent high degeneracy of solutions and warrant improvements to the models for further study (e.g., \ion{Na}{1} versus [\ion{Co}{3}]; \citealt{Dessart13}). 

For example, the debate over whether or not hydrogen and/or helium are detected in some early Ibc spectra has been difficult to navigate on account of the wavelength separation of observed weak features and the number of plausible interpretations \citep{Deng00,Branch02,Anumpama05,Elmhamdi06,Parrent07,Ketchum08,Soderberg08,James10,Benetti11,Chornock11,Dessart12,Danmil1311ei,Danmil13,Takaki13}. Historically, the term ``conspicuous'' has defined whether or not a supernova belongs to a particular spectroscopic class. By way of illustration, \emph{photographic spectrograms of type II events reveal conspicuous emission bands of hydrogen while type I events do not} \citep{Minkowski41}. With the advent of CCD cameras in modern astronomy, it has been determined that 6300 \AA\ absorption features (however weak) in the early spectra of some type Ibc supernovae are often no less conspicuous than 6100 \AA\ \ion{Si}{2} $\lambda$6355 absorption features in SN~Ia spectra, where some 6300 \AA\ features produced by SN Ibc may be due to \ion{Si}{2} and/or higher velocity H$\alpha$ \citep{Filippenko88,Filippenko90,Filippenko92Ic}. That is, while SN Ibc are of the type I class, they do not necessarily lack hydrogen and/or helium within their outer-most layers of ejecta, hence the conservative definition of type I supernovae as ``hydrogen/helium-poor'' events.

This conundrum of which ion signatures construct each observed spectral feature rests proportionately on the signal-to-noise ratio (S/N) of the data. However, resolving this spectroscopic dilemma is primarily dependent on the wavelength and temporal coverage of the observations and traces back to the pioneering work of \citet{McLaughlin63} who studied spectra of the type Ib supernova, SN~1954A, in NGC 4214 \citep{Wellmann55,Branch72,Blaylock00,Casebeer00}. Contrary to previous interpretations that supernova spectra were the result of broad, overlapping emission features \citep{Payne36,Humason36,Baade36,Walter37,Minkowski39,Payne40,Zwicky42rates,Baade56}, it was D. B. McLaughlin who first began to repeatedly entertain the idea that ``absorption-like'' features were present\footnote[12]{Admittedly \citet{Minkowski41} had previously mentioned ``absorptions and broad emission bands are developed [in the spectra of supernovae].'' Although, this was primarily within the context of early epoch observations that revealed a featureless, blue continuum: ``Neither absorptions nor emission bands can be definitely seen but some emission is suspected in the region of H$\alpha$'' \citep{Minkowski40}.} in regions that ``lacked emission'' \citep{McLaughlin59,McLaughlin60,McLaughlin63}. 

The inherent difficulties in reading supernova spectra and the history of uncertain line identifications for both conspicuous and \emph{concealed} absorption signatures are almost as old as the supernova field itself \citep{Payne40,Dessart13}. Still, spectroscopic intuitions can only evolve as far as the data allow. Therefore it is both appropriate and informative to recall the progression of early discussions on the spectra of supernovae, during which spectroscopic designations of type~I and type~II were first introduced:

\begin{quotation}
There appears to be a general opinion that the evidence concerning the spectrum of the most luminous nova of modern times was so contradictory that conclusions as to its spectra nature are impossible. This view is expressed, for example, by Miss Cannon: ``With the testimony apparently so conflicting, it is difficult to form any conception of the class of this spectrum'' \citep{Payne36}.
\end{quotation}

\begin{quotation}
It also seems ill advised to conclude anything regarding the distribution of temperature in super-novae from the character of their visible spectra as long as a satisfactory explanation of some of the most important features of these spectra is completely lacking \citep{Zwicky36}.
\end{quotation}

\begin{quotation}
The spectrum is not easy to interpret, as true boundaries of the wide emission lines are difficult to determine \citep{Humason36}. 
\end{quotation}

\begin{quotation}
Those [emission] bands with distinct maxima and a fairly sharp redward or violetward edge, excepting edges due to a drop in plate spectral sensitivity, may give an indication of expansion velocity \citep{Popper37}.
\end{quotation}

\begin{quotation}
Instead of the typical pattern of broad, diffuse emissions dominated by a band about 4600 \AA, it appeared like a continuum with a few deep and several shallow absorption-like minima. Two of the strongest ``absorption lines,'' when provisionally interpreted as $\lambda\lambda$4026, 4472 \ion{He}{1}, give velocities near $-$5000 km~s$^{-1}$ [...] The author is grateful to N. U. Mayall and R. Minkowski for the use of spectrograms, and for helpful discussions. However, this does not imply agreement with the author's interpretations \citep{McLaughlin59}. 
\end{quotation}

\begin{quotation}
It is hardly necessary to emphasize in detail the difficulties of establishing the correct interpretation of a spectrum which may reflect unusual chemical composition, whose features may represent emission, absorption, or both mixed, and whose details are too ill-defined to admit precise measures of wavelengths \citep{Minkowski63}. 
\end{quotation}

Given that our general understanding of blended spectral lines remains in a continual state of improvement, the frequently recurrent part of ``the supernova problem'' is pairing observed features with select elements of the periodic table \citep{Hummer76,Axelrod80,Jeffery90,HatanoAtlas,Branch00}. In fact, it was not until nearly a half-century after \citet{Minkowski63}, with the discovery and prompt spectroscopic follow-up of SN~2011fe (Figure~\ref{fig:pcygfiga} and \S4.1) that the loose self-similarity of SN~Ia time series spectra from the perceived beginning of the event to near maximum light was roundly confirmed (\citealt{Nugent11}, see also \citealt{Garavini05,Foley12c,Silverman12a,Childress13,Zheng13}). 

While SN~2011fe may not have revealed a direct confirmation on its progenitor system \citep{WLi11a}, daily spectroscopic records at optical wavelengths were finally achieved, establishing the most efficient approach for observing ill-defined features over time \citep{Pereira13}. This is important given that UV to NIR line identifications of all observed complexes are highly time-dependent, are sensitive to most physically relevant effects, continuously vary between subtypes, and rely on minimal constraint for all observed events\footnote[13]{See \citet{Foley12b} for ``The First Maximum-light Ultraviolet through Near-infrared Spectrum of a Type Ia Supernova.''}.

\begin{figure}
\begin{center}
\includegraphics[width=3in]{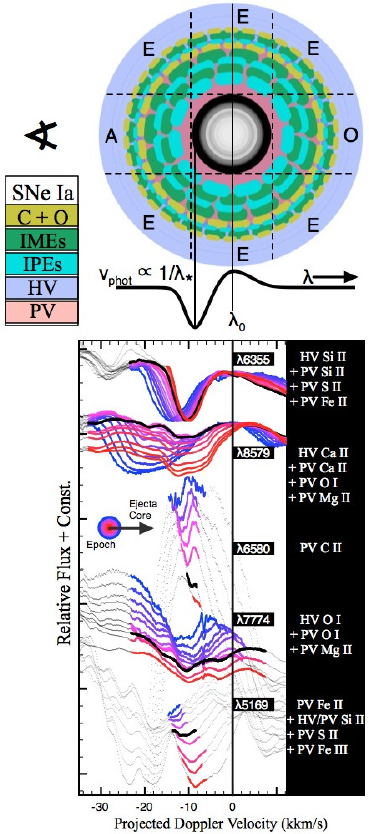}
\end{center}
\caption{{\it Top}: A schematic representation of how an assumed spherically sharp and embedded photosphere amounts to a pure line-resonance P Cygni profile under the conditions of Sobolev line transfer within a geometry of Absorbing, Emitting, and Occulted regions of material \citep{Jeffery90,Branch05}. The approximate photospheric velocity, $v_{phot}$, is proportional to the blue ward shift of an unblended absorption minimum. {\it Bottom}: Application of the above P Cygni diagram to SN~Ia spectra in terms of which species dominate and what other species are known to influence the temporal behavior \citep{Bongard08}, each of which are constrainable from complete spectroscopic coverage. For each series of spectra, the black line in bold represents maximum light.}
\label{fig:pcygfigb} 
\end{figure}

Even so, this rarely attainable observing strategy does not necessarily illuminate nor eliminate all degeneracies in spectral feature interpretations. However the advantage of complimentary high frequency follow-up observations is that the spectrum solution associated with any proposed explosion scenario can at least be consistently tested and constrained by the observed rapid changes over time (``abundance tomography'' goals, e.g., \citealt{Hauschildt99,Stehle05,Sauer06,Kasen06,Hillier12,Pauldrach13}). It then follows that hundreds of well-observed spectrophotometric datasets serve to carve out the characteristic information, $f(\lambda; t)$, for each SN~Ia between subtypes, in addition to establishing the perceived boundaries of the SN~Ia diversity problem (see Fig.~11 of \citealt{Blondin12} for this concept at maximum light). 

For supernovae in general, Figure~\ref{fig:pcygfiga} also serves as a reminder that all relative strengths evolve continuously over time, where entire features are always red-shifting across wavelength (line velocity space) during the rise and fall in brightness. A corollary of this situation is that prescriptions for taking measurements of spectroscopic behaviors (whereby interpretations rely on a subjective ``goodness of fit'') and robustly associating with any number of physical causes do not exist. Instead there are two primary means for interpreting SN~Ia spectra and taking measurements of features for the purposes of extracting physical properties. 

\emph{Indirect} analysis assumes a detailed explosion model and is primarily tasked with assessing the accuracy and flaws of the model. \emph{Direct} analysis seeks to manually measure via spectrum synthesis where one can either assume an initial post-explosion ejecta composition \emph{or} give up abundance information altogether to assess the associated uncertainties and consequences of supernova line blending via \emph{purposeful} high parameterizations. For the latter of these direct inference methods, the conclusions about spectroscopic interpretations$-$which are supported by remnants of inconsistencies throughout the literature$-$are summarized as follows.

For the most part, particularly at early epochs and as far as anyone can tell with current limiting datasets, the features in SN~Ia spectra are due to IMEs and IPEs formed by resonance scattering of
continuum and decay-chain photons, and have P Cygni-type profiles overall (\citealt{Pskovskii69,Mustel71,Branch73,Kirshner73}; see Figure~\ref{fig:pcygfigb}). Emission components peak at or near the rest wavelength and absorption components are blue-shifted according
to the opacity profile of matter at and above the photospheric line forming region. The combination of these effects can often lead to ``trumped'' emission features \citep{Jeffery90}, giving SN~Ia spectra their familiar shapes. 

Essentially all \emph{relevant} atomic species (isotope plus ionization state) are present somewhere within the ejecta, each with its own 3-dimensional abundance profile.  At optical wavelengths, conditions and abundance tomographies of the ejecta maintain the dominance of select singly$-$triply ionized subsets of C$+$O, IMEs, and IPEs \citep{HatanoAtlas}. From shortly after the onset of the explosion to around the time of maximum light, the optical$-$NIR spectrum of a normal
SN~Ia consists of a continuum level with superimposed features
that are primarily consistent with strong permitted lines of ions such as \ion{O}{1}, \ion{Mg}{2}, \ion{Si}{2}, \ion{Si}{3}, \ion{S}{2}, \ion{Ca}{2}, \ion{Fe}{2}, \ion{Fe}{3}, and trace signatures of \ion{C}{1} and \ion{C}{2} \citep{Branch06,Thomas07,Bongard08,Nugent11,Parrent12,Hsiao13,Mazzali13,Dessart13models}. After the pre-maximum light phase, blends of \ion{Fe}{2} (in addition to other IPEs) begin to dominate or influence the temporal behavior of many optical$-$NIR features over timescales from weeks to months (see \citealt{Branch08} and references therein).

With the above mentioned approximated view of line formation in mind (Figure~\ref{fig:pcygfigb}), the real truth is that the time-dependent state of the ejecta and radiation field \emph{at\ all\ locations} dictates how the material presence within the line forming regions will be imparted onto the spectral continuum, i.e. the radiation field and the matter are said to be ``coupled.'' With the additional condition of near-relativistic expansion velocities ($\sim$0.1$c$), line identifications themselves can also be thought of as coupled to the abundance tomography of ejected material, which includes the projected Doppler velocities spanned by the recipe of absorbing material. Subsequently, while spectra can be used for constraining limits of some model parameters, it comes with a cost of certainty on account of \emph{natural} uncertainties imparted by the large expansion velocities and associated expansion opacities. 

\begin{figure*}[htbp]
\begin{center}
\includegraphics[width=6.8in]{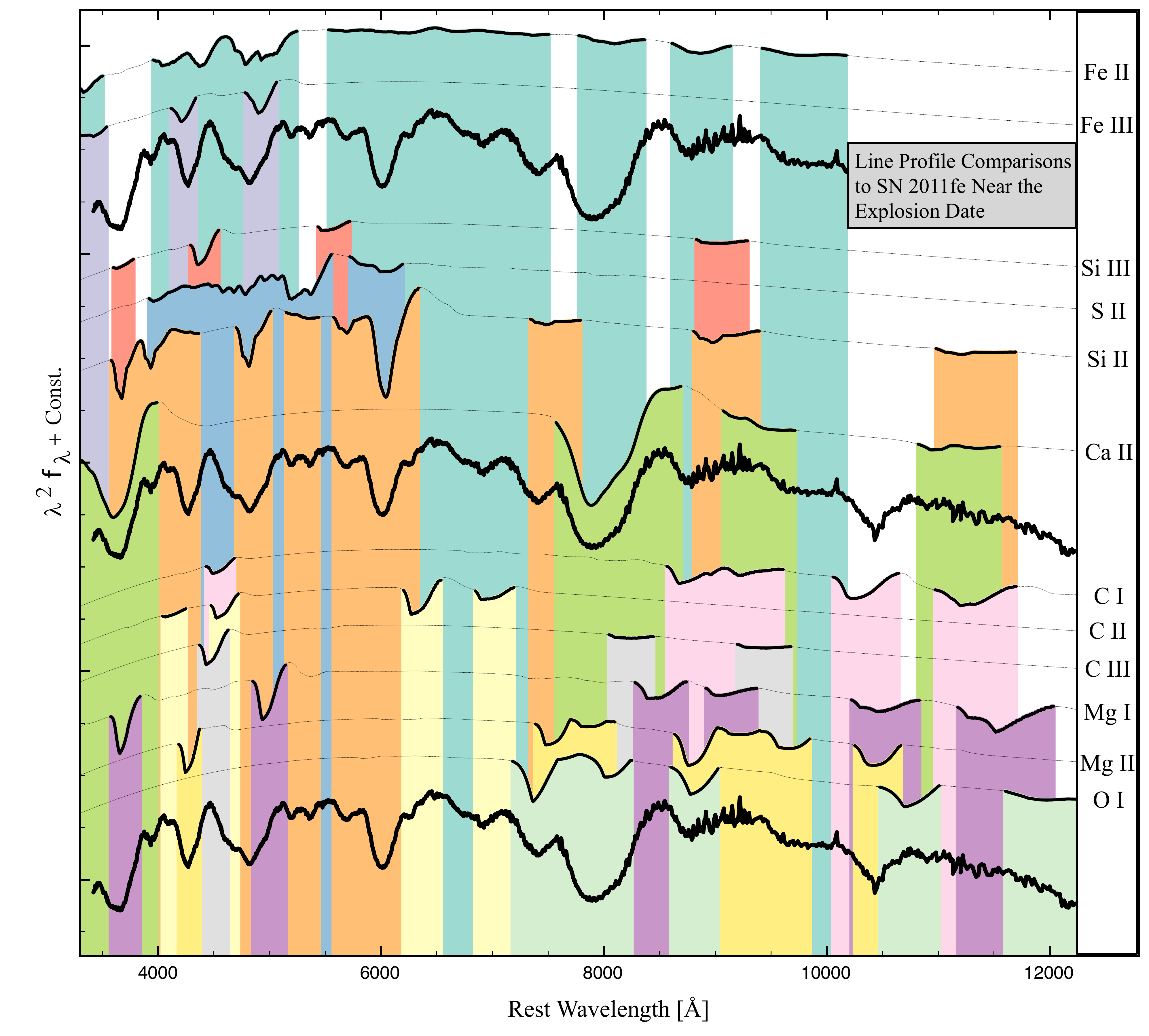}
\end{center}
\caption{\texttt{SYN++} calculation comparisons to the early optical$-$NIR spectrum of SN~2011fe \citep{Hsiao13,Pereira13}. Calculations are based on an optical set of photospheric phase spectra (see \citealt{Parrent12}) and are true-to-scale. Bands of color are intended to show overlap between lines under the simplified however informative assumption of permitted line scattering under homologous expansion. Some of the weaker lines have not been highlighted for clarity.}
\label{fig:lineblending101} 
\end{figure*}

As an exercise in this point, in Figure~\ref{fig:lineblending101} we have constructed an early epoch set of toy model line profiles that are representative of normal SN~Ia line identification procedures (e.g., \citealt{Branch05,Parrent11}) and over-plot them with an early optical$-$NIR spectrum (the observed outermost layers, sans UV) of SN~2011fe. We summarize the take away points of Figure~\ref{fig:lineblending101} as follows.

\begin{itemize}

\item Even without considering weak contributions, at no place along the (UV$-$) optical$-$NIR spectrum is any observed feature removed from being due to less than 2 sources (more precisely, see also \citealt{Bongard08}). That is, under the basic assumptions of pure resonance line scattering and homologous expansion (Figure~\ref{fig:pcygfigb}), all features are complex blends of at least 2$+$ ions and are universally influenced by multiple regions of emitting and/or absorbing material (e.g., ``high[-er] velocity'' and ``photospheric velocity'' intervals of material, see also \citealt{Marion13}). 

\item For supernovae, the components of the spectrum are most easily constrained via spectrum synthesis, and subsequently measurable (not the converse), when the bounds of wavelength coverage, $\lambda$$_{a}$ and $\lambda$$_{b}$, are between $\sim$2000$-$3500 and 12000 \AA, respectively. If $\lambda$$_{b}$ $<$ 7500$-$9500 \AA, then the velocities and relative strengths of several physically relevant ions (e.g., \ion{C}{1}, \ion{O}{1}, \ion{Mg}{2}, and \ion{Ca}{2}) are said to be devoid of useful constraint and provide a null (or uncertain) measurement for every other overlapping spectral line signature (i.e. all features). That is, in order to viably ``identify'' and measure a single feature, the entire spectrum must be reproduced. While empirical measurements of certain absorption features are extremely useful for identifying trends in the observed behavior of SN~Ia, these methods do not suffice to measure the truest underlying atomic recipe and its time-dependent behavior, much less the ``strength'' of contributing lines (e.g., multiple velocity components of \ion{Si}{2} in SN~2012fr, \S4.2.2). Specifically, empirical feature strength measurements at least require a proper modeling of the non-blackbody, IPE-dominated pseudo continuum level \citep{Bongard08} or the use of standardized relative strength parameters (e.g., \citealt{Childress13HVF}).

\item Therefore, as in Figure~\ref{fig:pcygfigb}, employing stacked Doppler velocity scaled time series spectra provides useful and timely first-order comparative estimates for when (epoch) and where (projected Doppler velocity) contributing ions appear, disappear, and span as the photospheric region recedes inward over time. 

\end{itemize}

We speak on this only to point out that even simple questions$-$particularly for homogeneous SN Ia$-$are awash in detection/non-detection ambiguities. However, it should be noted that a powerful exercise in testing uncertain line identifications and resolving complex blends can be done, in part, without the use of additional synthetic spectrum calculations. That is, by comparing a single observed spectrum to that of other well-observed SN~Ia, where the analysis of the latter offers a greater context for interpretation than the single spectrum itself, one can deduce whether or not a ``mystery'' absorption feature is common to most SN~Ia in general. On the other hand, if a matching absorption feature is not found, then one can infer the presence of either a newly identified, compositionally consistent ion or the unblended line of an already accounted for species (resulting from forbidden line emission, non-LTE effects, and/or when line strengths or expansion velocities differ between subtypes). Given also the intrinsic dispersion of expansion opacities between SN~Ia, it is likely that an ``unidentified'' feature is that of a previously known ion at higher and/or lower velocities. It is this interplay between expansion opacities and blended absorption features that keep normal and some peculiar SN~Ia within the description of a homogenous set of objects, however different they may appear. 

In fact, when one compares the time series spectra of a broad sample of SN~Ia subtypes, however blended, there is little room for degeneracy among plausible ion assignments (sans IPEs, e.g., \ion{Fe}{2} versus \ion{Cr}{2} during post-maximum phases). 
In other words, there exists a unique set of ions, common to most SN~Ia atmospheres, that make up the resulting spectrum, where differences in subtype are associated with differences in temperature and/or the abundance tomography of the outermost layers \citep{Tanaka08}. The atomic species listed in Figure~\ref{fig:lineblending101} do not so much represent a complete account of the composition, or the ``correct'' answer, as they are consistent with the subsequent time evolution of the spectrum toward maximum light, and therefore serve to construct characteristic standards for direct comparative diversity assessments. 

Said another way, it is the full time series dataset that enables the best initial spectrum solution hypothesis, which can be further tested and refined for the approximate measurement of SN~Ia features \citep{Branch07a}. Therefore, this idea of a unique set of ions remains open since$-$with current limiting datasets$-$species with minimal constraint \emph{and} competing line transfer processes can be ambiguously present\footnote[14]{See Fig.~9 of \citet{Stritzinger13} to see clear detections of permitted \ion{Co}{2} lines in the NIR spectra of the peculiar and faint SN~2010ae.}, even for data with an infinite S/N (i.e. sources with few strong lines, or lines predominately found blue ward of $\sim$6100 \AA, e.g., \ion{C}{3}, \ion{O}{3}, \ion{Si}{4}, \ion{Fe}{1}, \ion{Co}{2}, \ion{Ni}{2}). One can still circumvent these uncertainties of direct analysis by either using dense time series observations (e.g., \citealt{Parrent12}) or by ruling out spurious inferred detections by including adjacent wavelength regions into the spectroscopic analysis (UV$-$optical$-$NIR; see \citealt{Foley12b,Hsiao13,Mazzali13}).

\subsection{Models}\label{ss:spectramodels}

A detailed account of SN~Ia models is beyond the scope of our general review of SN~Ia spectra (for the latest discussions, see \citealt{WangB12,Nomoto13,Hillebrandt13,Calder13,Maoz13}). However, in order to understand the context by which observations are taken and synthetic comparisons made, here we only mention the surface layer of matters relating to observed spectra. For some additional recent modeling work, see \citet{Fryer08}, \citet{Bravo09}, \citet{Jordan09}, \citet{Kromer10}, \citet{Blondin11a}, \citet{Hachisu12a}, \citet{Jordan12}, \citet{Pakmor13}, \citet{Seitenzahl13}, \citet{Dan13}, \citet{Kromer13SN2010lp}, \citet{Moll13}, and \citet{Raskin13mergers}.

Realistic models are not yet fully ready because of the complicated physical
conditions in the binary stellar evolution that leads up to an expanding SN~Ia atmosphere. For instance, the explosive conditions of the SN~Ia problem take place over a large dynamic range of relevant length-scales (R$_{WD}$ $\sim$ 1R$_{\earth}$ and flame-thicknesses of $\sim$ 0.1 cm; \citealt{Timmes92,Gamezo99}), involve turbulent flames that are fundamentally multi-dimensional \citep{Khokhlov95,Khokhlov00,Reinecke02b,Reinecke02a,Gamezo03,Gamezo05,Seitenzahl13}, and consist of uncertainties in both the detonation velocity \citep{Dominguez11} and certain nuclear reaction rates, especially $^{12}$C$+$$^{12}$C (\citealt{Bravo11}, however see also \citealt{Bravo12,Chen13}).

Most synthetic spectra are angle-averaged representations of higher-dimensional detailed models. Overall, the observed spectra of normal SN~Ia have differed less amongst themselves than that of some detailed models compared to the data of \emph{normal} SN~Ia. This is not from a lack of efforts, but is simply telling of the inherent difficulty of the problem and limiting assumptions and interests of various calculations. \citet{Kasen08} reviewed previous work done of N-dimensional SN~Ia models and presented the first high-resolution 3D calculation of a SN~Ia spectrum at maximum light. Their results are still in a state of infancy, however they represent the first step toward the ultimate goal of SN~Ia modeling, i.e. to trace observed SN~Ia properties and infer the details of the progenitor and its subsequent disruption by comparing 3D model spectra and light curves of 3D explosion simulations with the best observed temporal datasets. 

Still, progress has been made in understanding general observed properties of SN~Ia and their relation to predictions of simulated explosion models. For example, one-dimensional (1D) numerical models of SN~Ia have been used in the past
 to test
the possible explosion mechanisms such as subsonic flame or supersonic detonation models, as well as conjoined delayed-detonations (e.g., \citealt{Arnett68,Nomoto84,Lentz01a}). The one-dimensional models disfavor the
route of a pure thermonuclear detonation as the mechanism to explain
most SN~Ia events \citep{Hansen69,Arnett69,Axelrod80}. Such a mechanism produces mostly $^{56}$Ni and almost
none of the IMEs observed in the spectra of all SN~Ia (e.g., \citealt{Branch82,Flipper97,Gamezo99,Pastorello07a}).

However, one-dimensional models have shown that a detonation \emph{can}
produce intermediate mass elements if it propagates through 
a Chandrasekhar-mass WD that has pre-expanded during an
initial deflagration stage \citep{Khokhlov91,Yamaoka92,Khokhlov93,Arnett94a,Arnett94b,Wheeler95,Hoflich95,Khokhlov97}. To their advantage, these deflagration-to-detonation transition
(DDT) and pulsating delayed-detonation (PDD) models \emph{are} able to reproduce the observed characteristics of
SN~Ia, however not without the use of an artificially-set transition density between stages of burning \citep{Khokhlov91PRD,Hoflich95,Lentz01a,Lentz01b,Baron08,Bravo09,Dessart13models}. Subsequently, a bulk of the efforts within the modeling community has been the pursuit of conditions or mechanisms which cause the burning front to naturally transition from a sub-sonic deflagration to a super-sonic detonation, e.g., gravitationally confined detonations \citep{Jordan09}, prompt detonations of merging WDs, a.k.a. ``peri-mergers'' \citep{Moll13}.

With the additional possibility that the effectively burned portion of the progenitor is enclosed or obscured by some body of circumstellar or envelope/disk of material (see \citealt{Sternberg11,Foley12dust,Forster12,Scalzo12,Raskin13,Silverman13IaCSM,Dan13,Dessart13models,Moll13}), the intrinsically multi-dimensional nature of the explosion itself is also expected to manifest signatures of asymmetric plumes of burned material and pockets of unburned material within a spheroidal debris field of flexible asymmetries (see \citealt{Khokhlov95,Niemeyer95,Gamezo04,WW08,Patat09,Kasen09}). Add to this the degeneracy of SN~Ia flux behaviors, i.e. colors are sensitive to dust/CSM extinction and intrinsic dispersions in the same direction \citep{Tripp99}, whether large or small redshift-color dependencies \citep{Saha99,Jha99,Parodi00,WangX08a,Goobar08,WangX09Subtype,Foley11,Mohlabeng13}, and we find the true difficulty in constraining SN~Ia models.

\citet{Blondin13} recently presented and discussed the photometric and spectroscopic properties at maximum light of a sequence of 1D DDT explosion models, with ranges of synthesized $^{56}$Ni masses between 0.18 and 0.81 M$_{\odot}$. In addition to showing broad consistencies with the diverse array of observed SN~Ia properties, the synthetic spectra of \citet{Blondin13} predict weaker absorption features of unburned oxygen (\ion{O}{1} $\lambda$7774) at maximum light, in proportion to the amount of $^{56}$Ni produced. This is to be expected \citep{Hoflich95}, however constraints on the remaining amount of unburned material, in addition to its temporal behavior, are more readily seen during the earliest epochs (within the outermost layers of ejecta) via \ion{C}{2} $\lambda$6580 and \ion{O}{1} $\lambda$7774 \citep{Thomas07,Parrent11,Nugent11}. Consequently, temporal spectrum calculations of detailed explosion models are needed for the purposes of understanding why the properties of SN~Ia are most divergent well before maximum light \citep{Branch06,Dessart13models}. 

Nucleosynthesis in two-dimensional (2D) delayed detonation models of SN~Ia were explored by \citet{Maeda10}. In particular, they focused on the distribution of species in an off-center DDT model and found the abundance tomography to be stratified, with an inner region of $^{56}$Ni surrounded by an off-center shell of electron-capture elements (e.g., Fe$^{54}$, Ni$^{58}$). Later, \citet{Maedanature} investigated the late time emission profiles associated with this off-center inner-shell of material within several observed SN~Ia and found a correlation between \emph{possible} nebular-line Doppler shifts along the line-of-sight and the rate-of-decline of \ion{Si}{2} velocities at earlier epochs. Their interpretation is to suggest that some SN~Ia subtypes may represent two different hemispheres of the ``same'' SN~Ia (LVG vs. HVG subtypes; see \S3.2). Moreover, the findings of \citet{Maedanature} and \citet{Maund10a} remain largely consistent with the additional early and late time observations of the well-observed SN~2011fe \citep{Smith11,McClelland13} and those of larger SN~Ia samples \citep{Blondin12,Silverman13latetime}. However, even the results of \citet{Maedanature} and others that rely on spectroscopic measurements at all epochs are not without reservation given that late time emission profiles are subject to more than line-shifts due to Doppler velocities and ionization balance \citep{Bongard08,Friesen12}.


\citet{Seitenzahl13} presented 14 3-dimensional (3D) high resolution Chandrasekhar-mass delayed-detonations that produce a range of $^{56}$Ni (depending on the location of ignition points) between $\sim$ 0.3 and 1.1 M$_{\odot}$. For this set of models, unburned carbon extends down to 4000 km~s$^{-1}$ while oxygen is not present below 10,000 km~s$^{-1}$. \citet{Seitenzahl13} conclude that if delayed-detonations are to viably produce normal SN~Ia brightnesses, the region of ignition cannot be far off-center so as to avoid the over-production of $^{56}$Ni. As noted by \citet{Seitenzahl13}, these models warrant tests via spectrum synthesis given their 3D nature and possible predictive relations to the WLR, spectropolarimetry, and C$+$O ``footprints'' \citep{Howell01a,Baron03,Thomas07,WW08}.

\citet{Dessart13models} recently compared synthetic light curves and spectra of a suite of DDT and PDD models. Based on comparisons to SN~2002bo and SN~2011fe, two SN Ia of different spectroscopic subtypes, and based on poor to moderate agreement between recent DDT models and observed SN Ia diversity \citep{Blondin11a}, \citet{Dessart13models} convincingly argue that these two SN Ia varieties (LVG vs. HVG, as above) are dissimilar enough to be explained by different explosion scenarios and/or progenitor systems \citep{WangX13}. For SN Ia in general, delineating spectroscopic diversity has been a difficult issue \citep{Benetti05,Branch09}, and has only recently been made clear with the belated release of decades-worth of unpublished data \citep{Blondin12,Silverman12spectra}.

\section{Spectroscopic Diversity of SN~Ia}\label{s:diversity}

Observationally and particularly at optical wavelengths, SN~Ia increase in brightness over $\sim$13 to 23 days before reaching maximum light ($\overline t_{rise}$ = 17.38 $\pm$ 0.17;  \citealt{Hayden10}). However, it is not until $\sim$1 year later that the period of observation is said to be ``complete.'' From the time of the explosion our perspective as outside observers begins at the outermost layers if the SN~Ia is caught early enough. In the approximate sense, this is because the line-forming region (the ``photosphere'') recedes as the ejecta expand outward, which in turn means that the characteristic information for each explosion mechanism and progenitor channel is specified by the temporal spectrophotometric attributes of the ``inner'' and ``outer'' layers of freshly synthesized and remaining primordial material. In addition, because the expanding material cools as it expands, the net flux of photons samples different layers (of different states and distributions) over time. And since the density profile of the material roughly declines from the center outward, significant changes within the spectra for an individual SN~Ia take place daily before or near maximum light, and weekly to monthly thereafter. 

Documenting the breadth of temporal spectroscopic properties for each SN~Ia is not only useful for theoretical purposes, but is also necessary for efficiently typing and estimating the epoch of newly found possible supernova candidates before they reach maximum light. Several supernova identification tools have been made that allow for fair estimates of both subtype and epoch (e.g., \texttt{SNID}; \citealt{BT07}, \texttt{Gelato}; \citealt{Harutyunyan08}, \texttt{Superfit}; \citealt{Howell05}). In addition, the spectroscopic goodness-of-fit methods of \citet{Jeffery07} allow one to find the ``nearest neighbors'' of any particular SN~Ia within a sample of objects, enabling the study of so called ``transitional subtype'' SN~Ia (those attributed with contrasting characteristics of two or more subtypes).

\subsection{Data}

\begin{figure*}[htbp]
\begin{center}
\includegraphics[width=4.in]{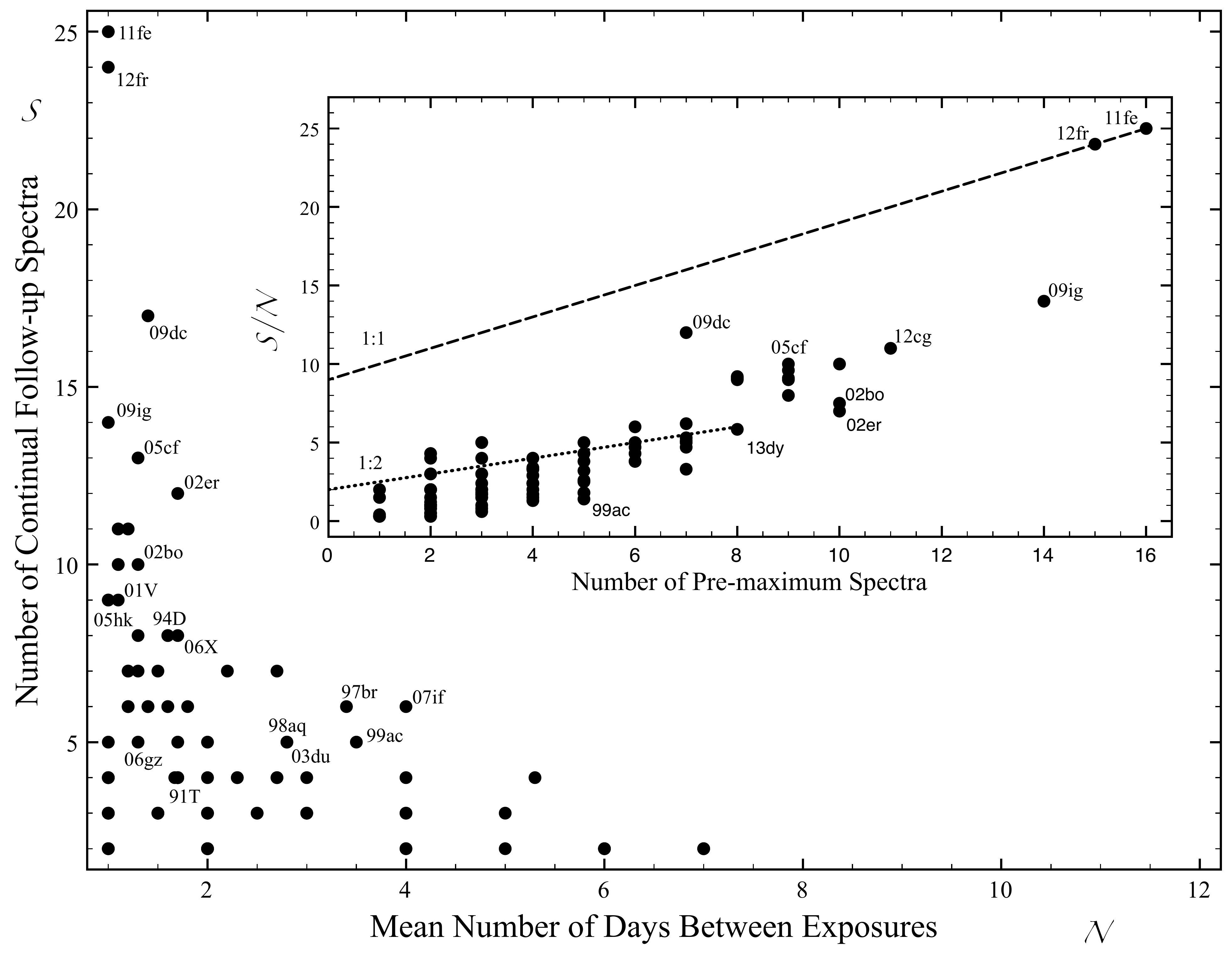}
\end{center}
\caption{Continual spectroscopic follow-up efficiencies for the most ``well-observed'' SN~Ia at early phases (not counting multiple spectra per day). Some of the values reported may be slightly lower for instances of unpublished data. Dashed lines represent the upper-limit spectroscopic efficiencies and peak number of pre-maximum light spectra for one and two day follow-up cadences during the first 25 days post-explosion. See \S3.1.}
\label{fig:SNratings} 
\end{figure*}

One of the major limitations of spectroscopic studies has been data quality. For example, the signal-to-noise ratio, S/N, of a spectrum signifies the quality across wavelength and is usually moderate to high for high-z events. Similarly, and at least for low-z SN~Ia, there should exist a quantity that specifies the density of spectra within a time series dataset. We suggest $\mathcal{S}$/$\mathcal{N}$$\bullet$($\mathcal{P}$) $\equiv$ the number of \emph{continual} follow-up spectra~/ the mean number of nights passed between exposures $\bullet$ (total number of spectra \emph{prior} to maximum light). In Figure~\ref{fig:SNratings} we apply this quantity to literature data.

An ideal dataset consisting of 25 spectra during the first 25 days post-explosion would yield $\mathcal{S}$/$\mathcal{N}$$\bullet$($\mathcal{P}$)~=~25 (16) (e.g., SN~2011fe), whereas a dataset of spectra at days $-$12, $-$10, $-$7, $-$4, $+$0, $+$3, $+$8, $+$21, $+$48, $+$119 (a common occurrence) would be said to have $\mathcal{S}$/$\mathcal{N}$$\bullet$($\mathcal{P}$)~=~3.3 (4) plus follow-up at days $+$21, $+$48, and $+$119. By including the total number of spectra prior to maximum light in parentheses, we are anticipating those cases where $\mathcal{S}$/$\mathcal{N}$ = 1, but with $\mathcal{P}$ = 3, e.g., a dataset with days $-$12, $-$9, and $-$6 observed. It may serve a purpose to also add second and third terms to this quantity that take into account the number of post-maximum light and late time spectra.

Regardless of moniker and definition, a quantity that specifies the density of spectra observed during the earliest epochs would aid in determining, quantitatively, which datasets are most valuable for various SN~Ia diversity studies. Clearly such a high follow-up rate for slow-evolving events (e.g., SN~2009dc) or events caught at maximum light are not as imperative. However, when SN~Ia are found and typed early, a high $\mathcal{S}$/$\mathcal{N}$ ensures no loss of highly time sensitive information, e.g., when high velocity features and C$+$O signatures dissipate. Since most datasets are less than ideal for detailed temporal inspections of many events (by default), astronomers have instead relied upon comparative studies (\S3.2); those that maximize sample sizes by prioritizing the most commonly available spectroscopic observables, e.g., line velocities of 6100 \AA\ absorption minima near maximum light. 

Another limitation of spectroscopic studies has been the localized release of all published data. The Online Supernova Spectrum Archive (SuSpect\footnote[15]{\url{http://suspect.nhn.ou.edu/~suspect/}}; \citealt{Richardson01}) carried the weight of addressing data foraging during the past decade, collecting a total of 867 SN~Ia spectra (1741 SN spectra in all). Many of these were either at the request of or donation to SuSpect, while some other spectra were digitized from original publications in addition to original photographic plates \citep{Casebeer98,Casebeer00}. Prior to and concurrent with SuSpect, D. Jeffery managed a collection of SUpernova spectra PENDing further analysis (SUSPEND\footnote[16]{\url{http://nhn.nhn.ou.edu/~jeffery/astro/sne/spectra/spectra.html}}).

With the growing need for a manageable influx of data, the Weizmann Interactive Supernova Data Repository (WISeREP\footnote[17]{\url{www.weizmann.ac.il/astrophysics/wiserep/}}; \citealt{WISEREP}) has since served as a replacement and ideal central data hub, and has increased the number of SN~Ia spectra to 7661 (with 7933 publicly available SN spectra out of 13,334 in all). We encourage all groups to upload published data to WISeREP, whether or not made available elsewhere.

\subsubsection{Samples}

By far the largest data releases occurred during the past five years, and are available on WISeREP and their affiliated archives. \citet{Matheson08} and \citet{Blondin12} presented 2603 optical spectra ($\sim$3700$-$7500 \AA\ on average) of 462 nearby SN~Ia ($\tilde{z}$ = 0.02; $\sim$ 85 Mpc) obtained by the Center for Astrophysics (CfA) SN group with the F. L. Whipple Observatory from 1993 to 2008. They note that, of the SN~Ia with more than two spectra, 313 SN~Ia have eight spectra on average. \citet{Silverman12distances} and the Berkeley SuperNova Ia Program (BSNIP) presented 1298 optical spectra ($\sim$3300$-$10,400 \AA\ on average) of 582 low-redshift SN~Ia (z $<$ 0.2; $\sim$ 800 Mpc) observed from 1989 to 2008. Their dataset includes spectra of nearly 90 spectroscopically peculiar SN~Ia. \citet{Folatelli13} released 569 optical spectra of 93 low-redshift SN~Ia ($\tilde{z}$ $\sim$ 0.04; $\sim$ 170 Mpc) obtained by the Carnegie Supernova Project (CSP) between 2004 and 2009. Notably, 72 CSP SN~Ia have spectra earlier than 5 days prior to maximum light, however only three SN~Ia have spectra as early as day $-$12. 

These samples provide a substantial improvement and crux by which to explore particular issues of SN~Ia diversity. However, the remaining limitation is that our routine data collection efforts continue to yield several thousand SN~Ia with few to several spectra by which to dissect and compare SN~Ia atmospheres.

\subsubsection{Comparisons of ``Well-Observed'' SN~Ia}

\begin{figure}
\begin{center}
\includegraphics[width=\linewidth]{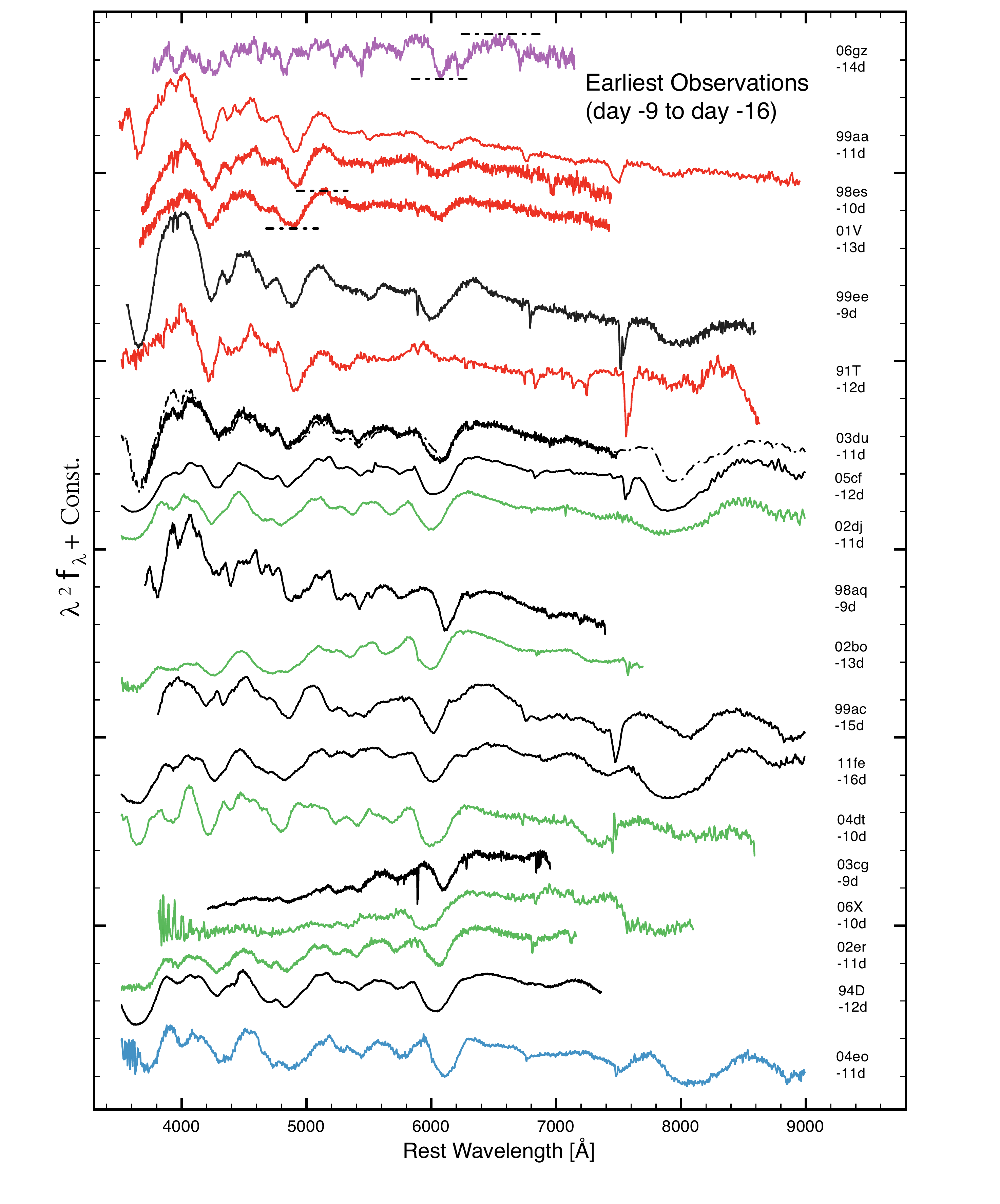}
\end{center}
\caption{Early pre-maximum light, rest frame optical spectra of some of the most well-observed and often referenced SN~Ia are plotted, loosely in order of increasing $\Delta$m$_{15}$(\emph{B}) (top-down). Subtypes shown include bright SN~2006gz, 2009dc-like super-Chandrasekhar candidate (SCC; purple), high-ionization, shallow-silicon SN~1991T-like (SS; red), normal SN~1994D, 2005cf, 2011fe-like (CN; black), broad-lined SN~1984A, 2002bo-like (BL; green), and sub-luminous, low-ionization SN~1991bg, 2004eo-like (CL; blue) SN~Ia. The horizontal dashed lines represent our normalization bounds that were applied to each spectrum. This ensures a fair comparison of all relevant spectroscopic features, sans continuum differences. For the SS SN~Ia, in Figure~\ref{fig:fig5} and Figure~\ref{fig:fig6} only, we have normalized to the \ion{Fe}{3} feature as indicated. For the purposes of this review, we have only included SN~Ia that have received particular attention within the literature (see \S4 and the Appendix). Many other time series observations can be found in \citet{Matheson08}, \citet{Silverman12spectra}, and \citet{Blondin12}. The peculiar PTF09dav is shown in Figure~\ref{fig:fig8} for comparison, as it is not a prototypical SN~Ia, however appearing similar to SN~1991bg-like events \citep{Sullivan10,Kasliwal12}.}
\label{fig:fig5} 
\end{figure}

\begin{figure}
\begin{center}
\includegraphics[width=\linewidth]{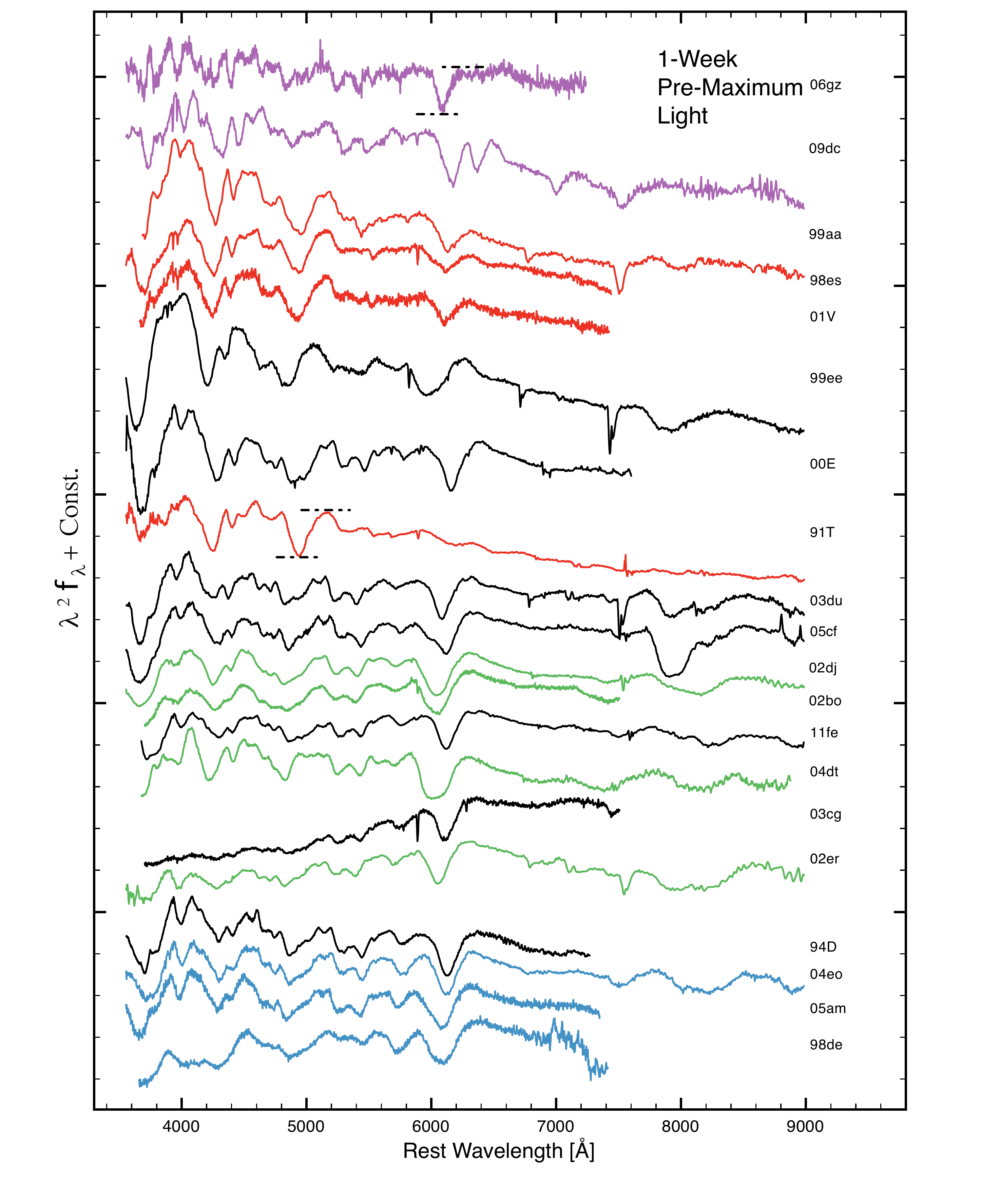}
\end{center}
\caption{1-week pre-maximum light optical spectroscopic comparisons. See Figure~\ref{fig:fig5} caption. }
\label{fig:fig6} 
\end{figure}

\begin{figure}
\begin{center}
\includegraphics[width=\linewidth]{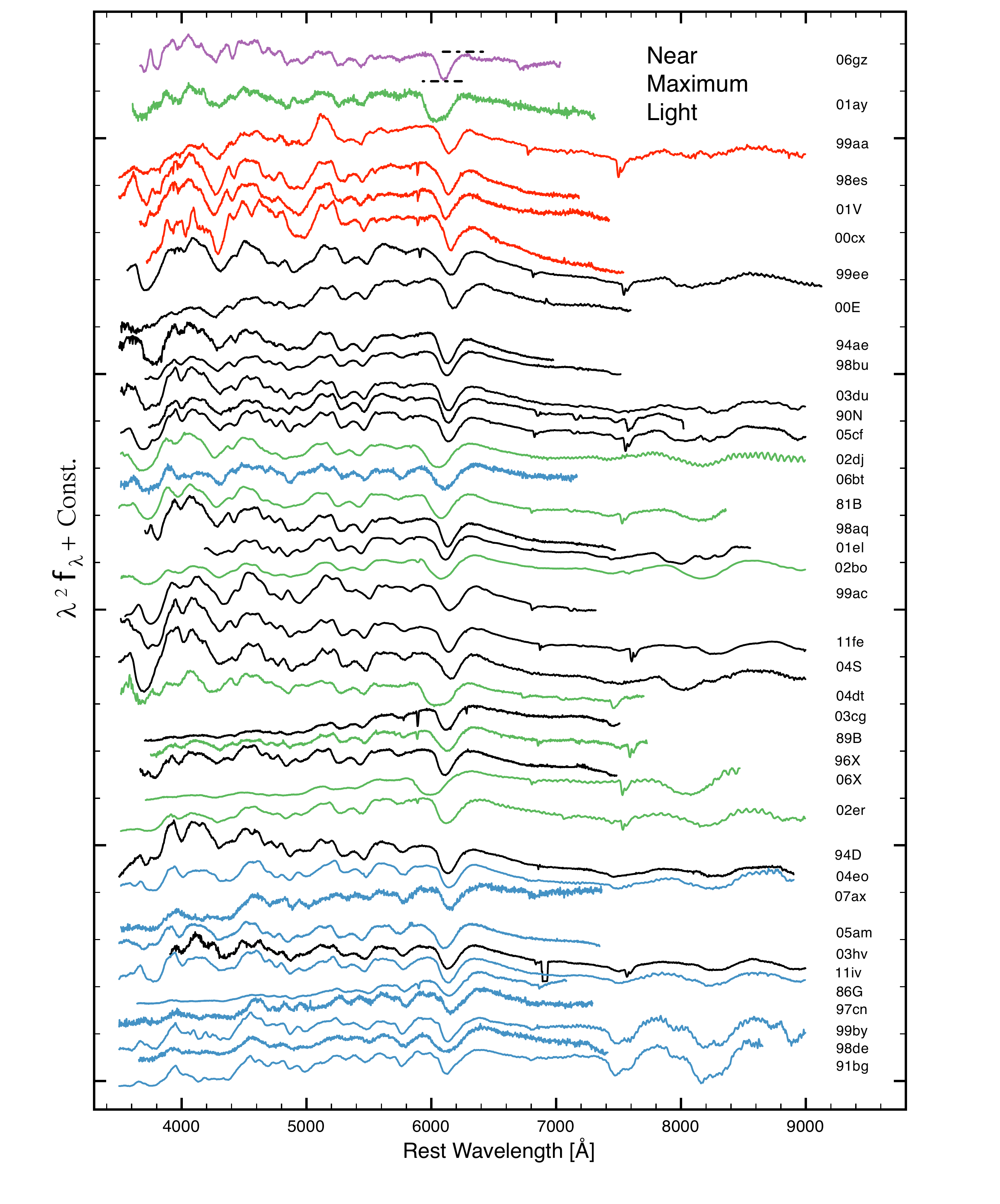}
\end{center}
\caption{Maximum light optical spectroscopic comparisons. See Figure~\ref{fig:fig5} caption. }
\label{fig:fig7} 
\end{figure}

\begin{figure}
\begin{center}
\includegraphics[width=\linewidth]{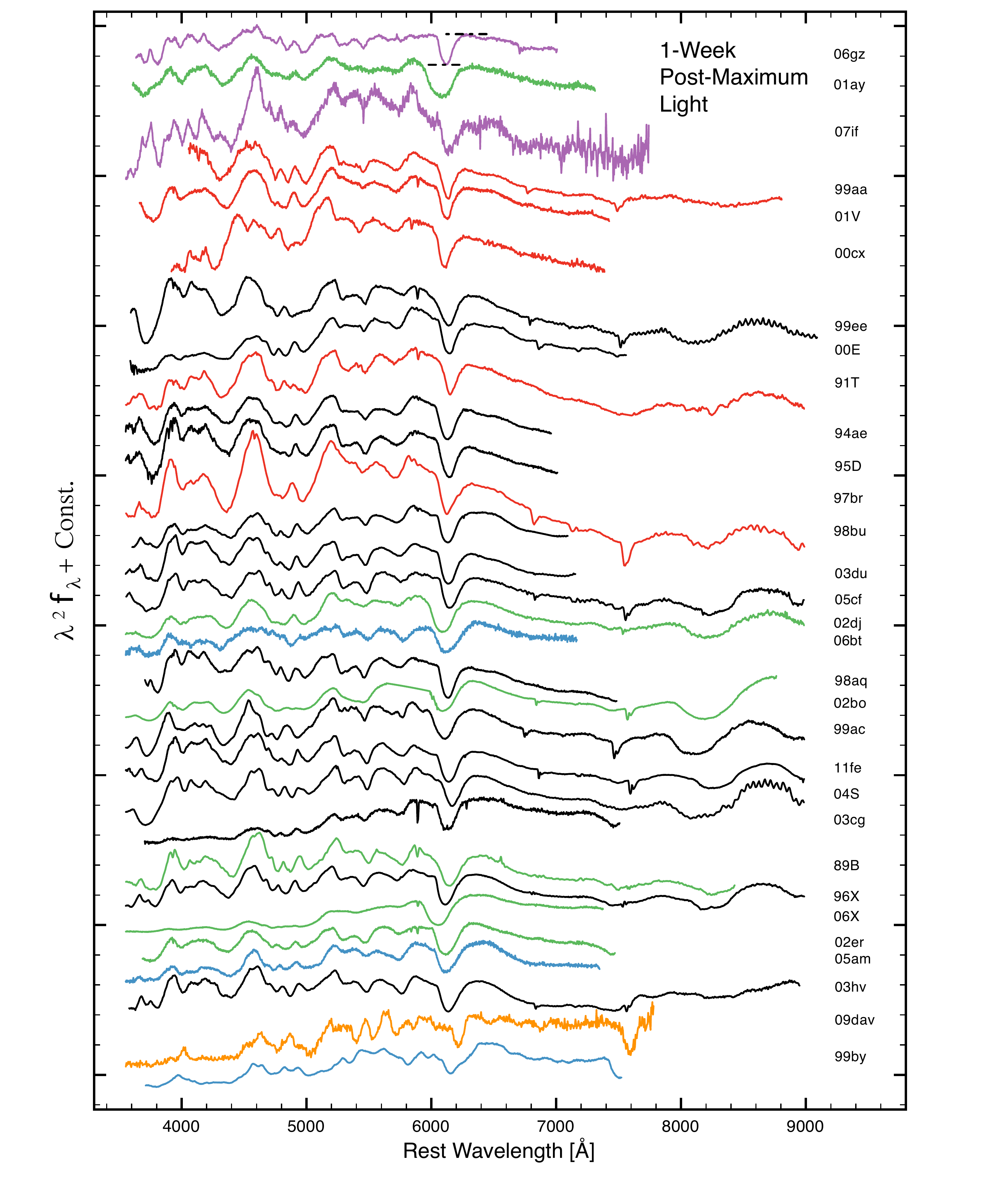}
\end{center}
\caption{1-week post-maximum light optical spectroscopic comparisons. See Figure~\ref{fig:fig5} caption.}
\label{fig:fig8} 
\end{figure}

\begin{figure}
\begin{center}
\includegraphics[width=\linewidth]{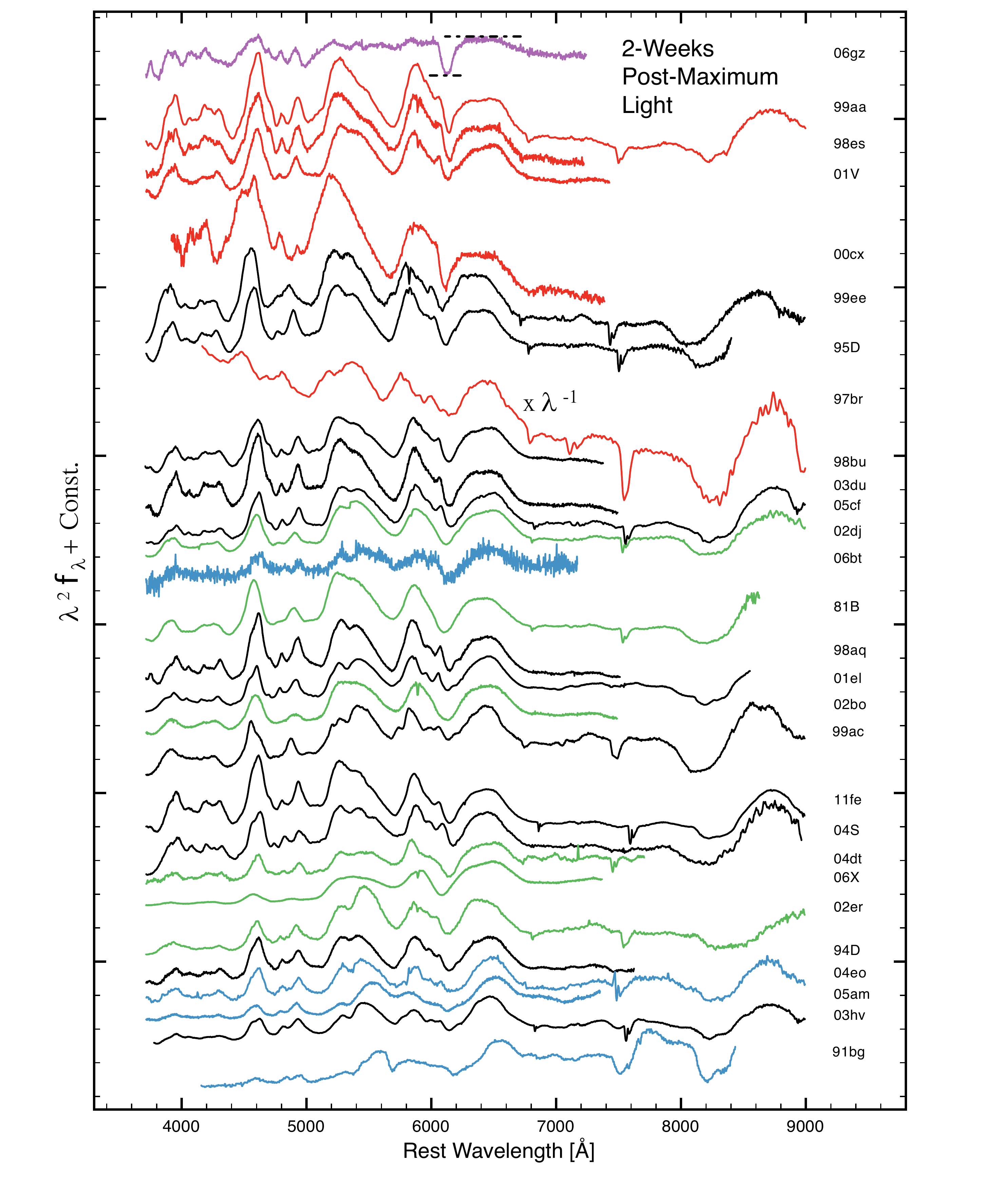}
\end{center}
\caption{two weeks post-maximum light optical spectroscopic comparisons. See Figure~\ref{fig:fig5} caption. }
\label{fig:fig9} 
\end{figure}

\begin{figure}
\begin{center}
\includegraphics[width=\linewidth]{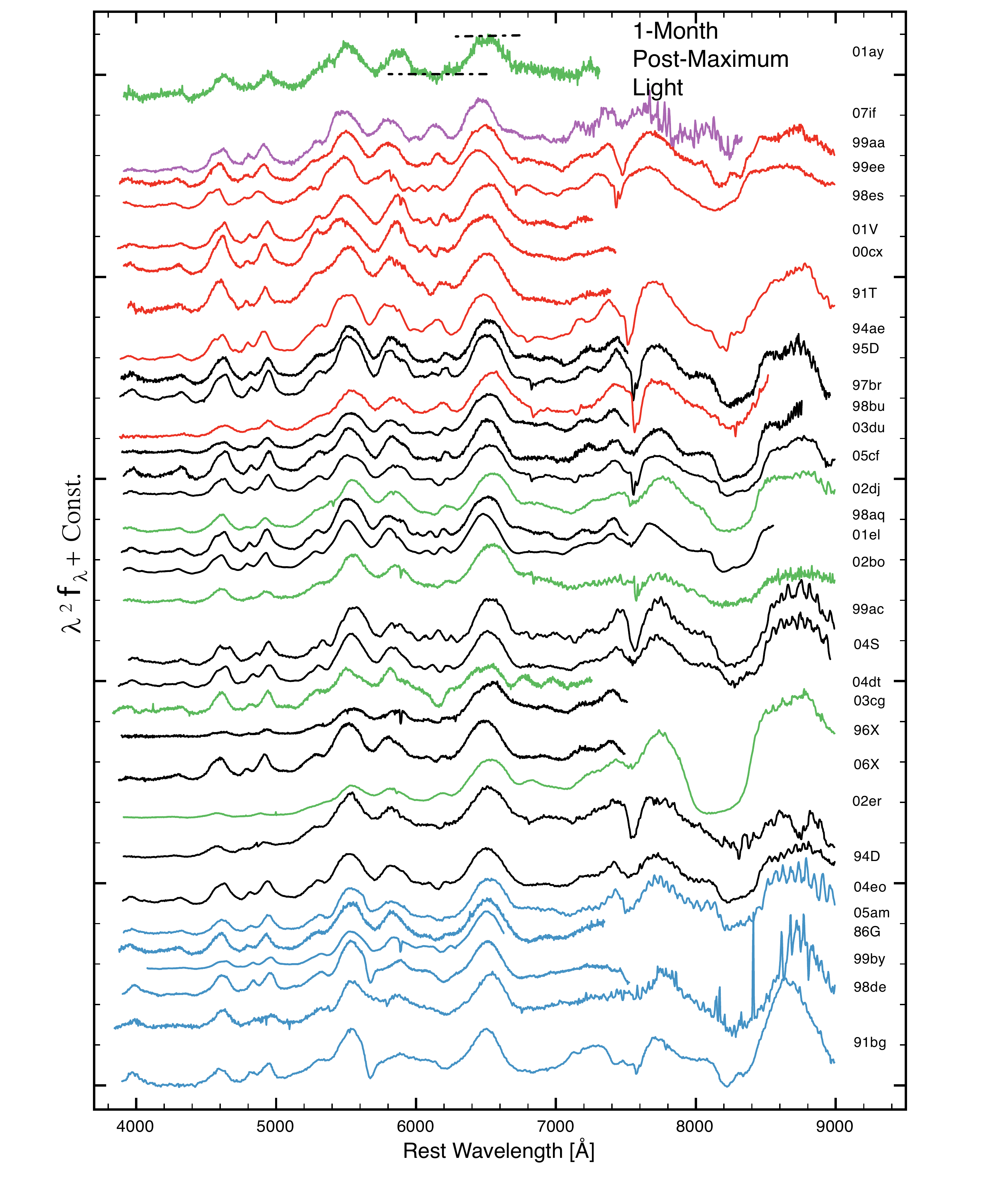}
\end{center}
\caption{1-month post-maximum light optical spectroscopic comparisons. See Figure~\ref{fig:fig5} caption. }
\label{fig:fig10} 
\end{figure}

\begin{figure}
\begin{center}
\includegraphics[width=\linewidth]{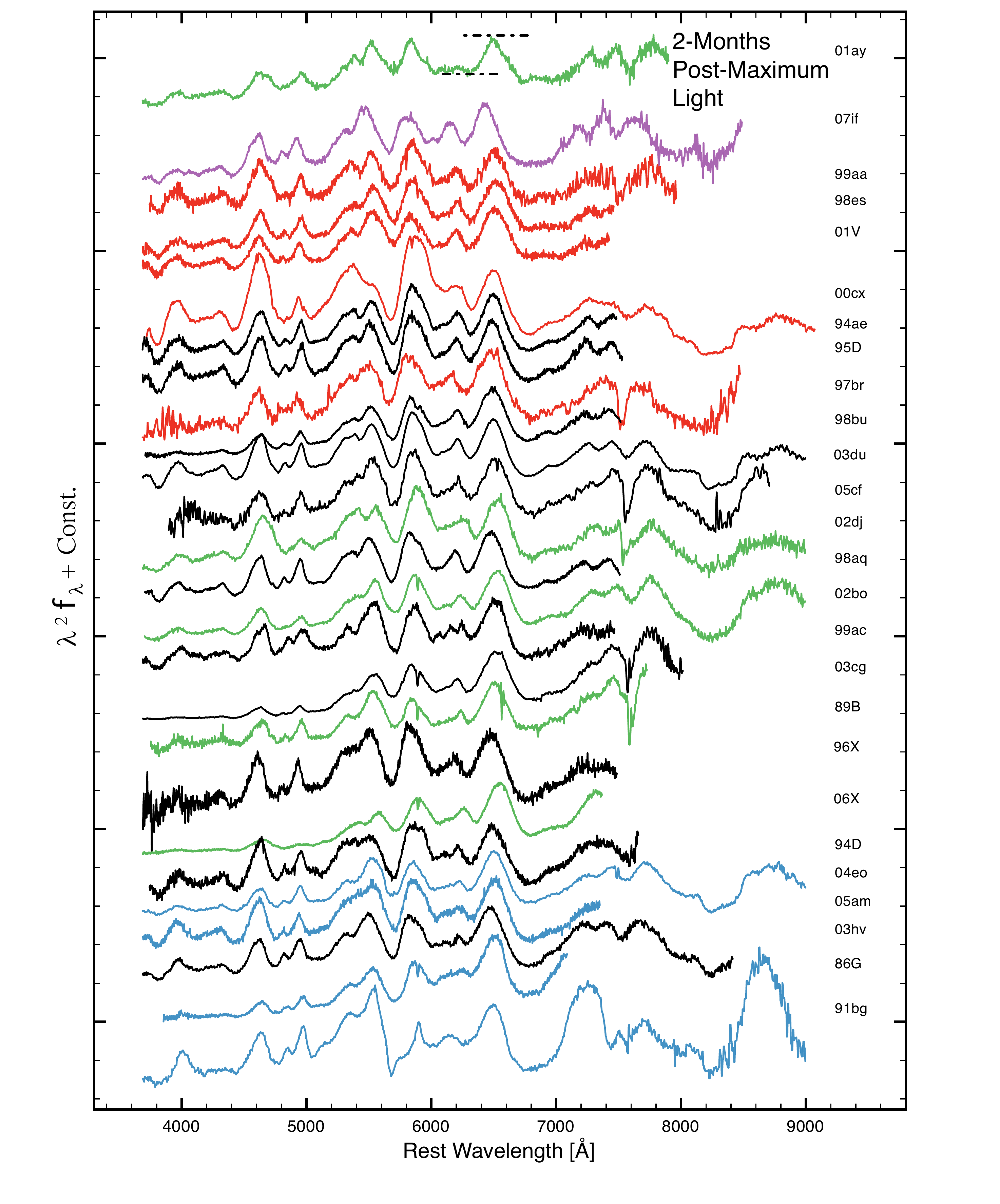}
\end{center}
\caption{2-months post-maximum light optical spectroscopic comparisons. See Figure~\ref{fig:fig5} caption. }
\label{fig:fig11} 
\end{figure}

\begin{figure}
\begin{center}
\includegraphics[width=\linewidth]{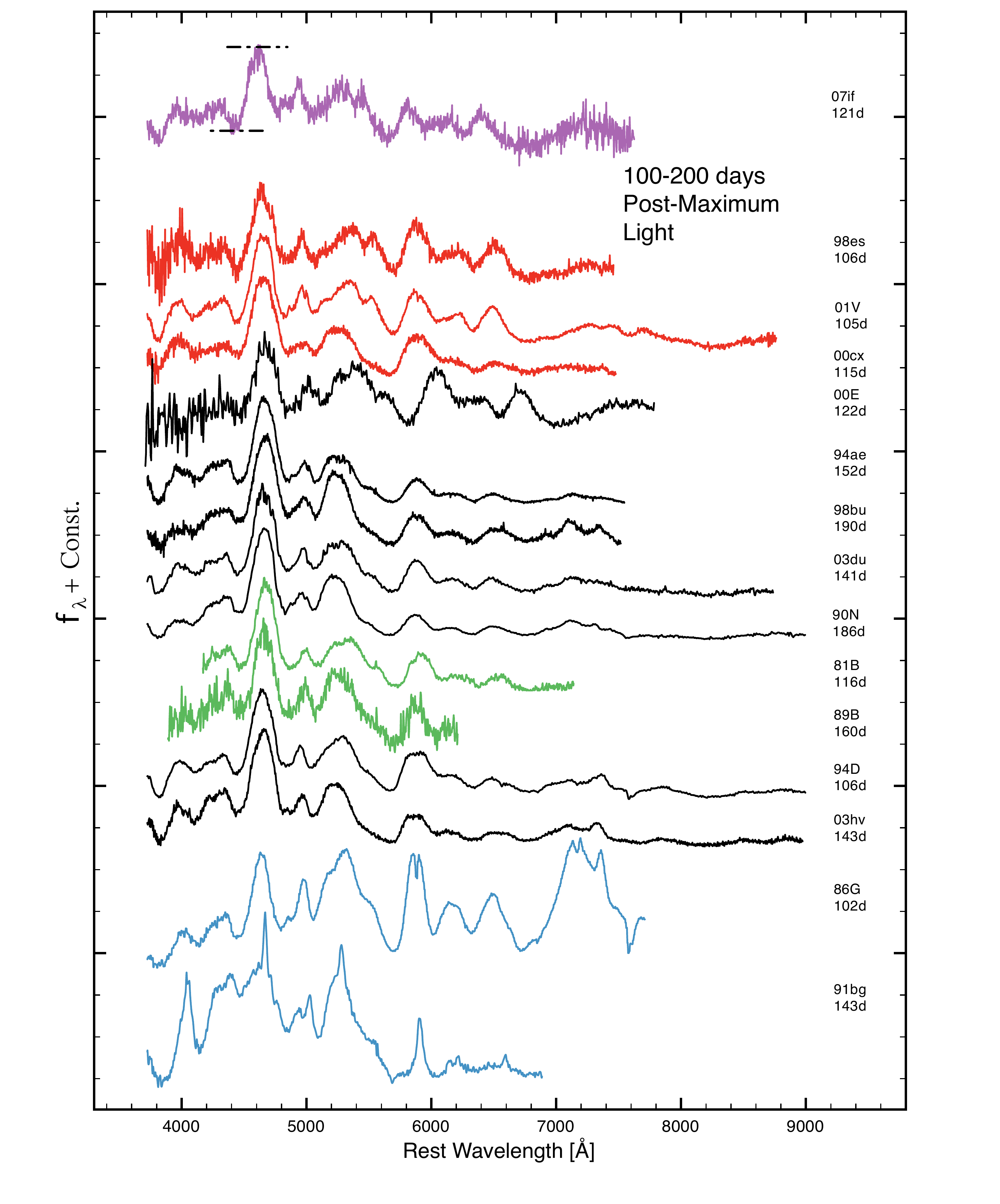}
\end{center}
\caption{100$+$ days post-maximum light optical spectroscopic comparisons. See Figure~\ref{fig:fig5} caption. }
\label{fig:fig12} 
\end{figure}

\begin{figure}
\begin{center}
\includegraphics[width=\linewidth]{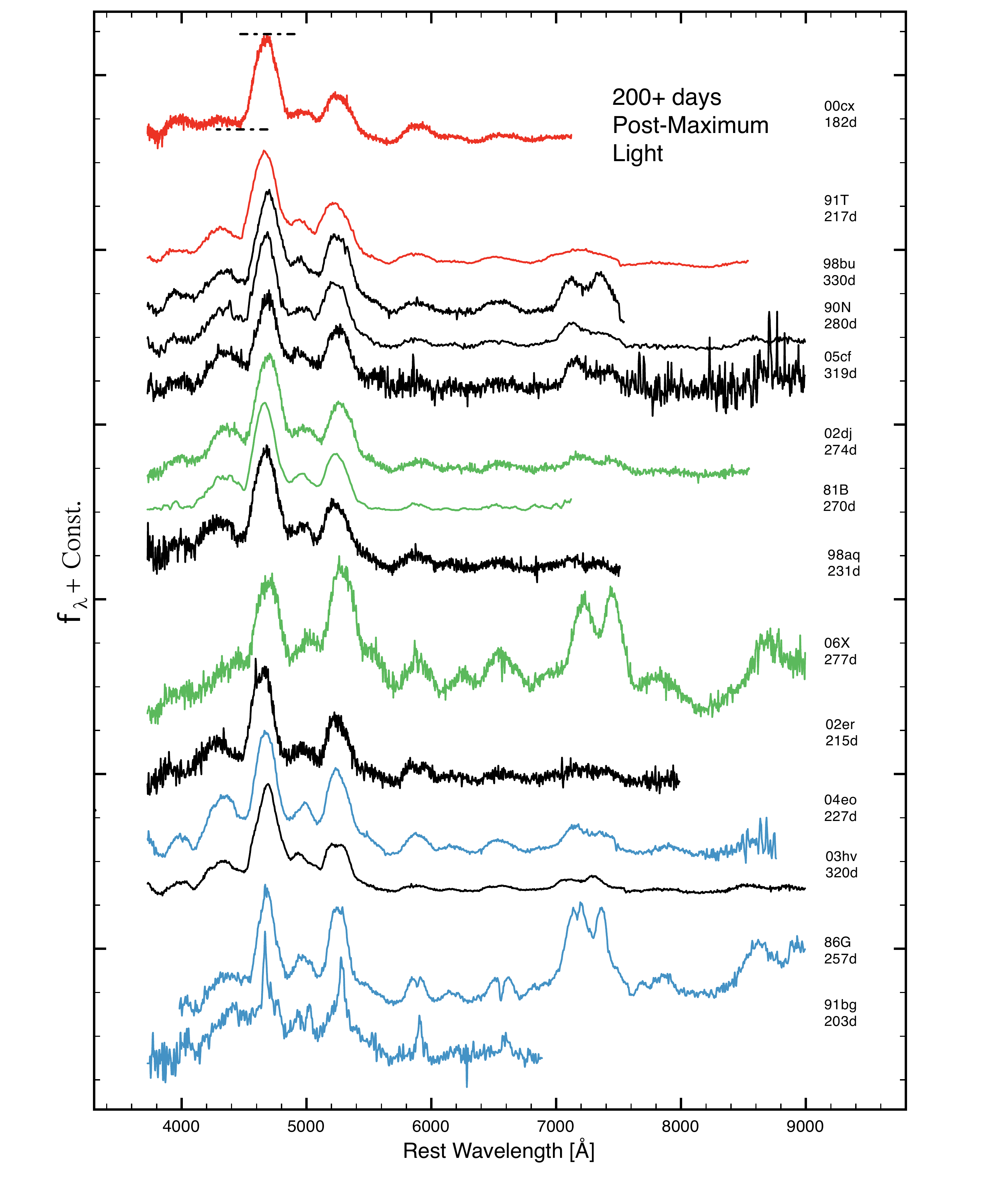}
\end{center}
\caption{Late-time optical spectroscopic comparisons. See Figure~\ref{fig:fig5} caption. }
\label{fig:fig13} 
\end{figure}

Given that both quantitative and qualitative spectrum comparisons are at the heart of SN~Ia diversity studies, in Figure~\ref{fig:fig5} - Figure~\ref{fig:fig13} we plot spectroscopic temporal snapshots for as many ``well-observed'' SN~Ia as are currently available on WISeREP (Tables 1 and 2). Because the decline parameter, $\Delta$m$_{15}$(\emph{B}), remains a useful parameter for probing differences of synthesized $^{56}$Ni mass, properties of the ejecta, limits of CSM interaction, etc., we have \emph{loosely} ordered the spectra with increasing $\Delta$m$_{15}$(\emph{B}) (top-down) based on average values found throughout the literature (Tables 3$-$6) and M$_{\emph{B}}$(peak) considerations for cases that are reported as having the same $\Delta$m$_{15}$(\emph{B}). The spectra have been normalized with respect to 6100 \AA\ line profiles in order to amplify relative strengths of the remaining features (see caption of Figure~\ref{fig:fig5}). We also denote the spectroscopic subtype for each object in color in order to show the overlap of these properties between particular SN~Ia subclasses (see \S3.2 and \citealt{Blondin12}). 

By inspection, the collected spectra show how altogether different and similar SN~Ia (both odd and normal varieties) have come to be since nearly 32 years ago. With regard to the recent modeling of \citet{Blondin13} and their accompanying synthetic spectra, we plot the spectra in Figure~\ref{fig:fig5} - Figure~\ref{fig:fig11} in the flux-representation of $\lambda^{2}$F$_{\lambda}$ for ease of future comparisons. These juxtapositions should reveal the severity of the SN~Ia diversity problem as well as the future of promising studies and work that lie ahead.  

\subsection{Deciphering 21$^{st}$ Century SN~Ia Subtypes}\label{ss:subtypeproperties}

Observationally, the whole of SN~Ia are hetero-, homogeneous events \citep{Oke74,Flipper97}; some of the observed differences in their spectra are clear, while other suspected differences are small enough to fall below associable certainty. Because of this, observational studies have concentrated on quantitatively organizing a mapping between the most peculiar and normal events. In this section our aim is to review SN~Ia subtypes. In all, three observational classification schemes will be discussed \citep{Benetti05,Branch06,WangX09Subtype}, as well as the recent additions of so-called over- and sub-luminous events (see \citealt{Scalzo12}, \citealt{Foley13}, \citealt{Silverman13IaCSM} and references therein). For other relatively new and truly peculiar subclasses of supernova transients, we refer the reader to \citet{Shen10}, \citet{Kasliwal12} and references therein.

\subsubsection{\citet{Benetti05} Classification}\label{sss:specphotclasses}

Understanding the origin of the WLR is a key issue for understanding the diversity of SN~Ia light curves and spectra, as well as their use as cosmological distance indicators. Brighter SN~Ia with broader light curves
tend to occur in late-type spiral galaxies,
while dimmer, faster declining SN~Ia are preferentially located
in an older stellar population and thus the age and/or metallicity of the progenitor
system may be relevant factors affecting SN~Ia properties (\citealt{Hamuy95,Howell01b,Pan13}, see also \citet{Hicken09morphology}).

    With this in mind, \citet{Benetti05} studied the observational
properties of 26 well-observed SN~Ia (e.g., SN~1984A, 1991T, 1991bg, 1994D) with the intent of exploring
SN~Ia diversity. Based on the observed projected Doppler velocity evolution from the spectra\footnote[18]{The velocity gradient$-$the mean velocity decline rate $\Delta$v/$\Delta$t$-$of a particular absorption minimum (e.g., \.{v}$_{Si\ II}$) has been redefined to be measured over a fixed phase range~[t$_{0}$,~t$_{1}$] \citep{Blondin12}.}, in conjunction with characteristics of the light curve (M$_{B}$,~$\Delta$m$_{15}$), \citet{Benetti05} considered three different groups of SN~Ia: (1)~``FAINT'' SN~1991bg-likes, (2)~``low velocity gradient'' (LVG) SN~1991T/1994D-likes, and (3)~``high velocity gradient'' (HVG) SN~1984A-like events. The velocity gradient here is based on the time-evolution of 6100 (``6150'') \AA\ absorption minima as inferred from \ion{Si}{2} $\lambda$6355 line velocities. Overall, HVG SN~Ia have higher mean expansion velocities than FAINT and LVG SN~Ia, while LVG SN~Ia are brighter than FAINT and HVG SN~Ia on average \citep{Silverman12maxlight,Blondin12}. Given an apparent separation of SN~Ia subgroups from this sample of 26 objects, \citet{Benetti05} considered it as evidence that LVG, HVG, and FAINT classifications signify three distinct kinds of SN~Ia. 
 
 \subsubsection{\citet{Branch06} Classification}\label{sss:specclasses}
 
\citet{Branch05,Branch06,Branch07b,Branch08,Branch09} published a series of papers
based on systematic, comprehensive, and comparative direct analysis of normal and peculiar SN~Ia spectra at various epochs with the parameterized supernova synthetic spectrum code, \texttt{SYNOW}\footnote[19]{\texttt{SYNOW} is a simplified spectrum synthesis code used for the timely determination and measurement of all absorption features complexes. \texttt{SYNOW} has been updated (\texttt{SYN++}) and can be used as an automated spectrum fitter (\texttt{SYNAPPS}; see \citealt{ThomasSYNAPPS} and \url{https://c3.lbl.gov/es/}).} \citep{Fisher00,Branch07a}. From the systematic analysis of 26 spectra of SN~1994D, \citet{Branch05} infer a compositional structure that is radially stratified, overall. In addition, several features are consistent with being due to permitted lines well into the late post-maximum phases ($\sim$120 days, see \citealt{Branch08,Friesen12}). Another highlight of this work is that, barring the usual short comings of the model, \texttt{SYNOW} is shown to provide a necessary consistency in the \emph{direct} quantification of spectroscopic diversity \citep{Branch07a}. Consequently, the \texttt{SYNOW} model has been useful for assessing the basic limits of a spectroscopic ``goodness of fit'' (Figure~\ref{fig:lineblending101}), with room for clear and obvious improvements \citep{Friesen12}.

In their second paper of the series on
comparative direct analysis of SN~Ia spectra, \citet{Branch06} studied the spectra
of 24 SN~Ia close to maximum light. Based on empirical pEW measurements of 5750, 6100 \AA\ absorption features, in addition to spectroscopic modeling with \texttt{SYNOW}, \citet{Branch06} organized SN~Ia diversity by four spectroscopic patterns: (1)~``Core-Normal'' (CN) SN~1994D-likes, (2)~``Broad-line'' (BL), where one of the most extreme cases is
SN~1984A, (3)~``Cool'' (CL) SN~1991bg-likes, and (4)~``Shallow-Silicon'' (SS) SN~1991T-likes. In this manner, a particular SN~Ia is defined by its spectroscopic similarity to one or more SN~Ia prototype via 5750, 6100 \AA\ features. These spectroscopic subclasses also materialized from analysis of pre-maximum light spectra \citep{Branch07b}.

The overlap between both \citet{Benetti05} and \citet{Branch06} classifications schemes comes by comparing Table 1 in \citet{Benetti05} to Table 1 of \citet{Branch06}, and it reveals the following SN~Ia descriptors: HVG$-$BL, LVG$-$CN, LVG$-$SS, and FAINT$-$CL. This holds true throughout the subsequent literature \citep{Branch09,Folatelli12,Blondin12,Silverman12maxlight}. 

In contrast with \citet{Benetti05} who interpreted FAINT, LVG, and HVG to correspond to the ``discrete grouping'' of \emph{distinctly separate} SN Ia origins among these subtypes, \citet{Branch06} found a continuous distribution of properties between the four subclasses defined above. We should point out that this classification scheme of \citet{Branch06} is primarily tied to the notion that SN~Ia spectroscopic diversity is related to the temperature sequence found by \citealt{Nugent95a}. That is, despite the contrast with \citet{Benetti05} (continuous versus discrete subgrouping of SN~Ia), so far these classifications say more about the state of the ejecta than the various number of possible progenitor systems and/or explosion mechanisms (see also \citealt{Dessart13models}). Furthermore, the existence of ``transitional'' subtype events support this notion (e.g., SN~2004eo, 2006bt, 2009ig, 2001ay, and PTF10ops; see appendix).

\citet{Branch09} later analyzed a larger sample of SN~Ia spectra. They found that
SN~1991bg-likes are not a physically distinct subgroup \citep{Doull11}, and that there are probably many SN~1999aa-like events (A.5) that similarly may not constitute a physically distinct variety of SN~Ia. 

With regard to the fainter variety of SN~Ia, \citet{Doull11} made detailed comparative analysis of spectra of peculiar
SN~1991bg-likes.  They also
studied the intermediates, such as SN~2004eo (A.23), and discussed the
spectroscopic subgroup distribution of SN~Ia. The CL SN~Ia
are dim, undergo a rapid decline in luminosity, and produce significantly
less $^{56}$Ni than normal SN~Ia. They also have an unusually deep
and wide trough in their spectra around 4200 \AA\, suspected as due to \ion{Ti}{2} \citep{Flipper92}, in addition to a relatively
strong 5750 \AA\ absorption (due to more than \ion{Si}{2} $\lambda$5972; see \citealt{Bongard08}). \citet{Doull11}
analyzed the spectra of SN~1991bg, 1997cn, 1999by,
and 2005bl using \texttt{SYNOW}, and found this group of SN~Ia
to be fairly homogeneous, with many of the blue spectral features well
fit by \ion{Fe}{2}.

\subsubsection{\citet{WangX09Subtype} Classification}\label{sss:singleparamclasses}

Based on the maximum light expansion velocities inferred from \ion{Si}{2} $\lambda$6355 absorption minimum line velocities, \citet{WangX09Subtype} studied 158 SN~Ia, separating them into two groups called ``high velocity'' (HV) and ``normal velocity'' (NV). This classification scheme is similar to those previous of \citet{Benetti05} and \citet{Branch06}, where NV and HV SN~Ia are akin to LVG$-$CN and HVG$-$BL SN~Ia, respectively. That is, while the subtype notations differ among authors, memberships between these classification schemes are roughly equivalent (apart from outliers such as the HV-CN SN~2009ig, see \citealt{Blondin12}). 

Explicitly, \citet{Benetti05} and \citet{WangX09Subtype} subclassifications are based on empirically estimated mean expansion velocities near maximum light ($\pm$4 days; $\pm$500 $-$ 2000 km~s$^{-1}$) of 6100 \AA\ features produced by an assumed single broad component of \ion{Si}{2}. The notion of a single photospheric layer, much less a single-epoch snapshot, does not realistically account for the multilayered nature of spectrum formation \citep{Bongard08}, its subsequent evolution post-maximum light \citep{Patat96,Scalzo12}, and potential relations to line-of-sight considerations \citep{Maedanature,Blondin11a,Moll13}. In the strictest sense of SN~Ia sub-classification, ``normal'' refers to \emph{both} of these subtypes since they differ foremost by a continuum of inferred mean expansion velocities and the extent of expansion opacities, simultaneously.

Furthermore, note from a sample of 13 LVG and 8 HVG SN~Ia that \citet{Benetti05} found 10 $\lesssim$ \.{v}$_{Si\ II}$ (km~s$^{-1}$ day$^{-1}$) $\lesssim$ 67 ($\pm$7) and 75 $\lesssim$ \.{v}$_{Si\ II}$ $\lesssim$ 125 ($\pm$20) for each, respectively. Similarly, and from a sample of 14 LVG and 29 HVG SN~Ia, \citet{Silverman12maxlight} report that 10 $\lesssim$ \.{v}$_{Si\ II}$ $\lesssim$ 445 ($\pm$50) and 15 $\lesssim$ \.{v}$_{Si\ II}$ $\lesssim$ 290 ($\pm$140) for LVG and HVG events, respectively. Additionally, the pEW measurements of 5750, 6100 \AA\ absorption features (among others) are seen to share a common convergence in observed values \citep{Branch06,Hachinger06,Blondin12,Silverman12maxlight}. The continually consistent overlap between the measured properties for these two SN~Ia ``subtypes'' implies that the notion of a characteristic separation value for \.{v}$_{Si\ II}$ $\sim$ 70 km~s$^{-1}$ day$^{-1}$ (including the inferred maximum light separation velocity, v$_{0}$~$\gtrsim$ 12,000 km~s$^{-1}$) is still devoid of any physical significance beyond overlapping bimodal distributions of LVG$-$CN and HVG$-$BL SN~Ia properties (see \S5.3 of \citealt{Silverman12maxlight}, \S5.2 of \citealt{Blondin12}, and \citealt{Silverman12distances}). Rather, a continuum of \emph{empirically\ measured} properties exists between the extremities of these two particular \emph{historically-based} SN~Ia classes (e.g., SN~1984A and 1994D). Given also the natural likelihood for a physical continuum between NV and HV subgroups, considerable care needs to be taken when concluding on underlying connections to progenitor systems from under-observed, early epoch snapshots of blended 6100 \AA\ absorption minima.

Hence, the primary obstacle within SN~Ia diversity studies has been that it is not yet clear if the expanse of all observed characteristics of each subtype has been fully charted. For the observed properties of normal SN~Ia, it is at least true that \.{v}$_{Si\ II}$ resides between 10$-$445 km~s$^{-1}$ day$^{-1}$, with a median value of $\sim$ 60$-$120 km~s$^{-1}$ day$^{-1}$ \citep{Benetti05,Blondin12,Silverman12distances}, while the rise to peak \emph{B}-band brightness ranges from 16.3 to 19 days \citep{Ganeshalingam11,Mazzali13}.

Recently, \citet{WangX13} applied this NV and HV subgrouping to 123 ``Branch normal'' SN~Ia with known positions within their host galaxies and report that HV SN~Ia more often inhabit the central and brighter regions of their hosts than NV SN~Ia. This appears to suggest that a supernova with ``higher velocities at maximum light'' is primarily a consequence of a progenitor with larger than solar metallicities, or that PDD/HVG SN Ia are primarily found within the galactic distribution of DDT/LVG SN Ia (c.f. \citealt{Blondin11a,Blondin12,Dessart13models}). This is seemingly in contrast to interpretations of \citet{Maedanature} who propose, based on both early epoch and late time considerations, that LVG and HVG SN~Ia are possibly one in the same event where the LVG-to-HVG transition is ascribed to an off-center ignition. 

While it is true that increasing the C$+$O layer metallicity can affect the blueshift of the 6100 \AA\ absorption feature$-$in addition to lower temperatures and increased UV line-blocking$-$this is not primarily responsible for the shift in 6100 \AA\ absorption minima \citep{Lentz01a,Lentz01b}, where the dependence of this effect is not easily decoupled from changes in the temperature structure \citep{Lentz00}. However, it is also worthwhile to point out that, while the early epoch spectra of SN~2011fe (a NV event) are consistent with a DDT-like composition with a sub-solar C$+$O layer metallicity (``W7$^{+}$,'' \citealt{Mazzali13}) \emph{and} a PDD-like composition \citep{Dessart13models}, the outermost layers of SN~2010jn (a HV event; A.41) are practically void of unburned material and subsequently already overabundant in synthesized metals for progenitor metallicity to be well determined \citep{Hachinger13}. Therefore, discrepancies between NV and HV SN~Ia must still be largely dependent on more than a single parameter, e.g. differences in explosion mechanisms \citep{Dessart13models,Moll13}, where progenitor metallicity is likely to be only one of several factors influencing the dispersions of each subgroup \citep{Lentz00,Hoflich10,WangX12}.

It should be acknowledged again that metallicity-dependent aspects of stellar evolution are expected to contribute, in part, to the underlying variance of holistic SN~Ia characteristics. However thus far, the seen discrepancies from metallicities share similarly uncertain degrees of influence as for asymmetry and line-of-sight considerations of ejecta-CSM interactions for a wide variety of SN~Ia \citep{Lentz00,Kasen03,Leloudas13}. Similar to this route of interpretation for SN~Ia subtypes are active galactic nuclei and the significance of the broad absorption line quasi-stellar objects (BALQSOs, see \citealt{deKool95,Becker97,Elvis00,Branch02FeLowBAL,Hamann04,Casebeer08,Leighly09,Elvis12}).

\subsubsection{Additional Peculiar SN~Ia Subtypes}

Spectroscopically akin to some luminous SS SN~Ia are a growing group of events thought to be ``twice as massive,'' aka super-Chandrasekhar candidates (SCC, \citealt{Howell06,Jeffery06,Hillebrandt07,Hicken07,Maeda09,Chen09,Yamanaka09a,Scalzo10,Tanaka10,Yuan10,Silverman11,Taubenberger11,Kamiya12,Scalzo12,Hachinger12,Yamanaka13}). Little is known about this particular class of over-luminous events, which is partly due to there having been only a handful of events studied. Thus far, SCC SN~Ia are associated with metal-poor environments \citep{Childress11,Khan11metalpoor}. Spectroscopically, the differences that set these events apart from normal SN~Ia are fairly weak \ion{Si}{2}/\ion{Ca}{2} signatures and strong \ion{C}{2} absorption features relative to the strength of \ion{Si}{2} lines. Most other features are comparable in relative strengths to those of normal SN~Ia, if not muted by either top-lighting or effects of CSM interaction \citep{Branch00,Leloudas13}, and are less blended overall due to lower mean expansion velocities. In addition, there is little evidence to suggest that SCC SN~Ia spectra consist of contributions from physically separate high velocity regions of material ($\gtrsim$ 4000 km~s$^{-1}$ above photospheric). This range of low expansion velocities ($\sim$5000$-$18,000 km~s$^{-1}$), in conjunction with larger than normal \ion{C}{2} absorption signatures, are difficult to explain with some M$_{Ch}$ explosion models (\citealt{Scalzo12,Kamiya12}, however see also \citealt{Hachisu12a,Dessart13models,Moll13} for related discussions).

\citet{Silverman13IaCSM} recently searched the BSNIP and PTF datasets, in addition to the literature sample, and compiled a list of 16 strongly CSM interacting SN~Ia (referred to as ``Ia-CSM'' events). These supernovae obtain their name from a conspicuous signature of narrow hydrogen emission atop a weaker hydrogen P Cygni profile that together are superimposed on a loosely identifiable SS-like SN~Ia spectrum \citep{Aldering06,Prieto07,Leloudas13}. Apart from exhibiting similar properties to the recent PTF11kx (\S4.5.1) and SN~2005gj (\S4.5.3), \citet{Silverman13IaCSM} find that SN~Ia-CSM have a range of peak absolute magnitudes ($-$21.3 $\le$ M$_{\emph{R}}$ $\le$ $-$19), are a spectroscopically homogenous class, and all reside in late-type spiral and irregular host-galaxies.

As for peculiar sub-luminous events, \citet{Narayan11} and \citet{Foley13} discussed the heterogeneity of the SN~2002cx-like subclass of SN~Ia. Consisting of around 25 members spectroscopically similar to SN~2002cx \citep{WLi03}, these new events generally have lower maximum light velocities
 spanning from 2000 to 8000 km~s$^{-1}$ and a range of peak luminosities that are typically lower than those of FAINT SN~Ia ($-$14.2 to $-$18.9). In addition, this class of objects have ``hot'' temperature structures and$-$in contrast to SN~Ia that follow the WLR$-$have low luminosities for their
 light curve shape.  This suggests a distinct origin, such as a failed deflagration of a C$+$O white dwarf \citep{Foley09,Jordan12,Kromer13} or double detonations of a sub-Chandrasekhar mass white dwarf with non-degenerate helium star companion \citep{Fink10,Sim12,BoWang13}. It is  estimated that for every 100 SN~Ia, there are 31\ce{^{$+$17}_{$-$13}} peculiar SN~2002cx-like objects in a given volume \citep{Foley13}. 
 
\subsubsection{SN~Ia Subtype Summary}

\begin{figure}
\begin{center}
\includegraphics[width=\linewidth, trim = 0mm 40mm 235mm 50mm]{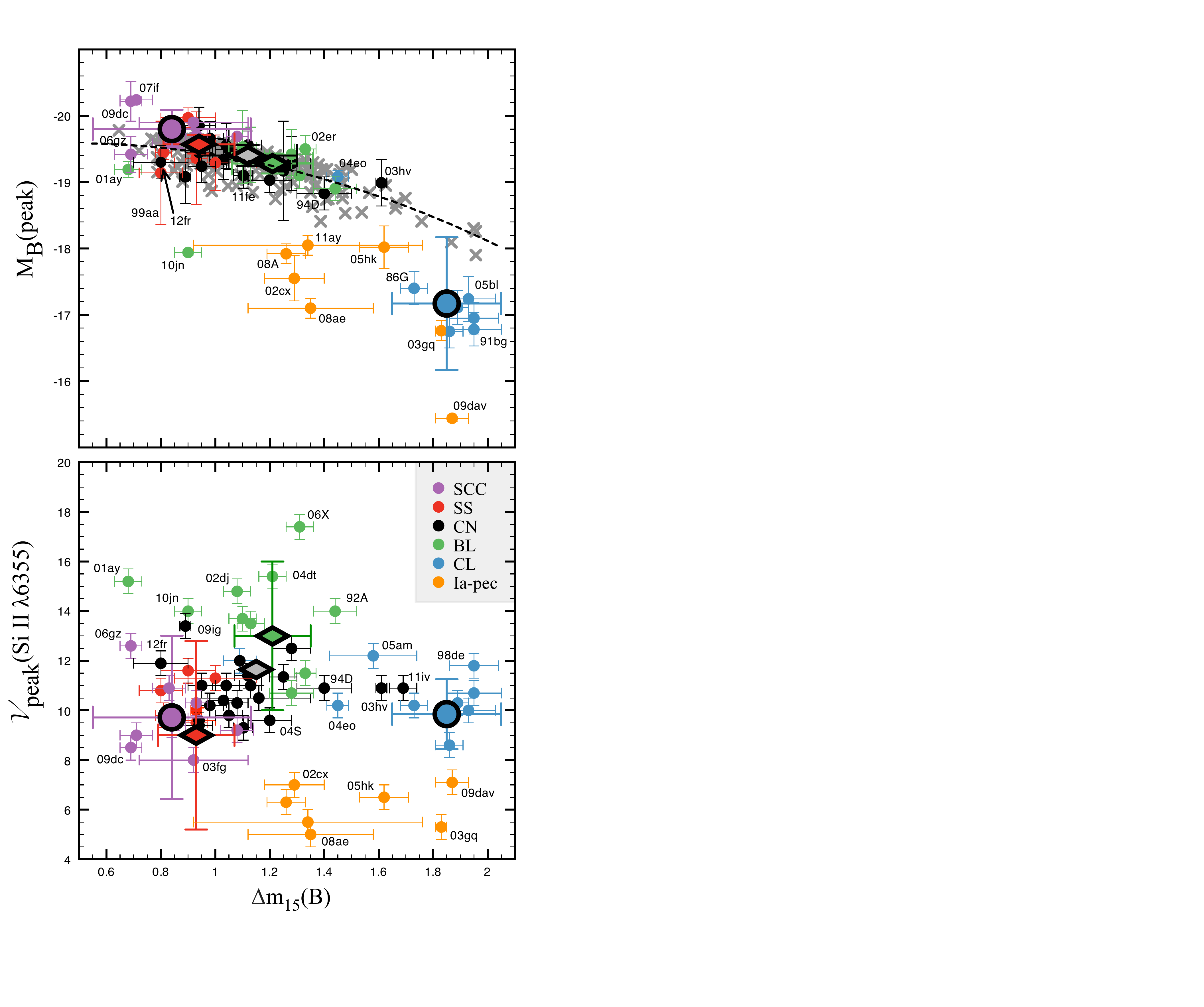}
\end{center}
\caption{{\it Top}: Peak absolute \emph{B}-band magnitudes versus $\Delta$m$_{15}$(\emph{B}) for most well-observed SN~Ia found in the literature. Additional data (grey) taken from \citet{Folatelli12}, \citet{Blondin12}, and additional points discussed in \citet{Pakmor13}. {\it Bottom}: Expansion velocities at maximum light ($\pm$ 3 days; via \ion{Si}{2} $\lambda$6355 line velocities) versus $\Delta$m$_{15}$(\emph{B}). All subtypes have been tagged in accordance with the same color-scheme as in Figure~\ref{fig:fig5} - Figure~\ref{fig:fig13}. Included for reference are the brightest, peculiar SN~2002cx-likes (light blue circles). Outliers for each subtype have been labeled for clarity and reference. We also plot mean values for the SCC and CL subtypes (larger circles), and include the mean values for SS, CN, and BL SN~Ia (large diamonds) as reported by \citet{Blondin12}.}
\label{fig:dm15} 
\end{figure}

In Figure~\ref{fig:dm15} we plot average literature values of M$_{\emph{B}}$(peak), $\Delta$m$_{15}$(\emph{B}), and $\mathcal{V}$$_{peak}$(\ion{Si}{2} $\lambda$6355) versus one another for all known SN~Ia subtypes. For M$_{\emph{B}}$(peak) versus $\Delta$m$_{15}$(\emph{B}), the WLR is apparent. We have included the brightest SN~2002cx-likes \citep{Foley13} for reference, as these events are suspected as having separate origins from the bulk of normal SN~Ia \citep{Hillebrandt13}. We have not included Ia-CSM events given that estimates of expansion velocities and luminosities, without detailed modeling, are obscured by CSM interaction. However, it suffices to say for Figure~\ref{fig:dm15} that Ia-CSM are nearest to SS and SCC SN~Ia in both projected Doppler velocities and peak M$_{\emph{R}}$ brightness \citep{Silverman13IaCSM}. At a separate end of these SN~Ia diversity plane(s), $\mathcal{V}$$_{peak}$(\ion{Si}{2} $\lambda$6355) versus $\Delta$m$_{15}$(\emph{B}) further separates FAINT$-$CL SN~Ia and peculiar events away from the pattern between SCC/SN~1991T-like over-luminous SN~Ia and normal subtypes, where the former tend to be slow-decliners (i.e. typically brighter) with slower average velocities. 

\begin{figure}[Htbp]
\begin{center}
\includegraphics[width=2in, trim = 55mm 5mm 55mm 0mm]{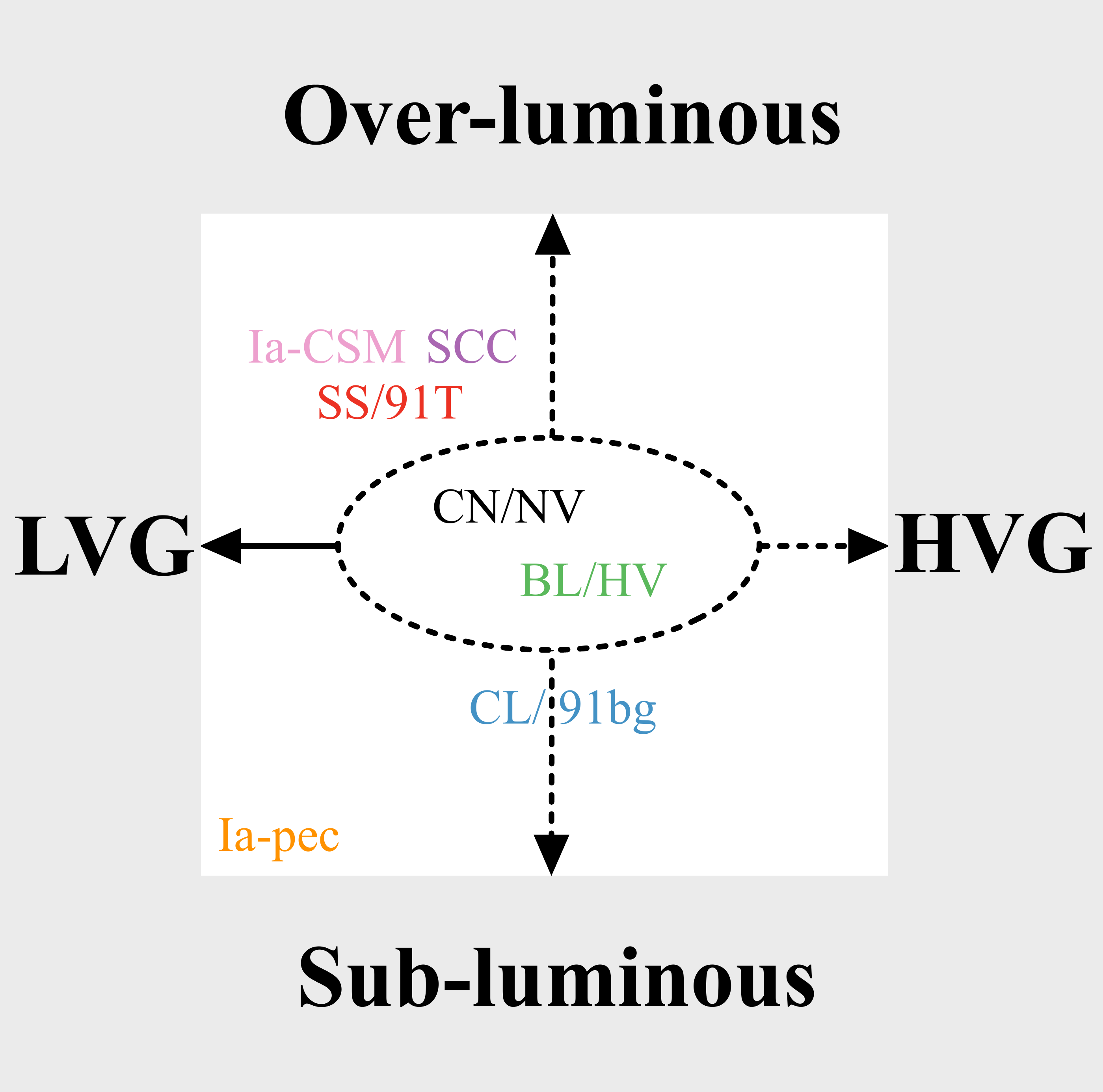}
\end{center}
\caption{Subtype reference diagram. Dashed lines denote an open transitional boundary between adjacent spectroscopic subtypes.}
\label{fig:aid} 
\end{figure}

To summarize the full extent of SN~Ia subtypes in terms of the qualitative luminosity and expansion velocity patterns, in Figure~\ref{fig:aid} we have outlined how SN~Ia relate to one another thus far (for quantitative assessments, see \citealt{Blondin12,Silverman12maxlight,Folatelli13}). Broadly speaking, the red ward evolution of SN~Ia features span low to high rates of decline for a large range of luminosities. Shallow Silicon and Super-Chandrasekhar Candidate SN~Ia are by far the brightest, while Ia-CSM SN exhibit bright H$\alpha$ emission features. These ``brightest'' SN~Ia also show low to moderate expansion velocities and \.{v}$_{Si\ II}$. From BL to CN to SS/SSC SN~Ia, mean peak absolute brightnesses scale up with an overall decrease in maximum light line velocities. Meanwhile, CL SN~Ia fall between low velocity and high velocity gradients, but lean toward HVG SN~Ia in terms of their photospheric velocity evolution. Comparatively, peculiar SN~2002cx-like and other sub-luminous events are by far the largest group of thermonuclear outliers.

Obtaining observations of SN~Ia that lie outside the statistical norm is important for gauging the largest degree by which SN~Ia properties diverge in nature. However, just as imperative for the cause remains filling the gaps of observed SN~Ia properties (e.g., \.{v}$_{Si\ II}$, v$_{neb}$, v$_{C}$(t), v$_{Ca}$(t), M$_{\emph{B}}$(peak), $\Delta$m$_{15}$(\emph{B}-band), t$_{rise}$, color evolution) with well-observed SN~Ia. This is especially true for those SN~Ia most similar to one another, aka ``nearest neighbors'' \citep{Jeffery07}, and transitional-type SN~Ia.

\subsection{Signatures of C$+$O Progenitor Material}\label{ss:COmaterial}

If the primary star of most SN~Ia is a C$+$O WD, and if the observed range of SN~Ia properties is primarily due to variances in the ejected mass or abundances of material synthesized in the explosion (e.g., $^{56}$Ni), then this should also be reflected in the remaining amount of carbon and oxygen if M$_{Ch}$ is a constant parameter (see \citealt{Maeda10,Blondin13,Dessart13models}). On the other hand, if one assumes that the progenitor system is the merger of two stars \citep{Webbink84,Iben84,Pakmor13,Moll13} or a rapidly rotating WD \citep{Hachisu12a}$-$both of which are effectively obscured by an amorphous region and/or disk of C$+$O material$-$then the properties of C and O absorption features will be sensitive to the interplay between ejecta and the remaining unburned envelope (see \citealt{Livio11}). 

Oxygen absorption features (unburned plus burned ejecta) are often present as \ion{O}{1}~$\lambda$7774 in the pre-maximum spectra of SN~Ia (Figure~\ref{fig:fig5}). They may exhibit similar behavior to those seen in SN~2011fe (\S4.1), however current datasets lack the proper temporal coverage of a large sample of events that would be necessary to confirm such claims. Still, comparisons of the blue-most wing in the earliest spectra of many SN~Ia to that of SN~2009ig (\S4.2.1), 2010jn (A.41), 2011fe (\S4.1), 2012cg (A.43), and 2012fr (\S4.2.2) may reveal some indication of HV \ion{O}{1} if present and if caught early enough (e.g., SN~1994D; \citealt{Branch05}).

Spectroscopic detections of carbon-rich material have been documented since the discovery of SN~1990N (see \citealt{Leibundgut91,Jeffery92,Branch07b,Tanaka08}) and have been primarily detected as singly ionized in the optical spectra of LVG$-$CN SN Ia \citep{Parrent11}. However, NIR spectra of some SN~Ia subtypes have been suspected of harboring \ion{C}{1} absorption features (\citealt{Hoflich02,Hsiao13}, see also \citealt{Marion06,Marion09}), while \ion{C}{3} has been tentatively identified in ``hotter'' SS/SN~1991T-like SN~Ia \citep{Hatano02,Garavini04,Chornock06}. 

Observations of the over-luminous SCC SN~2003fg suggested the presence of a larger than normal \ion{C}{2} $\lambda$6580 absorption signature \citep{Howell06}. Later in 2006, with the detection of a \emph{conspicuous} \ion{C}{2} $\lambda$6580 absorption ``notch'' in the early epoch observations of the normal SN~Ia, SN~2006D, \citet{Thomas07} reconsidered the question of whether or not spectroscopic signatures of carbon were a ubiquitous property of all or at least some SN~Ia subtypes. 

As follow-up investigations, \citet{Parrent11} and \citet{Folatelli12} presented studies of carbon features in SN~Ia spectra, particularly those of \ion{C}{2} $\lambda\lambda$6580, 7234 (which are easier to confirm than $\lambda\lambda$4267, 4745). However weak, conspicuous 6300 \AA\ absorption features were reported in several SN~Ia spectra obtained during the pre-maximum phase. It was shown that most of the objects that exhibit clear signatures are of the LVG$-$CN SN~Ia subtype, while HVG$-$BL SN~Ia may either be void of conspicuous signatures due to severe line blending, or lack carbon altogether, the latter of which is consistent with DDT models (e.g., \citealt{Hachinger13}) and could also be partially due to increased progenitor metallicities \citep{Lentz00,Meng11,Milne13}. This requires further study and spectrum synthesis from detailed models. 

     \citet{Thomas11} presented additional evidence of unburned carbon
     at photospheric velocities from observations of 5 SN~Ia obtained
     by the Nearby Supernova Factory. Detections were based on the presence of relatively strong
     \ion{C}{2} 6300 \AA\ absorption signatures in multiple spectra of each SN, supported
     by automated fitting with the \texttt{SYNAPPS} code \citep{ThomasSYNAPPS}.  They estimated that at least 22\ce{^{+10}_{-6}}\% of SN~Ia exhibit spectroscopic \ion{C}{2} signatures as late as day $-$5, i.e. carbon features, whether or not present in all SN~Ia, are not often seen even as early as day $-$5. 
     
     \citet{Folatelli12} later searched through the Carnegie Supernova Project (CSP) sample and found at least 30\%
       of the objects show an absorption feature that can be
       attributed to \ion{C}{2} $\lambda$6580. \citet{Silverman12b} searched for carbon
	 in the BSNIP sample and found that $\sim$ 11\%
	 of the SN~Ia studied show carbon absorption features, while $\sim$
	 25\% show some indication of weak 6300 \AA\ absorption. From their sample, they
	 find that if the spectra of SN~Ia are obtained before day $-$5, then the detection percentage is higher than $\sim$ 30\%. Recently it has also been confirmed that ``carbon-positive'' SN~Ia tend to have bluer near-UV colors than those without conspicuous \ion{C}{2} $\lambda$6580 signatures \citep{Thomas11,Silverman12b,Milne13}. 
	 
	  \citet{Silverman12b} estimate
	 the range of carbon masses in normal SN~Ia ejecta to be (2~$-$~30)~x 10$^{-3}$~M$_{\odot}$. For SN~2006D, \citealt{Thomas07} estimated 0.007 M$_{\odot}$ of carbon between 10,000 and 14,000 km~s$^{-1}$ as a lower limit. \citealt{Thomas07} also note that the most vigorous model of \citet{Ropke06} left behind 0.085 M$_{\odot}$ of carbon in the same velocity interval. However, we are not aware of any subsequent spectrum synthesis for this particular model that details the state of an associated 6300 \AA\ signature.
	 
In the recent detailed study on SN~Ia spectroscopic diversity, \citet{Blondin12} searched for signatures of \ion{C}{2} $\lambda$6580 in a sample of 2603 spectra of 462 nearby SN~Ia and found 23 additional ``carbon-positive'' SN~Ia. Given that seven of the nine CN SN~Ia reported by \citet{Blondin12} with spectra prior to day $-$10 clearly exhibit signatures of \ion{C}{2}, and that $\sim$30$-$40\% of the SN~Ia within their sample are of the CN subtype, it is likely that at least 30$-$40\% of all SN~Ia leave behind some amount of carbon-rich material, spanning velocities between 8000 $-$ 18,000 km~s$^{-1}$ \citep{Parrent11,Pereira13,Cartier13}. 

Considering the volume-limited percentage of Branch normal SN~Ia estimated by \citet{WLi11b}, roughly 50\% or more are expected to contain detectable carbon-rich material in the outermost layers. If this is true, then it implies that explosion scenarios that do not naturally leave behind at least a detectable amount (pEWs $\sim$ 5$-$25 \AA) of unprocessed carbon can only explain half of all SN~Ia or less (sans considerations of Ia-CSM progenitors, subtype-ejecta hemisphere dualities, and effects of varying metallicities; see below).
	  
Historically, time-evolving signatures of \ion{C}{2} $\lambda\lambda$6580, 7234 from the \emph{computed\ spectra\ of\ some\ detailed\ models} have not revealed themselves to be consistent with the current interpretations of the observations. This could be due to an inadequate lower-extent of carbon within the models \citep{Thomas07,Parrent11,Blondin12} or the limits of the resolution for the computed spectra \citep{Blondin11a}.

It should be noted that 6300 \AA\ features \emph{are} present in the non-LTE pre-maximum light spectra of \citet{Lentz00} who assessed metallicity effects on the spectrum for a pure deflagration model (see their Fig.~7). Overall, \citet{Lentz00} find that an increase in C$+$O layer metallicities results in a decreased flux (primarily UV) in addition to a blue ward shift of absorption minima (primarily the \ion{Si}{2} 6100 \AA\ feature). While \citet{Lentz00} did not discuss whether or not the weak 6300 \AA\ absorption signatures are due to \ion{C}{2} $\lambda$6580, it is likely the case given that an increase in C$+$O layer metallicities is responsible for the seen decrease in the strength of the 6300 \AA\ feature. However, it should be emphasized that the ``strength'' of this supposed \ion{C}{2} $\lambda$6580 feature appears to be a consequence of how \citet{Lentz00} renormalized abundances for metallicity enhancements in the C$+$O layer. In other words, even though the preponderance of normal SN~Ia with detectable \ion{C}{2} $\lambda$6580 notches are of the NV subgroup, the fact that HV SN~Ia are thus far void of 6300 \AA\ notches \emph{does\ not} imply robust consistency with the idea that nearest neighbor HV SN~Ia properties are solely the result of a progenitor with relatively higher metallicities \citep{Lentz01b}. Such a claim would need to be verified by exploring a grid of models with accompanying synthetic spectra.

Additionally, carbon absorption features could signify an origin that is separate from explosion nucleosynthesis if most SN~Ia are the result of a merger. For example, \citet{Moll13} recently presented angle-averaged synthetic spectra for a few ``peri-merger'' detonation scenarios. In particular, they find a causal connection between ``normal'' \ion{C}{2} $\lambda$6580 signatures and the secondary star for both sub- and super-Chandrasekhar mass cases (c.f. \citealt{Hicken07,Zheng13,Dessart13models}). With constraints from UV spectra \citep{Milne13} and high velocity features (\S3.4), peri-megers can be used to explore the expanse of their spectroscopic influence within the broader picture of SN Ia diversity. 
	 
Coincident with understanding the relevance of remaining carbon-rich material is the additional goal of grasping the spectroscopic role of species that arise from carbon-burning below the outermost layers, e.g., magnesium \citep{Wheeler98}. While signatures of \ion{Mg}{2} $\lambda\lambda$4481, 7890 are frequently observed at optical wavelengths during the earliest phases prior to maximum light, these wavelength regions undergo severe line-blending compared to the NIR signatures of \ion{Mg}{2}. Consequently, \ion{Mg}{2} $\lambda$10927 has served as a better investment for measuring the lower regional extent and conditions during which a DDT is thought to have taken place (e.g., \citealt{Rudy02,Marion03,Marion06,Marion09,Hsiao13}, however see our \S4.1). 

\subsection{High Velocity ($>$16,000 km~s$^{-1}$) Features}\label{ss:HVFs}

   The spectra of many SN~Ia have shown evidence for high-velocity absorption lines of the \ion{Ca}{2} NIR triplet (IR3) in addition to an often concurrent signature of high-velocity \ion{Si}{2} $\lambda$6355 \citep{Mazzali05b,Mazzali05a}. Most recently, high velocity features (HVFs) have also been seen in SN~2009ig \citep{Foley12c}, SN~2012fr \citep{Childress13}, and the SN~2000cx-like SN~2013bh \citep{Silverman13SN2013bh}.  Overall, HVFs are more common before maximum light, display a rich diversity of behaviors \citep{Childress13HVF}, tend to be concurrent with polarization signatures \citep{Leonard05,Tanaka10}, and may be due to an intrinsically clumpy distribution of material \citep{Howell01a,Kasen03,Thomas04,Tanaka06,Hole10}. 
   
   \citet{Maund10a} showed that the \ion{Si}{2} $\lambda$6355 line velocity decline rate, \.{v}$_{Si\ II}$, is correlated with the polarization of the same line at day $-$5, p$_{Si\ II}$, and is consistent with an asymmetric distribution of IMEs. This interpretation is also complimentary with a previous finding that  \.{v}$_{Si\ II}$ is correlated with v$_{neb}$, the apparent Doppler line shift of [\ion{Fe}{2}] 7155 emitted from the ``core'' during late times \citep{Maedanature,Silverman13latetime}. For the recent SN~2012fr, high velocity features of \ion{Ca}{2} IR3 and \ion{Si}{2} $\lambda$6355 at day $-$11 show concurrent polarization signatures that decline in strength during post-maximum light phases \citep{Maund13}.

As for the \emph{origin} of HVFs, they may be the result of abundance
and/or density enhancements due to the presence of a circumstellar medium \citep{Gerardy04,Quimby06}. If abundance enhancements are responsible, it could be explained by an overabundant, outer region of Si and Ca synthesized during a pre-explosion simmering phase (see \citealt{Piro11} and \citealt{Zingale11}). On the other hand, the HVFs in
LVG SN~Ia spectra could indicate the presence of an opaque disk. For example, it is plausible that HVFs are due to magnetically induced merger outflows (\citealt{Ji13}, pending abundance calculations of a successful detonation), or interaction with a tidal tail and/or secondary star (e.g., \citealt{Raskin13,Moll13}). 

Most recently, \citet{Childress13HVF} studied 58 low-z SN~Ia (z $<$ 0.03) with well-sampled light curves and spectra near maximum light in order to access potential relationships between light curve decline rates and empirical relative strength measurements of \ion{Si}{2} and \ion{Ca}{2} HVFs. They find a consistent agreement with \citet{Maguire12} in that the \ion{Ca}{2} velocity profiles assume a variety of characteristics for a given $\Delta$m$_{15}$(\emph{B}) solely because of the overlapping presence of HVFs. In addition, \citet{Childress13HVF} show for their sample that the presence of HVFs is not strongly related to the overall intrinsic (\emph{B}~$-$~\emph{V})$_{max}$ colors. It is also seen that SN~Ia with $\Delta$m$_{15}$(\emph{B}) $>$ 1.4 continue to be void of conspicuous HVFs, while the strength of HVFs in normal SN~Ia is generally larger for objects with broader light curves. Finally, and most importantly, the strength of HVFs at maximum light does not uniquely characterize HVF pre-maximum light behavior. 

Notably, \citet{Silverman13latetime} find no correlation
between nebular velocity and $\Delta$m$_{15}$(\emph{B}), and for a given light-curve shape there is a large
range of observed nebular velocities. Similarly \citet{Blondin12} found no relation between the FWHM of late time 4700 \AA\ iron emission features and $\Delta$m$_{15}$(\emph{B}). This implies the peak brightness of these events \emph{do\ not} translate toward uniquely specifying their late time characteristics, however the data do indicate a correlation between observed (B~$-$~V)$_{max}$
and this particular measure of line-of-sight nebular velocities.

We should also note that while HVG SN~Ia do not clearly come with HVFs in the same sense as for LVG SN~Ia, the entire 6100 \AA\ absorption feature for HVG SN~Ia spans across velocity intervals for HVFs detected in LVG SN~Ia. This makes it difficult to regard LVG and HVG subtypes as two separately distinct explosion scenarios. Instead we can only conclude that HVFs are a natural component of all normal SN~Ia, whether conspicuously separate from a photospheric region \emph{or} concealed as an extended region of absorbing material in the radial direction.

\subsection{Empirical Diversity Diagnostics}\label{ss:diversitydiagnostics}

The depth ratio between 5750 and 6100 \AA\ absorption features, $\mathcal{R}$(``\ion{Si}{2}'') \citep{Nugent95a}, has been found to correlate with components of the WLR. In addition, \citet{Benetti05} find a rich diversity of $\mathcal{R}$(\ion{Si}{2}) pre-maximum evolution among LVG and HVG SN~Ia. 

As for some observables \emph{not} directly related to the decline rate parameter, \citet{Patat96} studied a small sample of well observed SN~Ia
and found no apparent correlation between the blue-shift of the 6100 \AA\
absorption feature at the time of maximum and $\Delta$m$_{15}$(\emph{B}). Similarly, \citet{Hatano00} showed
that $\mathcal{R}$(\ion{Si}{2}) does not correlate well with v$_{10}$(\ion{Si}{2}), the photospheric velocity
derived from the \ion{Si}{2} $\lambda$6355 Doppler line velocities 10 days after maximum light. This could arise from two or more explosion mechanisms, however \citet{Hatano00} note that their interpretation is ``rudimentary'' on account of model uncertainties and the limited number of temporal datasets available at that time. In the future, it would be worthwhile to re-access these trends with the latest detailed modeling. 

\citet{Hachinger06} made empirical measurements of spectroscopic feature pEWs, flux ratios, and projected Doppler velocities for 28 well-observed SN~Ia, which include LVG, HVG, and FAINT subtypes. For normal LVG SN~Ia they find similar observed maximum light velocities (via \ion{Si}{2} $\lambda$6355; $\sim$ 9000$-$10,600 km~s$^{-1}$). Meanwhile the HVG SN~Ia in their sample revealed a large spread of maximum light velocities ($\sim$ 10,300$-$12,500 km~s$^{-1}$), regardless of the value of $\Delta$m$_{15}$(\emph{B}). This overlap in maximum light velocities implies a natural continuum between LVG and HVG SN~Ia (enabling unification through asymmetrical contexts; \citealt{Maedanature,Maund10a}). They also note that FAINT SN~Ia tend to show slightly smaller velocities at \emph{B}-band maximum for larger values of $\Delta$m$_{15}$(\emph{B}), however no overreaching trend of maximum light expansion velocities from LVG to HVG to FAINT SN~Ia was apparent from this particular sample of SN~Ia. 

\citet{Hachinger06} did find several flux ratios to correlate with $\Delta$m$_{15}$({\it B}). In particular, they confirm that the flux ratio, $\mathcal{R}$(``\ion{S}{2} $\lambda$5454, 5640''/``\ion{Si}{2} $\lambda$5972''), is a fairly reliable spectroscopic luminosity indicator in addition to $\mathcal{R}$(\ion{Si}{2}).  \citet{Hachinger06} conclude that these and other flux ratio comparisons are the result of changes in relative abundances across the three main SN~Ia subtypes. In a follow-up investigation, \citet{Hachinger09} argue that the correlation with luminosity is a result of ionization balance, where dimmer objects tend to have a larger value of $\mathcal{R}$(\ion{Si}{2}). \citealt{Silverman12maxlight} later studied correlations between these and other flux ratios of SN~Ia from the BSNIP sample and find evidence to suggest that CSM-associated events tend to have larger 6100 \AA\ blue-shifts in addition to broader absorption features at the time of maximum light (see also \citealt{Arsenijevic11,Foley12dust}). 

\citet{Altavilla09} studied the $\mathcal{R}$(\ion{Si}{2}) ratio
and expansion velocities of intermediate-redshift supernovae. They find that the comparison
of intermediate-redshift SN~Ia spectra with high S/N
spectra of nearby SN~Ia \emph{do\ not} reveal significant differences
in the optical features and the expansion velocities
derived from the \ion{Si}{2} and \ion{Ca}{2} lines that are within the range observed
for nearby SN~Ia. This agreement is also found in the color and
decline of the light curve (see also \citealt{Mohlabeng13}). 
 
While the use of empirically determined single-parameter descriptions of SN~Ia have proved to be useful in practice, they do not fully account for the observed diversity of SN~Ia \citep{Hatano00,Benetti04,Pignata04}. With regard to SN~Ia diversity, it should be reemphasized that special care needs to be taken with the implementation of flux ratios and pEWs. Detailed modeling is needed when attempting to draw connections between solitary characteristics of the observed spectrum and the underlying radiative environment, where a photon-ray's route crosses many radiative contributions that form the spectrum's various shapes, from UV to IR wavelengths. For example, the relied upon 5750, 6100 \AA\ features used for $\mathcal{R}$(\ion{Si}{2}) have been shown to be influenced by more than simply \ion{Si}{2}, as well as from more locations (and therefore various temperatures) than a single region of line formation \citep{Bongard08}. In fact, it is likely that a number of effects are at play, e.g., line blending and phase evolution effects. Furthermore, v$_{10}$(\ion{Si}{2}) is a measure of the 6100 \AA\ absorption minimum during a phase of intense line blending with no less than \ion{Fe}{2}, which imparts a bewildering array of lines throughout the optical bands \citep{Baron95,Baron96}. Still, parameters such as $\mathcal{R}$(\ion{Si}{2}) have served as useful tools for SN~Ia diversity studies in that they often correlate with luminosity \citep{Bongard06} and are relatively accessible empirical measurements for large samples of under-observed SN~Ia. A detailed study on the selection of global spectral indicators can be found in \citealt{Bailey09}.

\subsection{The Adjacent Counterparts of Optical Wavelengths}\label{ss:morethanoptical}

\subsubsection{Ultraviolet Spectra}\label{sss:UVspec}

SN~Ia are known as relatively ``weak emitters'' at UV wavelengths ($<$~3500 \AA; \citealt{Panagia07}). It has been shown that UV flux deficits are influenced by line-blanketing effects from IPEs within the outermost layers of ejecta \citep{Sauer08,Hachinger13,Mazzali13}, overall higher expansion velocities \citep{Foley11,WangX12}, progenitor metallicity \citep{Hoflich98,Lentz00,Sim10}, viewing angle effects (e.g., \citealt{Blondin11a}), or a combination of these \citep{Moll13}. Although, it is not certain which of these play the dominant role(s) in controlling UV flux behaviors among all SN~Ia. 

For SN~1990N and SN~1992A, two extensively studied SN~Ia, pre-maximum light UV observations were made and presented by \citet{Leibundgut91} and \citet{Kirshner93}, respectively. These observations revealed their expected sensitivity to the source temperature and opacity at UV wavelengths. 

It was not until recently when a larger UV campaign of high S/N, multi-epoch spectroscopy of distant SN~Ia was presented and compared to that of local SN~Ia (\citealt{Ellis08}, see also \citealt{Milne13}). Most notably, \citet{Ellis08} found a larger intrinsic dispersion of UV properties than could be accounted for by the span of effects seen in the latest models (e.g., metallicity of the progenitor, see \citealt{Hoflich98,Lentz00}).

As a follow-up investigation, \citet{Cooke11} utilized and presented data from the STIS spectrograph onboard \emph{Hubble Space Telescope} (HST) with the intent of studying
near-UV, near-maximum light spectra (day $-$0.32 to day $+$4) of nearby SN~Ia. Between a high-z and low-z sample, they find a noticeable difference between the mean UV spectrum of each, suggesting that the cause may be related to different metallicities between the statistical norm of each sample. Said another way, their UV observations suggest a plausible measure of two different populations of progenitors (or constituent scenarios) that could also be dependent on the metallically thereof, including potentially larger dependencies such as variable $^{56}$Ni mass and line blanketing due to enhanced burning within the outermost layers \citep{Marion13}. It should be noted, despite the phase selection criterion invoked by \citet{Cooke11}, it may not be enough to simply designate a phase range in order to avoid phase evolution effects (see Fig. 7 of \citealt{Childress13HVF}). 

In order to confirm spectroscopic trends at UV wavelengths, a better method of selection will be necessary as the largest UV difference found by \citet{Cooke11} and \citet{Maguire12} between the samples overlaps with the \ion{Si}{2}, \ion{Ca}{2} H\&K absorption features (3600$-$3900~\AA), i.e. a highly blended feature that is far too often a poorly understood SN~Ia variable, both observationally (across subtype and phase) and theoretically, within the context of line formation at UV$-$NIR wavelengths \citep{Mazzali00,Kasen03,Thomas04,Foley12d,Marion13,Childress13HVF}. While it is true that different radiative processes dominate within different wavelength regions, there are a multitude of explanations for such a difference between the \ion{Si}{2}$-$\ion{Ca}{2} blend near 3700 \AA. Furthermore, the STIS UV spectra do not offer a look at either the state of the 6100 \AA\ absorption feature (is it completely photospheric?$-$the answer requires spectrum synthesis even for maximum light phases), nor is it clear if the same is true for \ion{Ca}{2} in the NIR where high-velocity components thereof are most easily discernible \citep{Lentz00}. 

It is important to further reemphasize that the time-dependent behavior of the sum total of radiative processes that generate a spectrum from a potentially axially-asymmetric (and as of yet unknown) progenitor system and explosion mechanism are not well understood, much less easily decipherable with an only recently obtained continuous dataset for how the spectrum itself evolves over time at optical wavelengths\footnote[20]{SN~2011fe. See \citealt{Pereira13} and \url{http://snfactory.lbl.gov/snf/data/}}. Which is only to say, given the current lack of certain predictability between particular observational characteristics of SN~Ia (e.g., spectroscopic phase transition times), time series observations at UV wavelengths would offer a beneficial route for the essential purposes of hand-selecting the `best' spectrum comparisons in order to ensure a complete lack of phase evolution effects. 

Recently, \citet{WangX12} presented HST multi-epoch, UV observations of SN~2004dt, 2004ef, 2005M, and 2005cf. Based on comparisons to the results of \citet{Lentz00} and \citet{Sauer08}, two studies that show a 0.3 magnitude span of UV flux for a change of two orders of magnitude in metallicity within the C$+$O layer of a pure deflagration model (W7; \citealt{Nomoto84}), \citet{WangX12} conclude that the UV excess for a HVG SN~Ia, SN~2004dt (A.22), \emph{cannot} be explained by metallicities or expansion velocities alone. Rather, the inclusion of asymmetry into a standard model picture of SN~Ia should be a relevant part of their observed diversity (e.g., \citealt{Kasen09,Blondin11a}).

More recently, \citet{Mazzali13} obtained 10 HST UV$-$NIR spectra of SN~2011fe, spanning $-$13.5 to $+$41 days relative to \emph{B}-band maximum. They analyzed the data along side spectrum synthesis results from three explosion models, namely a `fast deflagration' (W7), a low-energy delayed-detonation (WS15DD1; \citealt{Iwamoto99}), and a third model treated as an intermediary between the outer-layer density profiles of the other two models (``W7$^{+}$''). From the seen discrepancies between W7 and WS15DD1 during the early pre-maximum phase, in addition to optical flux excess for W7 and a mismatch in observed velocities for WS15DD1, \citet{Mazzali13} conclude that their modified W7$^{+}$ model is able to provide a better fit to the data because of the inclusion of a high velocity tail of low density material. In addition, and based on a spectroscopic rise time of $\sim$ 19 days, \citet{Mazzali13} infer a $\sim$ 1.4 day period of optical quiescence after the explosion (see \citealt{Piro13a,Chomiuk13}).

\subsubsection{Infrared Light Curves and Spectra}\label{sss:IRspec}

By comparing absolute magnitudes at maximum of two dozen SN~Ia, \citet{Krisciunas05} argue that SN~Ia can be best used as standard candles at NIR wavelengths (which was also suggested by \citealt{Elias81,Elias85}), even without correction for optical light curve shape. \citet{WoodVasey08} later confirmed this to be the case from the analysis of 1087 near-IR (JHK) measurements 
of 21 SN~Ia. Based on their data and
data from the literature, they derive absolute magnitudes of 41 SN~Ia in the \emph{H}-band
with rms scatter of 0.16 magnitudes. \citet{Folatelli10} find a weak dependence of \emph{J}-band luminosities on the decline rate from 9 NIR datasets, in addition to \emph{V}$-$\emph{J} corrected \emph{J}-band magnitudes with a dispersion of 0.12 magnitudes. \citet{Mandel11} constructed a statistical model for SN~Ia light curves across optical and NIR passbands and find that near-IR luminosities enable the most ideal use of SN~Ia as standard candles, and are less sensitive to dust extinction as well. \citet{Kattner12} analyzed the standardizability of SN~Ia in the near-IR
by investigating the correlation between observed peak near-IR absolute
magnitude and post-maximum $\Delta$m$_{15}$(\emph{B}). They confirm that there is a bimodal distribution in the near-IR
absolute magnitudes of fast-declining SN~Ia \citep{Krisciunas09} and suggest that applying a correction
to SN~Ia peak luminosities for decline rate is likely to be beneficial in the J and H bands,
making SN~Ia more precise distance indicators in the IR than at optical wavelengths \citep{BaroneNugent12}.

While optical spectra of SN~Ia have received a great deal of attention in the recent past, infrared datasets (e.g., \citealt{Kirshner73NIR,Meikle96,Bowers97,Rudy02,Hoflich02}) are either not obtained, or are not observed at the same epochs or rate as their optical counterparts. This has only recently begun to change. Thus far, the largest NIR datasets can be found in \citet{Marion03} and \citet{Marion09}. \citet{Marion03} obtained NIR spectra (0.8$-$2.5 $\mu$m) of 12
normal SN~Ia, with fairly early coverage. Later, \citet{Marion09} presented and studied a catalogue of NIR spectra (0.7 $-$ 2.5 $\mu$m) of 41 additional SN~Ia. In all, they report an absence of \emph{conspicuous} signatures of hydrogen and helium in the spectra, and no indications of carbon via \ion{C}{1} $\lambda$10693 (however, see our \S4.1). For an extensive review on IR observations, we refer the reader to \citet{Phillips12}.

\subsection{Drawing Conclusions about SN~Ia Diversity from SN~Ia Rates Studies}\label{ss:rates}

It has long been perceived that a supernova's local environment, rate of occurrence, and host galaxy properties (e.g., WD population) serve as powerful tools for uncovering solutions to SN~Ia origins \citep{Zwicky61,Hamuy95,vandenBergh05,Mannucci05,Leaman11,WLi11c,WLi11d}. After all, a variety of systems, both standard and exotic scenarios$-$all unconfirmed$-$offer potential for explaining ``oddball'' SN~Ia, as well as more normal events, at various distances (z; redshift) and associations with a particular host galaxy or WD population \citep{Yungelson00,Partha07,Hicken09morphology,Hachisu12a,Pakmor13,WangX13,Pan13,Kim14}.

Despite this broad extent of the progenitor problem, measurements of the total cosmic SN~Ia rate, R$_{SNIa}$(z), can be made to gauge the general underlying behavior of actively contributing systems \citep{Maoz12}. Further insight into how various progenitor populations impart their signature onto R$_{SNIa}$(z) comes about by considering which scenarios lead to a ``prompt'' (or a ``tardy'') stellar demise \citep{Scannapieco05,Mannucci05}. Whether or not mergers involve both a (``prompt'') helium-burning or (``tardy'') degenerate secondary star remains to be seen (\citealt{Woods11,Hillebrandt13,Dan13} and references therein). Because brighter SN~Ia prefer younger, metal-poor galaxies, and a linear relation exists between the SN~Ia light curve shape and gas-phase metallicity, the principle finding has been that the rate of the universally prompt component is proportional
to the star formation rate of the host galaxy, whereas the second delayed
component's rate is proportional to the stellar mass of the
galaxy \citep{Sullivan06,Howell07,Sullivan10,Zhang11,Pan13}. The SN~Ia galaxy morphology study of \citet{Hicken09morphology} has since progressed this discussion of linking certain observed SN~Ia properties with their individual environments. Overall, the trend of brighter/dimmer SN~Ia found in younger/older hosts remains, however now with indications that a continuous distribution of select SN~Ia subtypes exist in multiple host galaxy morphologies and projected distances within each host.  

To understand the full form of R$_{SNIa}$(z), taking into account the delay time distribution (DTD) for every candidate SN~Ia system is necessary (see \citealt{Bonaparte13,Claeys14}). \citet{Maoz10a} find that the DTD peaks prior to 2.2 Gyr and has a long tail out to $\sim$ 10 Gyr. They conclude that a DTD with a power-law t$^{-1.2}$ starting at time t = 400 Myr
to a Hubble time can satisfy both constraints of observed cluster SN rates and iron-to-stellar mass ratios, implying that
that half to a majority of all SN~Ia events occur within one Gyr of
star formation (see also \citealt{Strolger10,Meng11}). 

In general, the DTD may be the result of binary mergers
\citep{Ruiter09,Toonen12,Nelemans13} and/or a single-degenerate scenario \citep{Hachisu08,Hachisu12a,Chen13}, but with the consideration that evidence for delay times
as short as 100 Myr have been inferred from SN remnants in the Magellanic 
Clouds \citep{Badenes09, Maoz10b}. From a recent comparison of low/high-z SN~Ia rate measurements and DTDs of various binary population synthesis models, \citet{Graur13} argue that single-degenerate systems are ruled out between 1.8 $<$ z $<$ 2.4. Overall, their results
support the existence of a double-degenerate progenitor channel for SN~Ia if the
the number of double-degenerate systems predicted by binary population synthesis models can be ``aptly'' increased \citep{Maoz10a}. 

However, initial studies have primarily focussed upon deriving the DTD {\it without} taking into account the
possible effects of stellar metallicity on the SN~Ia rate in a given galaxy. Given that lower metallicity stars leave behind higher mass WD stars \citep{Umeda99,Timmes03}, \citet{Kistler13} and \citet{Meng11} argue that the effects of metallicity may serve to
significantly alter the SN~Ia rate (see also \citealt{Pan13}). In fact, models that include the effects of metallicity (e.g., \citealt{Kistler13}) find similar consistencies with the observed R$_{SNIa}$(z). Notably, recent spectroscopic studies \emph{do} indicate a stronger preference of low-metallicity hosts
for super-Chandrasekhar candidate SN~Ia \citep{Taubenberger11, Childress11}, which may just as well be explained by low metallicity single-degenerate systems \citep{Hachisu12a}. While there are not enough close binary WD systems in our own galaxy that would result in SCC DD scenarios \citep{Partha07}, sub-Chandrasekhar merging binaries may be able to account for discrepancies in the observed rate of SN~Ia \citep{Badenes12,Kromer13SN2010lp}. 

Although, we wish to remind the reader that since spectrophotometry of SN~Ia so far offer the best visual insight into these distant extragalactic events, and because there is no clear consensus on the origin of their observed spectrophotometric diversity, there is no clear certainty as to what distribution of progenitor scenarios connect with any kind of SN~Ia since none have been observed prior to the explosion. Furthermore, whether or not brighter or dimmer SN~Ia ``tend to'' correlate with any property of their hosts does not alleviate the discussion down to one or two progenitor systems (e.g., single- versus double-degenerate systems) since the most often used tool for probing SN~Ia diversity over all distance scales, i.e. the ``stretch'' of a light curve, does not necessarily uniquely determine the spectroscopic subtype. Rather, such correlations reveal the degree of an underlying effect from samples of uncertain and unknown SN~Ia subtype biases, i.e. dust extinction in star formation galaxies and progenitor ages also evolve along galaxy mass sequences \citep{Childress13Hosts} and the redshift-color evolution of SN Ia remains an open issue \citep{Mohlabeng13,Pan13,WangS13}.

While it is important to consider the full redshift range over which various hierarchies of progenitor and subtype sequences may dominate over others, such studies are rarely able to incorporate spectroscopic diversity as input (a ``serendipitous'' counter-example being \citealt{Krughoff11}). This is relevant given that the landscape of SN~Ia spectroscopic diversity has not yet been seen to be void of line-of-sight discrepancies for all progenitor scenarios (particularly so for double degenerate detonations/mergers, e.g., \citealt{Shen13,Pakmor13,Raskin13,Moll13,Raskin13mergers}). Ultimately, robust theories should be able to connect spectroscopic subtypes with individual or dual instances of particular progenitor systems, which requires detailed spectroscopic modeling. 

Thus, the consensus as to how many progenitor channels contribute to SN~Ia populations is still unclear. Broadly speaking, there are likely to be no less than two to three progenitor scenarios for normal SN~Ia so long as single-degenerate systems remain viable \citep{Hachisu12a}, if not restricted to explaining Ia-CSM SN alone (see \citealt{Han06,Silverman13IaCSM,Leloudas13}). Given also a low observed frequency of massive white dwarfs and massive double-degenerate binaries near the critical mass limit with orbital periods short enough to merge within a Hubble time, some normal SN~Ia are still perceived as originating from single-degenerate systems \citep{Partha07}. Meanwhile, some portion of events may also be the result of a core-degenerate merger \citep{Soker13}, while some merger phenomena are possibly accelerated within triple systems \citep{Thompson11,Kushnir13,Dong14}. It likewise remains unclear whether or not some double-degenerate mergers predominately result in the production of a neutron star instead of a SN~Ia \citep{Saio85,Nomoto91,Piersanti03,Saio04,Dan13,Tauris13}. At present, separately distinct origins for spectroscopically similar SN Ia cannot be ruled out by even one discovery of a progenitor system; the spectroscopic diversity is currently too great and too poorly understood to confirm without greater unanimity among explosion models and uniformity in data collection efforts. 

\section{Some Recent SN~Ia}

During the past decade, several normal, interesting, and peculiar
 SN~Ia have been discovered. For example, the recent SN~2009ig, 2011fe, and 2012fr are nearby SN~Ia that were discovered extremely young with respect to the onset of the explosion \citep{Nugent11,Foley12c,Childress13} and have been extensively studied at all wavelengths, yielding a clearer understanding of the time-dependent behavior of SN spectroscopic observations, in addition to a better context by which to compare. Below we briefly summarize some of the highlighted discoveries during the most recent decade, during which it has revealed a greater diversity of SN~Ia than was previously known. In the appendix we provide a guide to the recent literature of other noteworthy SN~Ia discoveries. We emphasize that these sections are not meant to replace reading the original publications, and are only summarized here as a navigation tool for the reader to investigate further. 
 
 \subsection{SN~2011fe in M101}\label{ss:11fe}

Thus far, the closest spectroscopically normal SN~Ia in the past 25 years, SN~2011fe (PTF11kly), has provided a great amount of advances, including testing SN~Ia distance measurement methods \citep{Matheson12,Vinko12,Lee12}. For example, the early spectroscopy of SN~2011fe showed a clear and certain time-evolving signature of high-velocity oxygen that varied on time
 scales of hours, indicating sizable
 overlap between C$+$O, Si, and Ca-rich material and newly synthesized IMEs within the outermost
 layers \citep{Nugent11}. 
 
 \citet{Parrent12} carried out analysis of 18 spectra of SN~2011fe during its first month. Consequently, they were able to follow
the evolution of \ion{C}{2} $\lambda$6580 absorption features from near the onset of the explosion until they diminished after maximum light, providing strong evidence for overlapping regions of burned and unburned
material between ejection velocities of at least 10,000 and 16,000 km~s$^{-1}$. At the same time, the evolution of a 7400 \AA\ absorption feature experienced a declining Doppler-shift until 5 days post-maximum light, with \ion{O}{1} $\lambda$7774 line velocities ranging 11,500 to 21,000
km~s$^{-1}$ \citep{Nugent11}. 
\citet{Parrent12} concluded that incomplete burning (in addition to progenitor
scenarios) is a relevant source of spectroscopic diversity among SN~Ia \citep{Tanaka08,Maeda10}.

\citet{Pereira13} presented high quality spectrophotometric observations of SN~2011fe, which span from day $-$15 to day $+$97, and discussed comparisons to other observations made by \citet{Brown12}, \citet{Richmond12}, \citet{Vinko12}, and \citet{Munari13}. From an observed peak bolometric luminosity of 1.17 $\pm$ 0.04 x 10$^{43}$ erg s$^{-1}$, they estimate SN~2011fe to have produced between $\sim$ 0.44 $\pm$ 0.08 $-$ 0.53 $\pm$ 0.11 M$_{\odot}$ of $^{56}$Ni. 

By contrast, \citet{Pastorello07a} and \citet{WangX09} estimate $^{56}$Ni production for the normal SN~2005cf (A.26) to be $\sim$ 0.7 M$_{\odot}$. It is also interesting to note that SN~2011fe and the fast-declining SN~2004eo produced similar amount of radioactive nickel, however lower for SN~2004eo ($\sim$ 0.4 M$_{\odot}$; \citealt{Mazzali08}). \citet{Pereira13} also made comparisons between SN~2011fe, a SNFactory normal SN~Ia (SNF20080514-002) and the broad-lined HV-CN SN~2009ig \citep{Foley12c}. \citet{Pereira13} note similarities (sans the UV) and notable contrast with respect to high-velocity features, respectively. 

\citet{Pereira13} calculated \.{v}$_{Si\ II}$ for SN~2011fe to be $\sim$ 60 ($\pm$ 3) km~s$^{-1}$ day$^{-1}$, near the high end of low-velocity gradient SN~Ia events (see \citealt{Benetti05,Blondin12}). Given their high S/N, time series dataset, \citet{Pereira13} were also able to place tighter constraints on the velocities over which \ion{C}{2} $\lambda$6580 is observed to be present in SN~2011fe. They conclude that \ion{C}{2} is present down to at least as low as 8000 km~s$^{-1}$, which is 2000 km~s$^{-1}$ lower than that estimated by \citet{Parrent12}, and is also $\sim$ 4000$-$6000 km~s$^{-1}$ (or more) lower than what is predicted by some past and presently favored SN~Ia abundance models (e.g., W7; \citealt{Nomoto84}, and the delayed detonations of \citealt{Hoflich06} and \citealt{Roepke12}). 

\citet{Hsiao13} presented and discussed NIR time series spectra of SN~2011fe that span between day $-$15 and day $+$17. In particular, they report a detection of \ion{C}{1} $\lambda$10693 on the blue-most side of a blended \ion{Mg}{2} $\lambda$10927 absorption feature at roughly the same velocities and epochs as \ion{C}{2} $\lambda$6580 found by \citet{Parrent11} and \citet{Pereira13}, which itself is blended on its \emph{blue-most} side with \ion{Si}{2} $\lambda$6355. While searches and studies of \ion{C}{1} $\lambda$10693 are extremely useful for understanding the significance of C-rich material from normal to cooler sub-luminous SN~Ia within the greater context of all \ion{C}{1}, \ion{C}{2}, \ion{C}{3}, \ion{O}{1} absorption features (\ion{C}{3} for ``hotter'' SN~1991T-likes), blended \ion{C}{1} $\lambda$10693 absorption shoulders are certainly no more (nor no less) useful for probing lower velocity boundaries than \ion{C}{2} $\lambda$6580 absorption notches. This is especially true given that \ion{C}{1} $\lambda$10693 absorption features are blended from the \emph{red-most} side (lower velocities) by the neighboring \ion{Mg}{2} line, which will only serve to obscure the lower velocity information of the \ion{C}{1} profile for the non-extreme cases (e.g., SN~1999by, \citealt{Hoflich02}). 

\citet{Hsiao13} used the observed temporal behavior, and later velocity-plateau, of \ion{Mg}{2} $\lambda$10927 to estimate a lower extent of $\sim$ 11,200 km~s$^{-1}$ for carbon-burning products within SN~2011fe. Given that this in contrast to the refined lower extent of \ion{C}{2} at $\sim$~8000 km~s$^{-1}$ by \citet{Pereira13}, this \emph{could} imply (i.e. assuming negligible temperature differences and/or non-LTE effects) that either some unburned material has been churned below the boundary of carbon-burning products via turbulent instabilities \citep{Gamezo99Cells,Gamezo99,Gamezo04} and/or the distribution of emitting and absorbing carbon-rich material is truly globally lopsided \citep{Kasen09,Maedanature,Blondin11a}, and may indicate the remains of a degenerate secondary star \citep{Moll13}.

Detailed studies of this nearby, normal, and
unreddenned SN~2011fe have given strong \emph{support} for double-degenerate scenarios (assuming low environmental abundances of hydrogen) and have placed strong \emph{constraint} on single-degenerate scenarios, i.e. MS and RG companion stars have been strongly constrained for SN~2011fe (see \citealt{Shappee13} and references therein, and also \citealt{Hayden10SDSS,Bianco11}). \citet{Nugent11}, \citet{WLi11a} and \citet{Bloom12} confirm that the primary star was a compact star (R$_{*}$~$\lesssim$~0.1~R$_{\odot}$, c.f., \citealt{Bloom12,Piro13a,Chomiuk13}). From the lack of evidence for an early shock outbreak \citep{Kasen10,Nakar12}, non-detections of radio and X-ray emissions \citep{Horesh12,Chomiuk12,Margutti12}, non-detections of narrow \ion{Ca}{2} H\&K or Na~D lines or pre-existing dust that could be associated with the event \citep{Patat13,Johansson13}, and low upper-limits on hydrogen-rich gas \citep{Lundqvist13}, the paucity of evidence for an environment dusted in CSM from a non-degenerate secondary strongly supports the double degenerate scenario for SN~2011fe. Plus, this inferred ambient environment is consistent with that of recent merger simulations \citep{Dan12}, and could signal an avenue of interpretation for signatures of carbon-rich material as well \citep{Branch05,Dan13,Moll13}. Specifically, the remaining amount of carbon-rich material predicted by some explosion models may already be accounted for, and more so than would be required by the existence of low velocity detections of \ion{C}{1} and \ion{C}{2}. If this turns out to be the case, spectroscopic signatures of both C and O could tap into understanding (i) the sizes of merger C$+$O common envelopes, (ii) potential downward mixing effects between the envelope and the underlying ejecta, and/or (iii) test theories on possible asymmetries of C$+$O material within the post-explosion ejecta of the primary and secondary stars \citep{Livio11}, which is expected to depend on the degree of coalescence \citep{Moll13,Raskin13}. 
 
 Of course, this all rests on the assumptions that (i)~the surrounding environment of a single degenerate scenario just prior to the explosion ought to be contaminated with some amount of CSM, above which it would be detected \citep{Justham11,Brown12}, and (ii)~the surrounding environment of a merger remains relatively ``clean'' \citep{Shen13,Raskin13}. In this instance, and assuming similarly above that current DDT-like models roughly fit the outcome of the explosion, absorption signatures of C ($+$ HV \ion{O}{1}) may point to super-massive single-degenerate progenitors with variable enclosed envelopes and/or disks of material (e.g., \citealt{Yoon04,Yoon05,Hachisu12a,Scalzo12,Tornambe13,Dan13}) or sub-Chandrasekhar mass ``peri-mergers'' for resolve (see \citealt{Moll13} and references therein).

 \subsection{Other Early Discoveries}
 
 \subsubsection{SN~2009ig in NGC 1015}\label{ss:09ig}

   \citet{Foley12c} obtained well-sampled, early UV and optical spectra of SN~2009ig as it was discovered 17 hr after the event \citep{Kleiser09,Navasardyan09}. SN~2009ig is found to be a normal
 SN~Ia, rising to \emph{B}-band maximum in $\sim$ 17.3 days. From the earliest 
 spectra, \citet{Foley12c}
 find \ion{Si}{2} $\lambda$6355 line velocities around 23,000 km~s$^{-1}$,
 which is exceptionally high for such a spectroscopically normal SN~Ia (see also \citealt{Blondin12}). SN~2009ig possess either an overall shallower density profile than other CN SN~Ia, or a buildup of IMEs is present at high velocities. 
  
\citet{Marion13} recently analyzed the photospheric to post-maximum light phase spectra of SN~2009ig, arguing for the presence of additional high-velocity absorption signatures from not only \ion{Si}{2}, \ion{Ca}{2}, but also \ion{Si}{3}, \ion{S}{2} and \ion{Fe}{2}. Whether or not two separate but compositionally equal regions of line formation is a ubiquitous property of similar SN~Ia remains to be seen. However, it should not be unlikely for primordial amounts of said atomic species to be present (in addition to singly-ionized silicon and calcium) on account of possible density and/or abundance enhancements within the outermost layers \citep{Thomas04,Mazzali05a,Mazzali05b}. For example, simmering effects during convective phases prior to the explosion may be responsible for dredging up IMEs later seen as HVFs, which would give favorability to single-degenerate progenitor scenarios (see \citealt{Piro11,Zingale11}). Similarly, it is worthwhile to access the versatility of mergers in producing high-velocity features. 
   
\subsubsection{SN~2012fr in NGC 1365}\label{ss:12fr}

\citet{Childress13} report on their time series spectroscopic observations of SN~2012fr \citep{Klotz12,Childress12CBET,Buil12}, complete with 65 spectra that cover between $\sim$ 15 days before and 40 days after it reached a peak \emph{B}-band brightness of $-$19.3. In addition to the simultaneous spectropolarimetric observations of \citet{Maund13}, the early to maximum light phase spectra of SN~2012fr reveal one of the clearest indications that SN~Ia of similar type (e.g., SN~1994D, 2001el, 2009ig, 2011fe, and many others; Mazzali05a) tend to have two distinctly separate regions of Si-, Ca-based material that differ by a range of separation velocities \citep{Childress13HVF}. 

\citet{Childress13} and \citet{Maund13} discussed the various interpretations that have been presented in the past, however no firm conclusions on the origin of HVFs could be realized given the uncertainties of current explosion models. Despite this, the most recent advance toward understanding HVFs is the continual detection of polarization signatures due to the high-velocity \ion{Si}{2} and \ion{Ca}{2} absorption features, indicating a departure from a radially stratified, spherically symmetric geometry at some layer near or above the ``photospheric region'' of IMEs. 
 
\subsection{Super-Chandrasekhar Candidate SN~Ia}

\subsubsection{Over-luminous SN~2003fg (SNLS-03D3bb)}\label{ss:03fg}

SN~2003fg was discovered as part of the 
Supernova Legacy Survey (SNLS); z = 0.2440 \citep{Howell06}. Its peak absolute magnitude was estimated to be $-$19.94 in \emph{V}-band, placing SN~2003fg
completely outside the M$_{\emph{V}}$-distribution of normal low-z SN~Ia (2.2~times brighter). Assuming Arnett's rule, such a high luminosity corresponds to $\sim$ 1.3 M$_{\odot}$ of $^{56}$Ni, which would be in conflict with SN~2003fg's spectra since only $\sim$~60\% of a Chandrasekhar pure detonation ends up as radioactive nickel (\citealt{Steinmetz92}, however see also \citealt{Pfannes10}). Given also the lower mean expansion velocities, this builds upon the picture of a super-Chandrasekhar mass progenitor for SN~2003fg and others like it \citep{Howell06,Jeffery06}. 

\citet{Yoon05}
proposed the formation of super-Chandrasekhar mass WD stars as a
result of rapid rotation. \citet{Pfannes10} later reworked these models and found that the ``prompt'' detonation of a super-Chandrasekhar mass WD produces enough nickel, as well as a remainder of IMEs in the outer layers (in contrast to \citealt{Steinmetz92}), to explain over-luminous SN~Ia. 

\citet{Hachisu12a} added to this model by taking into account processes of binary evolution. Namely, with the inclusion of mass-striping, optically thick winds of a differentially rotating primary star, \citet{Hachisu12a} find three critical mass ranges that are each separated according to the spin-down time of the accreting WD. All three of these single-degenerate scenarios may explain a majority of events from sub-luminous to over-luminous SN~Ia. So far no super-Chandrasekhar mass WD stars that would result in a SN~Ia have been found in the sample of known WD stars in our Galaxy (\citealt{Saffer98}, see also \citealt{Kilic12}). However, this does not so much rule out super-Chandrasekhar mass models as it suggests that these systems are rare in the immediate vicinity within our own galaxy. 

    \citet{Hillebrandt07} proposed an alternative scenario involving
only a Chandrasekhar-mass WD progenitor to explain the
SN~2003fg event. They demonstrated that an off-center explosion
of a Chandrasekhar-mass WD could explain the super-bright
SN~Ia. However, in this off-center explosion model it is not easy to account
for the high Ni mass in the outer layers, in addition to the special viewing direction.

\subsubsection{Over-luminous SN~2009dc in UGC 10064}\label{ss:09dc}

  \citet{Yamanaka09a} presented early phase optical
and NIR observations for SN~2009dc \citep{Puckett09CBET,Harutyunyan09CBET,Marion09CBET,Nicolas09CBET}. From the peak \emph{V}-band absolute magnitude they conclude
that SN~2009dc belongs to the most luminous class of SN~Ia ($\Delta$m$_{15}$(\emph{B}) = 0.65), and estimate the
$^{56}$Ni mass to be 1.2 to 1.6 M$_{\odot}$. Based on the JHK photometry \citet{Yamanaka09a} also find SN~2009dc had an unusually high NIR luminosity with enhanced fading after $\sim$ day $+$200 \citep{Maeda09,Silverman11,Taubenberger11}. The spectra of SN~2009dc
also show strong, long lasting 6300 \AA\ absorption features (until $\sim$ two weeks post-maximum light) Based on the observed spectropolarimetric indicators,
in combination with photometric and spectroscopic properties, \citet{Tanaka10}
similarly conclude that the progenitor mass of SN~2009dc was of super-Chandrasekhar origin and
that the explosion geometry was globally spherically symmetric, with a
clumpy distribution of IMEs.

  \citet{Silverman11} presented an analysis of 14 months of observations
of SN~2009dc and estimate a rise-time of $\sim$ 23 days and $\Delta$m$_{15}$(\emph{B}) = 0.72. They find a
lower limit of the peak
bolometric luminosity $\sim$ 2.4~x~10$^{43}$ erg s$^{-1}$ and caution that the actual value is likely almost 40\% larger. Based on the high luminosity and low
mean expansion velocities of SN~2009dc, \citet{Silverman11} derive a mass of more than
 2M$_{\odot}$ for the white
dwarf progenitor and a $^{56}$Ni mass of $\sim$ 1.4 to 1.7 M$_{\odot}$. \citet{Taubenberger11} find the minimum $^{56}$Ni mass to be 1.8 M$_{\odot}$,
 assuming the smallest possible rise-time of 22 days, and the ejecta mass to be
2.8 M$_{\odot}$.

\citet{Taubenberger13} compared photometric and spectroscopic observations of normal and SCC SN~Ia at late epochs, including SN~2009dc, and find a large diversity of properties spanning through normal, SS, and SCC SN~Ia. In particular the decline in the light curve ``radioactive tail'' for SCC SN~Ia is larger than normal, along with weaker than normal [\ion{Fe}{3}] emission in the nebular phase spectra. \citet{Taubenberger13} argue that the weak [\ion{Fe}{3}] emission is indicative of an ejecta environment with higher than normal densities. Previously, \citet{Hachinger12} carried out spectroscopic modeling for SN~2009dc and discussed the
 model alternatives, such as a 2 M$_{\odot}$ rotating WD, a core-collapse SN, and a CSM interaction scenario. Overall, \citet{Hachinger12} found the interaction scenario to be the most promising in that it does not require the progenitor to be super-massive. \citet{Taubenberger13} furthered this discussion in conjunction with their late time comparisons and conclude that the models of \citet{Hachinger12} do not simultaneously match the peak brightness and decline of SN~2009dc (see also \citealt{Yamanaka13}). Following the interaction scenario of \citet{Hachinger12}, \citet{Taubenberger13} propose a non-violent merger model that produces $\sim$1M$_{\odot}$ of $^{56}$Ni and is enshrouded by $\sim$0.6$-$0.7M$_{\odot}$ of C$+$O-rich material. In order to reconcile the low $^{56}$Ni production, \citet{Taubenberger13} note that additional luminosity from interaction with CSM is required during the first two months post-explosion. Further support for CSM interaction comes from the observed suppression of the double peak in the \emph{I}-band, which is thought to arise from a breaking of ejecta stratification in the outermost layers \citep{Kasen06NIR,Kamiya12}.
 
 It is not yet clear if SN~2003fg, 2006gz (A.32), 2007if (A.34), and SN~2009dc are the result of a single super-Chandrasekhar mass
WD star, given that even in our galaxy there is no observational evidence for the
existence 
of such a system. Likewise, there is no direct observational
evidence for the presence of very rapidly rotating massive WD stars, either single WDs or in binary systems as well. In fact, no double-degenerate close binary systems with a total mass amounting to             
super-Chandrasekhar mass configurations that can merge in Hubble-time have
been found \citep{Partha07}. Therefore, our current understanding of the origin of over-luminous SN~Ia is limited, and more observations are needed. For example, progress has been made with the recent discovery of 24 merging WD systems via the extremely low mass Survey (see \citealt{Kilic12} and references therein), however it is unclear if any are systems that would produce a normal SN~Ia.
 
\subsection{The Peculiar SN~2002cx-like Class of SN}

\subsubsection{SN~2002cx in CGCG-044-035}\label{ss:02cx}
 
\citet{WLi03} considered SN~2002cx as ``the most peculiar known
SN~Ia'' \citep{Wood-Vasey02CBET}. They obtained photometric and spectroscopic observations
which revealed it to be unique among all observed SN~Ia. \citet{WLi03} described SN~2002cx as having SN~1991T-like pre-maximum spectrum, a SN~1991bg-like luminosity, and 
expansion velocities roughly half those of normal SN~Ia.

    Photometrically, SN~2002cx has a broad peak in the \emph{R}-band, a plateau
phase in the \emph{I}-band, and a slow late time decline. The \emph{B} $-$ \emph{V} color evolution
are described as nearly normal, while the \emph{V} $-$ \emph{R} and \emph{V} $-$ \emph{I} colors are redder than normal. Spectra of SN~2002cx during early phases evolve rapidly and are dominated by lines
from IMEs and IPEs, but the features are weak overall. In addition, emission lines
are present around 7000 \AA\ during post-maximum light phases, while the late time nebular spectrum shows narrow lines of iron and cobalt. 

\citet{Jha06a} presented late time spectroscopy of SN~2002cx, which includes spectra at 227 and 277 days post-maximum light. They considered it as a prototype of a new subclass of SN~Ia. The spectra do \emph{not} appear to be dominated by the forbidden emission lines of iron, which is not 
expected during the ``nebular phase,'' where instead they find a number of P Cygni profiles
of \ion{Fe}{2} at exceptionally low expansion velocities of $\sim$ 700~km~s$^{-1}$ \citep{Branch04a}. A tentative
identification of \ion{O}{1}~$\lambda$7774 is also reported for SN~2002cx, suggesting the presence of oxygen-rich material.
Currently, it is difficult to explain all the observed photometric and spectroscopic properties
of SN~2002cx using the standard SN~Ia models (see \citealt{Foley13}). However, the spectral characteristics of SN~2002cx support 
pure deflagration or failed-detonation models that leave behind a bound remnant instead of delayed detonations \citep{Jordan12,Kromer13,Hillebrandt13}.
 
 \subsubsection{SN~2005hk in UGC 00272}\label{ss:05hk}

      \citet{Phillips07} presented extensive multi-color photometry
and optical spectroscopy of SN~2005hk \citep{Quimby05CBET}. \citet{Sahu08} also studied the spectrophotometric evolution SN
2005hk, covering pre-maximum phase to around 400 days after the event. These datasets reveal
that SN~2005hk is \emph{nearly} identical in its observed properties to SN
2002cx. Both supernovae exhibited high ionization SN~1991T-like
pre-maximum light spectra but with low peak luminosities like that of SN~1991bg.
The spectra reveal that SN~2005hk, like SN~2002cx, has expansion
velocities that are roughly half those of typical SN~Ia.

The \emph{R} and \emph{I}-band light curves of both supernovae are also peculiar for not
displaying the secondary maximum observed for normal SN~Ia. \citet{Phillips07}
constructed a bolometric light curve from 15 days before to
60 days after \emph{B}-band maximum.  They conclude that the shape and exceptionally
 low peak luminosity of
the bolometric light curve, low expansion velocities, and absence of a
secondary maximum in the NIR light curves are
in reasonable agreement with model calculations of a three-dimensional
deflagration that produces 0.25 M$_{\odot}$ of $^{56}$Ni. Note however that the low amount of continuum polarization observed for SN~2005hk ($\sim$~0.2\%$-$0.4\%) is far too similar to that of more normal SN~Ia to serve as an explanation for the spectroscopic peculiarity of SN~2005hk, and possibly other SN~2002cx-like events \citep{Chornock06,Maund10b}.
  
\subsubsection{Sub-luminous SN~2007qd}\label{ss:07qd}

\citet{McClelland10} obtained multi-band photometry and multi-epoch
spectroscopy of SN~2007qd \citep{Bassett07CBET}. Its observed properties place it broadly between those of the 
peculiar SN~2002cx and SN~2008ha (A.37). Optical
photometry indicate a fast rise-time and a
peak absolute \emph{B}-band magnitude of $-$15.4. \citet{McClelland10} carried out
spectroscopy of SN~2007qd near maximum brightness and detect signatures of IMEs. They find the
photospheric velocity to be 2800 km~s$^{-1}$ near maximum light,
and note that this is $\sim$ 4000 and 7000 km~s$^{-1}$ less than that inferred for SN~2002cx and
normal SN~Ia, respectively. \citet{McClelland10} find that the peak
luminosities of SN~2002cx-like objects are well correlated with
their light curve stretch and photospheric velocities. 
  
 \subsubsection{SN~2009ku}\label{ss:09ku}

   SN~2009ku was discovered by Pan-STARS-1 as a SN~Ia belonging
to the peculiar SN~2002cx class. \citet{Narayan11} studied
SN~2009ku and find that while its multi-band light curves are similar to that of SN~2002cx,
they are slightly broader and have a later rise to \emph{g}-band maximum.
Its peak brightness was found to be M$_{\emph{V}}$~=~$-$18.4 and the ejecta velocity
at 18 days after maximum brightness was found to be $\sim$ 2000 km~s$^{-1}$. Spectroscopically, SN~2009ku is similar to SN~2008ha (A.37).
\citet{Narayan11} note that the high luminosity and low ejecta velocity
for SN~2009ku is not in agreement with the trend seen for SN~2002cx class
of SN~Ia. The spectroscopic and photometric characteristics of SN~2009ku
indicate that the SN~2002cx class of SN~Ia are not homogeneous, and that the
SN~2002cx class of events may have a significant dispersion in their progenitor population
and/or explosion physics (see also \citealt{Kasliwal12} for differences between this class and sub-luminous ``calcium-rich'' transients).

\subsection{PTF11kx and the ``Ia-CSM'' Class of SN~Ia}
 
 \subsubsection{PTF11kx: A case for single-degenerate scenarios?}\label{ss:ptf11kx}

\citet{Dilday12} studied the photometric and spectroscopic properties of another unique SN~Ia event, PTF11kx. Using time series, high-resolution optical spectra, they find direct evidence supporting a single-degenerate progenitor system based on several narrow, temporal ($\sim$ 65 km~s$^{-1}$) spectroscopic features of the hydrogen Balmer series, \ion{He}{1}, \ion{Na}{1}, \ion{Ti}{2}, and \ion{Fe}{2}. In addition, and for the first time, PTF11kx observations reveal strong, narrow, highly time-dependent \ion{Ca}{2} absorption features that change from saturated absorption signatures to emission lines within $\sim$ 40 days. 

\citet{Dilday12} considered the details of these observations and concluded that the complex CSM environment that enshrouds PTF11kx is strongly indicative of mass loss or ``outflows,'' prior to the onset of the explosion of the progenitor system. Other SN~Ia have exhibited narrow, temporal Na D lines before (e.g., SN~2006X, 2007le; see \citealt{Simon09,Patat09,Patat10,Patat11,Sternberg11}), but none have been reported as having signatures of these particular ions, which are clearly present in the high-resolution spectra of PTF11kx. On the whole, and during the earliest epochs, \citet{Dilday12} find that the underlying SN~Ia spectroscopic component of PTF11kx most closely resembles that of SN~1991T \citep{Flipper91T,Gomez98} and 1999aa \citep{Garavini04}. 

As for the late time phases, \citet{Silverman13} studied spectroscopic observations of PTF11kx from 124 to 680 days post-maximum light and find that its nebular phase spectra are markedly different from those of normal SN~Ia. Specifically, the late time spectra of PTF11kx are void of the strong cobalt and iron emission features typically seen in other SN~1991T/1999aa-like and normal SN~Ia events (e.g., \citealt{Ruiz92,Salvo01,Branch03,Stehle05,Kotak05,McClelland13,Silverman13latetime}). For the most part, the late time spectra of PTF11kx are seen to be dominated by broad (FWHM $\sim$ 2000 km~s$^{-1}$) H$\alpha$ emission and strong \ion{Ca}{2} emission features that are superimposed onto a relatively blue, overly luminous continuum level that may be serving to wash out the underlying SN~Ia spectroscopic information. \citet{Silverman13} note that the H$\alpha$ emission increases in strength for $\sim$1 yr before decreasing. In addition, from the absence of strong H$\beta$, \ion{He}{1}, and \ion{O}{1} emission, as well as a larger than normal late time luminosity, \citet{Silverman13} conclude that PTF11kx indeed interacted with some form of CSM material; possibly of multiply thin shells, shocked into radiative modes of collisional excitation as the SN ejecta overtakes the slower-moving CSM. However, it should be noted that it is not yet clear if the CSM originates from a single-degenerate scenario or a H-rich layer of material that is ejected prior to a double-degenerate merger event (\citealt{Shen13}, see also \citealt{Soker13}).

\subsubsection{SN~2002ic}\label{ss:02ic}

    \citet{Hamuy03} detected a large H$\alpha$ emission in the spectra of
SN~2002ic \citep{Wood-Vasey02icCBET}. Seven days before to 48 days after maximum light, the optical
spectra of SN~2002ic exhibit normal SN~Ia spectral features in addition to the strong
H$\alpha$ emission. The H$\alpha$ emission line in the spectrum of SN~2002ic
consists of a narrow component atop a broad component (FWHM of about
1800 km~s$^{-1}$). \citet{Hamuy03} argue that the broad component
arose from ejecta$-$CSM interaction. By day $+$48, they find
that the spectrum is similar to that of SN~1990N. \citet{Hamuy03}
argue that the progenitor system contained a massive AGB star, associated with
a few solar masses of hydrogen-rich CSM.
    
   \citet{Kotak04} obtained the first high resolution, high S/N spectrum of SN~2002ic. The resolved H$\alpha$ line has a P Cygni-type
profile, indicating the presence of a dense, slow-moving outflow (about 100 km~s$^{-1}$). They also find a relatively large and unusual NIR excess and argue that this is the result of an infrared
light-echo originating from the presence of CSM. They
estimate the mass of CSM to be more than 0.3 M$_{\odot}$,
produced by a progenitor mass loss rate greater than 10$^{-4}$ M$_{\odot}$
yr$^{-1}$. For the progenitor, \citet{Kotak04} favor a single-degenerate
system with a post-AGB companion star.

   \citet{WoodVasey04} obtained pre-maximum and late time photometry
of SN~2002ic and find that a non-SN~Ia
component of the light curve becomes pronounced about 20 days post-explosion. They suggest
the non-SN~Ia component to be due to heating from a shock interaction
between SN ejecta and CSM. \citet{WoodVasey04} also suggest
that the progenitor system consisted of a WD and an AGB star
in the protoplanetary nebula phase. \citet{WoodVasey04} and \citet{Sokoloski06} proposed that a nova shell ejected
from a recurrent nova progenitor system, creating the evacuated region
around the explosion center of SN~2002ic.  They suggest that the periodic
shell ejections due to nova explosions on a WD sweep up     
the slow wind from the binary companion, creating density variations
and instabilities that lead to structure in the circumstellar medium.
 This type of phenomenon may occur in SN~Ia with recurrent nova
progenitors, however \citet{Schaefer11} recently reported on an ongoing observational campaign of recurrent novae (RN) orbital period changes between eruptions. For at least two objects (CI Aquilae and U Scorpii), he finds that the RN \emph{lose} mass, thus making RN unlikely progenitors for SN~Ia. 

  Nearly one year after the explosion, \citet{Wang04} found
that the supernova had become fainter overall, but H$\alpha$ emission
had brightened and broadened compared to earlier observations.
From their spectropolarimetry observations, \citet{Wang04} find that hydrogen-rich
matter is asymmetrically distributed. Likewise, \citet{Deng04} also found evidence
of a hydrogen-rich asymmetric circumstellar medium. From their observations of
SN~2002ic, \citet{Wang04} conclude that the event took
place within a ``dense, clumpy, disk-like'' circumstellar medium. They
suggest that the star responsible for SN~2002ic could either be a post-AGB
star or WD companion (see also \citealt{Hachisu99,Han06}).


\subsubsection{SN~2005gj}\label{ss:05gj}

     Similar to SN~2002ic,  \citet{Aldering06} argue that SN~2005gj is a SN~Ia in a
massive circumstellar envelope (see also \citealt{Prieto07}), which is located in a low metallicity host galaxy with a significant amount of star formation. Their first spectrum of SN~2005gj shows a blue continuum level with broad and
narrow H$\alpha$ emission. Subsequent spectra reveal muted SN~Ia features combined with broad and narrow
H$\gamma$, H$\beta$, H$\alpha$ and \ion{He}{1} $\lambda\lambda$5876, 7065 in emission, where high resolution spectra reveal narrow P Cygni profiles. An inverted
P Cygni profile for [\ion{O}{3}] $\lambda$5007 was also detected, indicating top-lighting effects from CSM interaction \citep{Branch00}. From their
early photometry of SN~2005gj, \citet{Aldering06} find that the interaction between the supernova
ejecta and CSM was much weaker for SN~2002ic. Notably, both \citet{Aldering06} and \citet{Prieto07} agree that a SN~1991T-like spectrum can account for many of the observed profiles with an assumed increase in continuum radiation from interaction with the hydrogen-rich material.

\citet{Aldering06}
also find that the light curve and measured velocity of the unshocked
CSM imply mass loss as recent as 1998. This is
in contrast to SN~2002ic, for which an inner cavity in the circumstellar
matter was inferred \citep{WoodVasey04}. Furthermore, SN~1997cy, SN~2002ic, and SN~2005gj all exhibit
large CSM interactions and are from low-luminosity
hosts. 

Consistent with this interpretation for CSM interactions is the recent report by \citet{Fox13} that a NIR re-brightening, possibly due to emission from ``warm'' dust, took place at late times for both SN~2002ic and 2005gj. Notably, and in contrast to SN~2002ic, \citet{Fox13} find that the mid-IR luminosity of SN~2005gj increased to $\sim$ twice its early epoch brightness.
 
\section{Summary and Concluding Remarks}\label{s:summary}

Observations of a significant number of
SN Ia during the last two decades have enabled us  
to document a larger expanse of their physical properties which is manifested through spectrophotometric diversity. While in general SN Ia have long been considered a homogeneous class, they
do exhibit up to 3.5 mag variations in the peak luminosity, whereas ``normal'' SN Ia dispersions are $\sim$1 mag, and constitute several marginally distinct subtypes \citep{Blondin12,Scalzo12,Silverman13IaCSM,Foley13,Dessart13models}. Consequently, the use
of normal SN Ia for cosmological purposes depends on empirical
calibration methods (e.g., \citealt{Bailey09}), where one of the most physically relevant
methods is the use of the width-luminosity relation \citep{Phillips93,Phillips99}. 

Understanding the physics and origin of the width-luminosity-relationship of SN Ia light curves is an important aspect
in the modeling of SN Ia \citep{Khokhlov93,Lentz00,Timmes03,Nomoto03,Kasen07,Kasen09,Meng11,Blondin11a}. Brighter SN Ia often have broad light curves that decline
slowly after peak brightness. Slightly less bright or dimmer SN Ia have
narrower and relatively rapidly declining light curves. In addition, several SN Ia do not follow the width-luminosity-relationship (e.g., SN~2001ay, 2004dt, 2010jn, SCC, CL and SN~2002cx-like SN Ia), which reinforces the notion that a significant number of physically relevant factors influence the diversity of SN Ia overall (see \citealt{WangX12,Baron12}). 

Despite the ever increasing number of caught-early supernovae, our perspective on their general properties and individual peculiarities undergoes a continual convergence toward a set of predictive standards with which models must be seen to comply. The most recent observational example is that of SN~2012fr \citep{Maund13,Childress13}, a normal/low-velocity-gradient SN Ia that has been added to the growing list of similar SN Ia that exhibit stark evidence for a distinctly separate region of ``high-velocity'' material ($>$16,000 km s$^{-1}$). While the origin of high velocity features in the spectra of SN Ia is not well understood, it is concurrent with polarization signatures in most cases which implies some amount of ejecta density asymmetries (e.g., \citealt{Kasen03,WW08,Smith11, Maund13}). Furthermore, since understanding the temporal behavior of high velocity \ion{Si}{2}/\ion{Ca}{2} depends on knowing the same for the photospheric component, studies that focus on velocity gradients and potential velocity-plateaus of the photospheric component could make clearer the significance of the physical separation between these two regions of material (see \citealt{Patat96,Kasen03,Tanaka08,Foley12c,Parrent12,Scalzo12,Childress13,Marion13}). However, it is at least certain that all viable models that encompass ``normal'' SN Ia conditions must account for the range of properties related to velocity evolution (see \citealt{Blondin12} and references therein), the occasionally observed however potentially under-detected signatures of C$+$O material at both low and high velocities \citep{Thomas07,Parrent11,Thomas11,Folatelli12,Silverman12b,Piro13a,Mazzali13}, a high-velocity region of either clumps or an amorphous plumage of opaque Si-, Ca-based material \citep{Gamezo04,Leonard05,Wang07,Maund10a,Piro11}, and the supposed blue/red-shift of nebular lines emitted from the inner IPE-rich material \citep{Maedanature,McClelland13,Silverman13latetime}. 

For at least normal SN Ia, there remain two viable explosion channels (with a few sub- and super-M$_{Ch}$ sub-channels) regardless of the hierarchical dominance of each at various redshifts and/or ages of galactic constituents (c.f., \citealt{Roepke12,Hachisu12a,Seitenzahl13,Pakmor13,Moll13,Claeys14}). Also, it may or may not be the case that some SN~Ia are 2$+$ subtypes viewed upon from various lines of sight amidst variable CSM interaction \citep{Maedanature,Foley12dust,Scalzo12,Leloudas13,Dan13,Moll13,Dessart13models}. However, with the current lack of complete observational coverage in wavelength, time, and mode (i.e. spectrophotometric and spectropolarimetric observations) for all SN~Ia subtypes and ``well-observed'' events, there is a limit for how much constraint can be placed on many of the proposed explosion models and progenitor scenarios. That is, despite observational indications for and theoretical consistencies with the supposition of multiple progenitor channels, the observed diversity of SN~Ia does not yet necessitate that each spectrophotometric subtype be from a distinctly separate explosive binary scenario than that of others within the SN~Ia family of observed events; particularly so for normal SN~Ia.

For the purposes of testing the multifaceted predictions of theoretical explosion models, time series spectroscopic observations of SN~Ia serve to visualize the post-explosion material of an unknown progenitor system. For example during the summer of 2011 astronomers bore witness to SN~2011fe, the best observed normal type Ia supernova of the modern era. The prompt discovery and follow-up of this nearby event uniquely allowed for a more complete record of observed properties than all previous well-observed events. More specifically, the full range (in wavelength and time) of rapid spectroscopic changes was documented with continual day-to-day follow-up into the object's post-maximum light phases and well beyond. However, the observational side of visualizing other SN Ia remains inefficient without the logistical coordination of many telescope networks (e.g., LCOGT; \citealt{LCOGT}), telescopes large enough to make nearly all SN ``nearby'' in terms of improved signal-to-noise ratios (e.g., The Thirty Meter Telescope, The Giant Magellan Telescope), or a space-based facility dedicated to the study of such time sensitive UV$-$optical$-$NIR transients. 

Existing SN~Ia surveys are currently acting toward optimizing a steady flow of discoveries, while other programs have produced a significant number of publicly available spectra \citep{Richardson01,Matheson08,Blondin12,WISEREP,Silverman12spectra,Folatelli13}. However, for the longterm future we believe it is imperative to begin a discussion of a larger (digital) network of international collaboration by way of (data-) cooperative competition like that done for both The Large Hadron Collider Experiment and Fermi Lab's Tevatron, with multiple competing experiments centered about mutual goals and mutual resources. Otherwise we feel the simultaneous collection of even very high quality temporal datasets by multiple groups will continue to create an inefficient pursuit of over-observing the most high profile event(s) of the year with a less than complete dataset. 

Such observational pursuits require an increasingly focused effort toward observing bright and nearby events. For example, 206 supernovae were reported in 1999 and 67 were brighter than 18th magnitude while only three reached $\sim$ 13 magnitude\footnote[21]{Quoted from the archives page of \url{http://www.rochesterastronomy.org/snimages/}.}. By 2012 the number of found supernovae increased to 1045 while 78 were brighter than 16th magnitude and five brighter than 13th magnitude. This clearly indicates that supernovae caught early are more prevalent than $\sim$ 15 years ago and it is worthwhile for multiple groups to continually increase collaborative efforts for the brightest events. Essentially this could be accomplished without interfering with spectrum-limited high[er]-z surveys by considering a distance threshold ($\lesssim$ 10$-$30 Mpc) as part of the public domain. Additionally, surveys that corroborate the immediate release of discoveries would further increase the number of well-observed events and could be supplemented and sustained with staggered observations given that there are two celestial hemispheres, unpredictable weather patterns, and caught-early opportunities nearly every week during active surveying.

In conclusion, to extract details of the spectroscopic behavior for all SN~Ia subtypes, during all phases, larger samples of {\it well-observed} events are essential, beginning from as close to the onset of the explosion as possible (e.g., SN~1999ac, 2009ig, 2011fe, 2012cg, 2012fr), where SN~Ia homogeneity diverges the most (see \citealt{Zheng13} for the most recent instance in SN 2013dy). Near-continuous temporal observations are most important for at least the first 1$-$2 months post-explosion and biweekly to monthly follow-up thereafter for $\sim$ 1 year. SN~Ia spectra are far too complicated to do so otherwise. Even normal SN~Ia deserve UV$-$optical$-$IR spectroscopic follow-up at a 1:1 to 2:1 ratio between days passed and spectrum taken, whenever possible, given that fine differences between normal SN~Ia detail the variance in explosion mechanism parameters and initial conditions of their unobserved progenitor systems. It is through such observing campaigns that the true diversity to the underlying nature of SN~Ia events will be better understood.


 \acknowledgments
 
This work was supported in part by NSF grant AST-0707704, and US DOE
Grant DE-FG02-
07ER41517, and by SFB 676, GRK 1354 from the DFG. Support for Program
number HST-GO-
12298.05-A was provided by NASA through a grant from the Space
Telescope Science Institute,
which is operated by the Association of Universities for Research in
Astronomy, Incorporated, under
NASA contract NAS5-26555. We wish to acknowledge the use of the \citet{Kurucz95} line list and \url{colorbrewer2.org} for the construction of Figure~\ref{fig:lineblending101}.

This review was made possible by collaborative discussions at the 2011 UC-UC-HIPACC International AstroComputing Summer School on Computational Explosive Astrophysics. We would like to thank Dan Kasen and Peter Nugent for organizing the program and providing the environment for a productive summit, and we hope that such programs for summer learning opportunities will continue in the future. We worked on this review during the visits of MP to the Homer L. Dodge
Department of Physics and Astronomy, University of Oklahoma,
Norman, OK, USA.,  McDonnell Center for the Space Science, Department of
Physics, Washington University in St. Louis, USA., National Astronomical
Observatory of Japan (NAOJ), Mitaka, Tokyo, Japan., and Inter-University
Centre for Astronomy and Astrophysics (IUCAA), Pune, India. 

MP is thankful to Prof. David Branch, Prof. Eddie Baron, Prof. Ramanath Cowsik,
Prof. Shoken Miyama, Prof. Masahiko Hayashi, Prof. Yoichi Takeda, Prof. Wako Aoki, Prof. Ajit Kembhavi, Prof. Kandaswamy Subramanian, and Prof. T. Padamanabhan for their kind support, encouragement, and hospitality. JTP would like to thank the University of Oklahoma Supernova Group, Rollin Thomas, and Alicia Soderberg for several years of support and many enlightening discussions on reading supernova spectra. JTP wishes to acknowledge helpful discussions with B. Dilday, R.~A. Fesen, R. Foley, M.~L. Graham, D.~A. Howell, G.~H. Marion, D. Milisavljevic, P. Milne, D. Sand, and S. Valenti, as well as S. Perlmutter for an intriguing conversation on ``characteristic information of SN~Ia'' at the 221st American Astronomical Society Meeting in Long Beach, CA. JTP is also indebted to Natalie Buckley-Medrano for influential comments on the text and figures presented here.

Finally, we would like to pay special tribute to our referee, Michael Childress, whose critical comments and suggestions were substantially helpful for the presentation of this review. 

\clearpage

\begin{table}
\centering
\caption{References for Spectra in Figures~\ref{fig:fig5}$-$\ref{fig:fig13}}
\begin{tabular}{ll}
\tableline\tableline
SN Name & References \\
\tableline
SN~1981B & \citealt{Branch83} \\
SN~1986G & \citealt{Cristiani92} \\
SN~1989B & \citealt{Barbon90}; \\
 & \citealt{Wells94} \\
SN~1990N & \citealt{Mazzali93}; \\
 & \citealt{Gomez98} \\
SN~1991T & \citealt{Flipper91T}; \\
 & \citealt{Gomez98} \\
SN~1991bg & \citealt{Flipper92}; \\
 & \citealt{Turatto96} \\
SN~1994D & \citealt{Patat96}; \\
 & \citealt{Ruiz97}; \\
& \citealt{Gomez98}; \\
 & \citealt{Blondin12} \\
SN~1994ae & \citealt{Blondin12} \\ 
SN~1995D & \citealt{Sadakane96}; \\
 & \citealt{Blondin12} \\
SN~1996X & \citealt{Salvo01}; \\
 & \citealt{Blondin12} \\
SN~1997br & \citealt{WLi99}; \\
 & \citealt{Blondin12} \\
SN~1997cn & \citealt{Turatto98} \\
SN~1998aq & \citealt{Branch03} \\
SN~1998bu & \citealt{Jha99}; \\ 
 & \citealt{Matheson08} \\
SN~1998de & \citealt{Modjaz01}; \\
 & \citealt{Matheson08} \\
SN~1998es & \citealt{Matheson08} \\
SN~1999aa & \citealt{Garavini04} \\
SN~1999ac & \citealt{Garavini05}; \\
 & \citealt{Phillips06}; \\
 & \citealt{Matheson08} \\
SN~1999by & \citealt{Garnavich04}; \\
 & \citealt{Matheson08} \\
SN~1999ee & \citealt{Hamuy02} \\
SN~2000E & \citealt{Valentini03} \\
SN~2000cx & \citealt{WLi01} \\
SN~2001V & \citealt{Matheson08} \\
SN~2001ay & \citealt{Krisciunas11} \\
SN~2001el & \citealt{Wang03a} \\
SN~2002bo & \citealt{Benetti04}; \\
 & \citealt{Blondin12} \\
SN~2002cx & \citealt{WLi03} \\

\tableline
\end{tabular}
\end{table}

\begin{table}
\centering
\caption{References for Spectra in Figures~\ref{fig:fig5}$-$\ref{fig:fig13}}
\begin{tabular}{ll}
\tableline\tableline
SN Name & References \\
\tableline
SN~2002dj & \citealt{Pignata08}; \\
 & \citealt{Blondin12} \\
SN~2002er & \citealt{Kotak05} \\
SN~2003cg & \citealt{Elias06}; \\
 & \citealt{Blondin12} \\
SN~2003du  & \citealt{Gerardy04}; \\
 & \citealt{Anupama05}; \\
 & \citealt{Leonard05}; \\
 & \citealt{Stanishev07}; \\
 & \citealt{Blondin12} \\
SN~2003hv  & \citealt{Leloudas09}; \\
 & \citealt{Blondin12} \\
SN~2004S & \citealt{Krisciunas07} \\
SN~2004dt  & \citealt{Leonard05}; \\
 & \citealt{Altavilla07}; \\
 & \citealt{Blondin12} \\
SN~2004eo  & \citealt{Pastorello07a} \\
SN~2005am & \citealt{Blondin12} \\
SN~2005cf  & \citealt{Garavini07}; \\
 & \citealt{WangX09}; \\
 & \citealt{Bufano09} \\
 SN~2005cg  & \citealt{Quimby06} \\
SN~2005hk & \citealt{Chornock06}; \\
 & \citealt{Phillips07}; \\
 & \citealt{Blondin12} \\
SN~2005hj  & \citealt{Quimby07} \\
SN~2006D & \citealt{Blondin12} \\
SN~2006X & \citealt{WangX08a}; \\
 & \citealt{Yamanaka09b}; \\
 & \citealt{Blondin12} \\
SN~2006bt & \citealt{Foley10b} \\
SN~2006gz & \citealt{Hicken07} \\
SN~2007ax & \citealt{Blondin12} \\
SN~2007if & \citealt{Silverman11}; \\
 & \citealt{Blondin12} \\
SN~2008J & \citealt{Taddia12} \\
SN~2008ha & \citealt{Foley09} \\
SN~2009dc & \citealt{Silverman11}; \\
 & \citealt{Taubenberger11} \\
PTF09dav & \citealt{Sullivan11} \\
SN~2011fe & \citealt{Parrent12} \\
SN~2011iv & \citealt{Foley12b} \\
\tableline
\end{tabular}
\end{table}

\clearpage
\newpage

\begin{table}
\centering
\caption{References for M$_{\emph{B}}$(Peak) and $\Delta$m$_{15}$(\emph{B}) plotted in Figure~\ref{fig:dm15}: 1981$-$1992}
\begin{tabular}{ll}
\tableline\tableline
SN Name & References \\
\tableline
SN~1981B & \citealt{Leibundgut93}; \\
 & \citealt{Saha96}; \\
 & \citealt{Hamuy96}; \\
 & \citealt{Saha01b} \\
SN~1984A & \citealt{Barbon89} \\
SN~1986G & \citealt{Flipper92}; \\
 & \citealt{Ruiz92}; \\
 & \citealt{Leibundgut93} \\
SN~1989B & \citealt{Barbon90}; \\
 & \citealt{Wells94}; \\
 & \citealt{Richmond95}; \\
 & \citealt{Saha99}; \\
 & \citealt{Contardo00}; \\
 & \citealt{Saha01a} \\
SN~1990N & \citealt{Saha97}; \\
 & \citealt{Lira98}; \\
 & \citealt{Saha01a} \\
SN~1991T & \citealt{Leibundgut93}; \\
 & \citealt{Lira98}; \\
 & \citealt{Phillips99}; \\
 & \citealt{Krisciunas04}; \\
 & \citealt{Contardo00}; \\
 & \citealt{Saha01b}; \\
 & \citealt{Tsvetkov11} \\
SN~1991bg & \citealt{Leibundgut93}; \\
 & \citealt{Turatto96}; \\
 & \citealt{Mazzali97}; \\
 & \citealt{Contardo00} \\
SN~1992A & \citealt{Leibundgut93}; \\
 & \citealt{Hamuy96}; \\
 & \citealt{Drenkhahn99}; \\
 & \citealt{Contardo00} \\
SN~1992K & \citealt{Hamuy94} \\
SN~1992al & \citealt{Misra05} \\
SN~1992bc & \citealt{Maza94}; \\
 & \citealt{Contardo00} \\
SN~1992bo & \citealt{Maza94}; \\
 & \citealt{Contardo00} \\
\tableline
\end{tabular}
\end{table}

\begin{table}
\centering
\caption{References for M$_{\emph{B}}$(Peak) and $\Delta$m$_{15}$(\emph{B}) plotted in Figure~\ref{fig:dm15}: 1994$-$1999}
\begin{tabular}{ll}
\tableline\tableline
SN Name & References \\
\tableline
SN~1994D & \citealt{Hoflich95}; \\
 & \citealt{Richmond95}; \\
 & \citealt{Patat96}; \\
 & \citealt{Vacca96}; \\
 & \citealt{Drenkhahn99}; \\
 & \citealt{Contardo00} \\
SN~1994ae & \citealt{Contardo00} \\
SN~1995D & \citealt{Sadakane96}; \\
 & \citealt{Contardo00} \\
SN~1996X & \citealt{Phillips99}; \\
 & \citealt{Salvo01} \\
SN~1997br & \citealt{WLi99} \\
SN~1997cn & \citealt{Turatto98} \\
SN~1998aq & \citealt{Riess99}; \\
 & \citealt{Saha01a} \\
SN~1998bu & \citealt{Jha99}; \\
 & \citealt{Hernandez00}; \\
 & \citealt{Saha01a} \\
SN~1998de & \citealt{Modjaz01} \\
SN~1998es & \citealt{Jha06b}; \\
 & \citealt{Tsvetkov11} \\
SN~1999aa & \citealt{Krisciunas00}; \\
 & \citealt{WLi03}; \\
 & \citealt{Tsvetkov11} \\
SN~1999ac & \citealt{Jha06b}; \\
 & \citealt{Phillips06} \\
SN~1999aw & \citealt{Strolger02} \\
SN~1999by & \citealt{Vinko01}; \\
 & \citealt{Howell01a}; \\
 & \citealt{Garnavich04}; \\
 & \citealt{Sullivan11} \\
SN~1999ee & \citealt{Stritzinger02}; \\
 & \citealt{Krisciunas04} \\
\tableline
\end{tabular}
\end{table}

\clearpage
\newpage

\begin{table}
\centering
\caption{References for M$_{\emph{B}}$(Peak) and $\Delta$m$_{15}$(\emph{B})  plotted in Figure~\ref{fig:dm15}: 2000$-$2005}
\begin{tabular}{ll}
\tableline\tableline
SN Name & References \\
\tableline
SN~2000E & \citealt{Valentini03} \\
SN~2000cx & \citealt{WLi01}; \\
 & \citealt{Candia03}; \\
 & \citealt{Sollerman04} \\
SN~2001V & \citealt{Vinko03} \\
SN~2001ay & \citealt{Krisciunas11} \\
SN~2001el & \citealt{Krisciunas03}\\
SN~2002bo & \citealt{Benetti04}; \\
 & \citealt{Stehle05} \\
SN~2002cv & \citealt{Elias08} \\
SN~2002cx & \citealt{WLi03} \\
SN~2002dj & \citealt{Pignata08} \\
SN~2002er & \citealt{Pignata04} \\
SN~2003cg & \citealt{Elias06} \\
SN~2003du & \citealt{Anupama05}; \\
 & \citealt{Stanishev07}; \\
 & \citealt{Tsvetkov11} \\
SN~2003fg & \citealt{Howell06}; \\
 & \citealt{Yamanaka09b} \\
 & \citealt{Scalzo10} \\
SN~2003hv & \citealt{Leloudas09} \\
SN~2004S & \citealt{Misra05}; \\
 & \citealt{Krisciunas07} \\
SN~2004dt & \citealt{Altavilla07} \\
SN~2004eo & \citealt{Pastorello07a} \\
SN~2005am & \citealt{Brown05} \\
SN~2005bl & \citealt{Taubenberger08}; \\
 & \citealt{Hachinger09} \\
SN~2005cf & \citealt{Pastorello07b}; \\
 & \citealt{WangX09} \\
SN~2005hk & \citealt{Phillips07} \\
\tableline
\end{tabular}
\end{table}

\begin{table}
\centering
\caption{References for M$_{\emph{B}}$(Peak) and $\Delta$m$_{15}$(\emph{B})  plotted in Figure~\ref{fig:dm15}: 2006$-$2012}
\begin{tabular}{ll}
\tableline\tableline
SN Name & References \\
\tableline
SN~2006X & \citealt{WangX08b} \\
SN~2006bt & \citealt{Hicken09}; \\
 & \citealt{Foley10b} \\
SN~2006gz & \citealt{Hicken07}; \\
 & \citealt{Scalzo10} \\
SN~2007ax & \citealt{Kasliwal08} \\
SN~2007if & \citealt{Scalzo10} \\
SN~2007qd & \citealt{McClelland10} \\
SN~2008J & \citealt{Taddia12} \\
SN~2008ha & \citealt{Foley09} \\
SN~2009dc & \citealt{Yamanaka09b}; \\
 & \citealt{Scalzo10}; \\
 & \citealt{Silverman11}; \\
 & \citealt{Taubenberger11} \\
SN~2009ig & \citealt{Foley12c} \\
SN~2009ku & \citealt{Narayan11} \\
SN~2009nr & \citealt{Khan11}; \\
 & \citealt{Tsvetkov11} \\
PTF09dav & \citealt{Sullivan11} \\
SN~2010jn & \citealt{Hachinger13} \\
SN~2011fe & \citealt{Richmond12}; \\
 & \citealt{Munari13}; \\
 & \citealt{Pereira13} \\
SN~20011iv & \citealt{Foley12b} \\
SN~2012cg & \citealt{Silverman12b}; \\
 & \citealt{Munari13} \\
SN~2012fr & \citealt{Childress13} \\
\tableline
\end{tabular}
\end{table}

\clearpage
\newpage

\begin{appendices}

\section{Some recent SN~Ia, Continued}

\subsection{Peculiar SN~1997br in ESO 576-G40}\label{ss:97br}

\citet{WLi99} presented observations of the peculiar SN~1991T-like,
SN~1997br \citep{Bao97CBET}.
\citet{Hatano02} analyzed the spectra of SN~1997br and raised the
question of whether or not \ion{Fe}{3} and \ion{Ni}{3} features in the early spectra
are produced by $^{54}$Fe and $^{58}$Ni rather than by $^{56}$Fe and
$^{56}$Ni. In addition, \citet{Hatano02} discussed the issue of
SN~1991T-like events as more powerful versions of normal SN~Ia, rather
than a physically distinct subgroup of events.

\subsection{SN~1997cn in NGC 5490}\label{ss:97cn}

   \citet{Turatto98} studied the faint SN~1997cn, which is located in an
   elliptical host galaxy \citep{Li97CBET,Turatto97CBET}. Like SN~1991bg, spectra of SN~1997cn show a
   deep \ion{Ti}{2} trough between 4000 and 5000 \AA, strong \ion{Ca}{2} IR3 absorption features, a large $\mathcal{R}$(``\ion{Si}{2}''), and slow
  mean expansion velocities.

\subsection{SN~1997ff and other ``farthest known'' SN~Ia}\label{ss:97ff}

   With a redshift of z = 1.7, SN~1997ff was the most distant SN~Ia
discovered at that time \citep{Riess01,Benitez02}. There have been $\sim$110 high-z (1~$<$~z~$<$~2) SN~Ia discoveries since SN~1997ff \citep{Riess04,Riess07,Suzuki12}, with the most recent and and one of the most distant SN~Ia known being ``SN UDS10Wil'' at z = 1.914 \citep{Jones13}. With these and future observations of ``highest-z'' SN~Ia, constraints on DTD timescales \citep{Strolger10,Graur13} and dark energy \citep{Rubin13} will certainly improve.
 
\subsection{SN~1998aq in NGC 3982}\label{ss:98aq}

  \citet{Branch03} used \texttt{SYNOW} to study 29 optical spectra of the normal SN~1998aq \citep{Hurst98CBET}, covering 9 days before to 241 days after maximum light (days $-$9 and $+$241, respectively). Notably, they find evidence for \ion{C}{2} down to 11,000 km~s$^{-1}$, $\sim$ 3000 km~s$^{-1}$ below the cutoff of carbon in the pure deflagration model, W7 \citep{Nomoto84}. 

\subsection{SN~1999aa in NGC 2595}

From day $-$11 to day $+$58, \citet{Garavini04} obtained 25 optical spectra of SN~1999aa \citep{Armstrong99CBET}. While SN~1999aa appears SN~1991T-like, \citet{Garavini04} note that the \ion{Ca}{2} absorption feature strengths are between those of SN~1991T  (SS) and the SN~1990N (CN), along with a phase transition to normal SN~Ia characteristics that sets in earlier than SN~1991T. Subsequently, they suggest SN~1999aa to be a link between SN~1991T-likes and spectroscopically normal SN~Ia. Evidence of carbon-rich material is also found in SN~1999aa; decisively as \ion{C}{2} $\lambda$6580, tentatively as \ion{C}{3} $\lambda$4649 (see \citealt{Parrent11}). A schematic representation of their \texttt{SYNOW} fitting results is also presented, showing the inference of \ion{Co}{2}, \ion{Ni}{2}, and \ion{Ni}{3} during the pre-maximum phases. These results deserve further study from more detailed models.

\subsection{SN~1999ac in NGC 2841}\label{ss:99ac}

Between day $-$15 and day $+$42, \citet{Garavini05} obtained spectroscopic observations of the
unusual SN~1999ac \citep{Modjaz99CBET}. The pre-maximum light spectra are
similar to that of SN~1999aa-like, while appearing spectroscopically normal during later epochs.
\citet{Garavini05} find evidence of a fairly conspicuous, heavily blended
\ion{C}{2} $\lambda$6580 feature in the day $-$15 spectrum with approximate ejection velocities 
$>$16,000 km~s$^{-1}$. By day $-$9, the \ion{C}{2} absorption feature is weak or absent amidst blending with the neighboring 6100 \AA\ feature. This alone indicates that studies cannot fully constrain SN Ia models without spectra prior to day $-$10. 

\subsection{SN~1999aw in a low luminosity host galaxy}\label{ss:99aw}

   \citet{Strolger02} find SN~1999aw to be a luminous,
slow-declining SN~Ia, similar to 1999aa-like
events. \citet{Strolger02} derive a peak luminosity of 1.51~x~10$^{43}$ and a $^{56}$Ni mass of 0.76 M$_{\odot}$.       

\subsection{SN~1999by in NGC 2841}\label{ss:99by}

\citet{Vinko01} presented and discussed the first three pre-maximum light spectra of SN~1999by \citep{Arbour99CBET}, where they find it to be a sub-luminous SN~Ia similar to SN~1991bg \citep{Flipper92,Leibundgut93,Turatto96}, SN~1992K \citep{Hamuy94}, SN~1997cn \citep{Turatto98}, and SN~1998de \citep{Modjaz01}; in addition, the list of sub-luminous SN~Ia include SN~1957A, 1960H, 1971I, 1980I, 1986G (see \citealt{Branch93,Doull11} and references therein) and several other recently discovered under-luminous SN~Ia \citep{Howell01b,McClelland10,Hachinger09}. Pre-maximum spectra of SN~1999by show relatively strong features due to O, Mg, and Si, which are
due to explosive carbon burning. In addition, blue wavelength regions reveal spectra dominated by \ion{Ti}{2} and some other IPEs.  

Meanwhile, \citet{Hoflich02} studied the infrared spectra of SN~1999by, covering from day $-$4 to day $+$14. Post-maximum spectra show features which
can be attributed to incomplete Si burning, while further support for incomplete burning comes from the detection of a pre-maximum \ion{C}{2} absorption feature \citep{Garnavich04}. \citet{Hoflich02} analyzed the spectra
 through the construction of an extended set of delayed detonation
models covering the entire range of normal to sub-luminous SN~Ia. They
estimate the $^{56}$Ni mass for SN~1999by to be on the order of 0.1 M$_{\odot}$. \citet{Garnavich04} obtained \emph{UBVRIJHK} light curves of SN~1999by. From the photometry of SN~1999by, the
recent Cepheid distance to NGC 2841 \citep{Macri01}, and minimal dust extinction along the line-of-sight, \citet{Garnavich04} derive a peak absolute magnitude
of M$_{\emph{B}}$ = $-$17.15.

In order to assess the role of material asymmetries as being responsible for the observed peculiarity of sub-luminous SN~Ia, \citet{Howell01a} obtained polarization spectra of SN~1999by near maximum light. They find relatively low levels of polarization (0.3\%$-$0.8\%), however significant enough to be consistent with a 20\% departure from spherical symmetry \citep{Maund10a}.

\subsection{SN~1999ee in IC 5179}\label{ss:99ee}

 From day $-$10 to day $+$53, \citet{Stritzinger02} obtained well-sampled \emph{UBVRIz} light curves
of SN~1999ee \citep{Maza99CBET}. They find the \emph{B}-band light curve is broader than normal SN~Ia, however sitting toward the over-luminous end of SN~Ia peak brightnesses, with M$_{\emph{B}}$ = $-$19.85 $\pm$ 0.28 and $\Delta$m$_{15}$ = 0.94. 

\citet{Hamuy02} obtained optical and infrared spectroscopy of
SN~1999ee between day $-$9 and day $+$42. Before maximum light, the spectra of SN~1999ee are normal, with relatively strong \ion{Si}{2} 6100 \AA\ absorption, however within the SS subtype \citep{Branch09}. \citet{Hamuy02} compared the infrared spectra of SN~1999ee
to that of other SN~Ia out to 60 days post-explosion, and find similar characteristics for SN~1999ee and 1994D \citep{Meikle96}.   

\subsection{SN~2000E in NGC 6951}\label{ss:00E}

     \citet{Valentini03} obtained \emph{UBVRIJHK} photometry and
optical spectra of SN~2000E, which is located in a spiral galaxy. Optical spectra
were obtained from 6 days before \emph{B}-band maximum to 122 days after \emph{B}-band maximum. The
photometric observations
span 230$+$ days, starting at day $-$16. 
The photometric light curves of SN~2000E are similar to other SS SN~Ia, however SN~2000E is classified as a slowly declining, spectroscopically ``normal'' SN~Ia
similar to SN~1990N. \citet{Valentini03} estimate the $^{56}$Ni mass to be
0.9 M$_{\odot}$ from the bolometric light curve.

\subsection{SN~2000cx in NGC 524}\label{ss:00cx}

One of the brightest supernovae observed in the year 2000 was the peculiar SN~2000cx, located in an S0 galaxy \citep{Yu00CBET,WLi01, Candia03}. It was classified as a
SN~Ia with a spectrum resembling that of the peculiar SN~1991T \citep{Chornock00}. \citet{Sollerman04} obtained late time \emph{BVRIJH} light curves
 of SN~2000cx covering 360 to 480 days after maximum. During these epochs, they find relatively constant NIR magnitudes,
indicating the increasing
 importance (with time) of the NIR contribution to the bolometric light curve. 
 
\citet{Branch04c} decomposed the photospheric-phase spectra of SN~2000cx with \texttt{SYNOW}. Apart from confirming HVFs of \ion{Ca}{2} IR3 (which are consistent with primordial abundances; \citealt{Thomas04}), \citet{Branch04c} also find HVFs of \ion{Ti}{2}. They attribute the odd behavior of SN~2000cx's \emph{B}-band light curve to the time-dependent behavior of these highly line blanketing \ion{Ti}{2} absorption signatures. \citet{Branch04c} find an absorption feature near 4530 \AA\
in the spectra of SN~2000cx that can be tentatively associated with H$\beta$ at high velocities, however this feature is more likely due to \ion{C}{3} $\lambda$4649 or \ion{S}{2}/\ion{Fe}{2} instead (see \citealt{Parrent11} and references therein). 

\citet{Rudy02} obtained 0.8$-$2.5 $\mu$m spectra of SN~2000cx at day $-$7 and day $-$8 before maximum light. From the $\lambda$10926 line of \ion{Mg}{2}, they find that carbon-burning has taken place up to $\sim$25,000 km~s$^{-1}$. Given the SS subtype nature of SN~2000cx, the early epoch IR spectra of \citet{Rudy02} are valuable for comparison with other SN~Ia IR datasets.

\subsection{SN~2001V in NGC 3987}\label{ss:01V}

\citet{Vinko03} presented photometry of SN~2001V \citep{Jha01CBET}. They find that SN~2001V is over-luminous, relative
to the majority of SN~Ia. Spectroscopic observations, spanning from day $-$14 to day $+$106, can be found in \citet{Matheson08} and reveal it to be a SS SN~Ia, consistent with its observed brightness. 

\subsection{Slowly declining SN~2001ay in IC 4423}\label{ss:01ay}

    \citet{Krisciunas11} obtained optical and near infrared photometry,
and optical and UV spectra of SN~2001ay \citep{Swift01CBET}. They find maximum light
 \ion{Si}{2} and \ion{Mg}{2} line velocities of $\sim$ 14,000 km~s$^{-1}$,
with \ion{Si}{3} and \ion{S}{2} near 9,000 km~s$^{-1}$. SN~2001ay is one of the
most slowly declining SN~Ia. However, a $\Delta$m$_{15}$(\emph{B}) =
0.68 is odd given it is not over-luminous like SCC SN~Ia slow decliners. In fact, the $^{56}$Ni yield of 0.58 is comparable to that of many normal SN~Ia. \citet{Baron12} note this apparent WLR violation is related to a decrease in $\gamma$-ray trapping deeper within the ejecta due to an overall outward shift of $^{56}$Ni, thus creating a fast rise in brightness followed by a slow decline caused by enhanced heating of the outer regions of material, which is a consequence of the larger expansion opacities. 

\subsection{SN~2001el in NGC 1448}\label{ss:01el}

    \citet{Krisciunas03} obtained well-sampled \emph{UBVRIJHK} light curves
of the nearby (about 18 Mpc) and normal SN~2001el \citep{Monard01}, from day $-$11 to day $+$142. Because \citet{Krisciunas03} obtained \emph{UBVRI} and \emph{JHK} light curves, they were able to measure a true optical$-$NIR reddening value (A$_{\emph{V}}$ = 0.57 mag along the line-of-sight) for the first time.
 
  \citet{Mattila05} obtained early time high resolution and low resolution 
optical spectra of SN~2001el. They estimate the mass loss rate (assuming 10$-$50 km~s$^{-1}$ wind velocities) from the
progenitor system of SN~2001el to be no greater than 9 x 10$^{-6}$ M$_{\odot}$
yr$^{-1}$ and 5~x~10$^{-5}$ M$_{\odot}$ yr$^{-1}$, respectively. The low resolution spectrum was obtained 400 days after maximum light with
no apparent signatures of hydrogen Balmer lines. High velocity \ion{Ca}{2} was detected out to 34,000 km~s$^{-1}$, while the 6100 \AA\ absorption feature is suspected of harboring high velocity \ion{Si}{2} (see also \citealt{Kasen03}). 

\subsection{SN~2002bo in NGC 3190}\label{ss:02bo}

Between day $-$13 and day $+$102, \citet{Benetti04} collected optical and NIR spectra and photometry of the BL SN~2002bo \citep{Cacella02CBET,Krisciunas04}. Estimates on host galaxy extinction from Na D equivalent width measurements are consistent with the inferred color excess determined by comparison to the Lira relation \citep{Lira95,Riess96,Phillips99}. From the time-evolution of the 6100 \AA\ absorption feature, \citet{Benetti04} find that SN~2002bo is an intermediary between the BL SN~1984A and the CN SN~1994D. \citet{Benetti04} also discuss SN~2002bo argue that some of the IME high velocity material may be primordial, while most is produced during the explosion and possibly by prolonged burning in a delayed detonation. This interpretation is also consistent with a lack of any clear signatures of unburned carbon. \citet{Stehle05} studied the abundance stratification by fitting a series of spectra with a Monte Carlo code and found that the elements
synthesized in different stages of burning are not completely mixed within the
ejecta. In the case of SN~2002bo, they derived the total mass of $^{56}$Ni to be 0.52 M$_{\odot}$. Similar to SN~2001ay's fast rise to maximum light \citep{Baron12}, \citet{Stehle05} attribute SN~2002bo's fast rise to outward mixing of $^{56}$Ni.

\subsection{SN~2002cv in NGC 3190}\label{ss:02cv}

 The NIR photometry of SN~2002cv reveal an obscured SN~Ia \citep{DiPaola02}; more than 8 magnitudes of visual extinction. Both optical and NIR spectroscopy indicate SN~2002cv is most similar to SN~1991T \citep{Meikle02CBET,Filippenko02CBET}. It should also be noted that the SS SN~2002cv and the BL SN~2002bo share the same host galaxy. \citet{Elias08} obtained and analyzed VRIJHK photometry, in addition to a sampling of optical and NIR spectroscopy near and after maximum light, and find a best fit value for the ratio between inferred extinction and reddening, R$_{{\it V}}$~=~1.59 $\pm$ 0.07 whereas 3.1 is often assumed for normal SN~Ia (however see \citealt{Tripp98,Astier06,Krisciunas06,Folatelli10}). They suggest this to indicate varying mean grain sizes for the dust along the line of sight toward SN~2002bo and 2002cv. 
  
  \subsection{SN~2002dj in NGC 5018}\label{ss:02dj}

For two years, and starting from day $-$11, \citet{Pignata08} monitored the optical and IR behaviors
of the SN~2002bo-like, high-velocity gradient SN~2002dj \citep{Hutchings02CBET}. The dataset presented make it one of the most well-observed SN~1984A-like SN~Ia and is a valuable tool for the discussion of SN~Ia diversity. 

\subsection{SN~2002er in UGC 10743}\label{ss:02er}

      From day $-$11 to day $+$215, \citet{Kotak05} carried out spectroscopic follow-up for the reddened, CN SN~2002er \citep{Wood-Vasey02erCBET}. By contrast with the photometric behavior seen for SN~1992A, 1994D, and 1996X, SN~2002er stands out for its slightly delayed second peak in the \emph{I}-band and similarly for \emph{V} and \emph{R}-bands. \citet{Pignata04,Kotak05} estimated the
mass of $^{56}$Ni to be on the order of 0.6 to 0.7 M$_{\odot}$, where the uncertainty
in the exact distance to SN~2002er was the primary limitation.

\subsection{SN~2003du in UGC 09391}\label{ss:03du}

For 480 days, and starting from day $-$13, \citet{Stanishev07} monitored the CN SN~2003du. From modeling of the bolometric light curve, \citet{Stanishev07} estimate the mass of $^{56}$Ni to between 0.6 and 0.8 M$_{\odot}$. Like other normal SN~Ia, the early spectra of SN~2003du contain HVFs of \ion{Ca}{2} and a 6100 \AA\ feature that departs from being only due to photospheric \ion{Si}{2}, suggesting either a distinctly separate region of HV \ion{Si}{2} or the radial extension of opacities from below.

   \citet{Tanaka11} studied the chemical composition distribution
in the ejecta of SN~2003du by modeling a one year extended time series of optical spectra. \citet{Tanaka11} do not find SN~2003du to be as fully mixed as a some 3D deflagration models.
Specifically, from their modeling \citet{Tanaka11} that the a core of stable IPEs supersedes $^{56}$Ni out to $\sim$ 3000~km~s$^{-1}$ ($\lesssim$~0.2 in mass coordinate). Atop this 0.65 M$_{\odot}$ of $^{56}$Ni are layers of IMEs, while the outermost layers consist of oxygen, some silicon, and no more than 0.016 M$_{\odot}$ of carbon above 10,500 km~s$^{-1}$. 

\subsection{SN~2003gs in NGC 936}\label{ss:03gs}

\citet{Krisciunas09} obtained near-maximum to late time optical and NIR observations of SN~2003gs, offering a chance to study the post-maximum light bolometric behavior of a fast declining SN~Ia that was sub-luminous at optical wavelengths, but of standard luminosity in NIR bands at maximum light. \citet{Krisciunas09} find $\Delta$m$_{15}$(\emph{B}) = 1.83 $\pm$ 0.02 and discuss comparisons to other fast decliners; namely, SN~2003hv \citep{Leloudas09}, SN~2004gs \citep{Folatelli10,Contreras10}, SN~2005bl \citep{Taubenberger08,Folatelli10,WoodVasey08}, and SN~2005ke, 2006gt, and 2006mr \citep{Folatelli10,Contreras10}. In particular, \citet{Krisciunas09} note that, in contrast to normal and over-luminous SN~Ia, the delay in the time of \emph{J}-band maximum from that of the \emph{B}-band, for fast decliners, is inversely proportional to the peak NIR magnitude. Furthermore, they discussed the possibility for two subsets of FAINT$-$CL fast decliners; those that do and do not show a \emph{J}-band peak before the \emph{B}-band (see also \citealt{Kattner12}); respectively, SN~1986G, 2003gs, 2003hv, and 2006gt, and SN~1991bg, 1999by, 2005bl, 2005ke, and 2006mr. 

\citet{Krisciunas09} conclude that differences in NIR opacity within the outer layers are responsible for dissimilar $\gamma$-ray trapping, and therefore longer \emph{J}-band than \emph{B}-band diffusion times for FAINT$-$CL SN~Ia that are fainter in the NIR. However, the origin (differences of explosion mechanism and/or progenitor systems) for this apparent `bimodal' difference in NIR opacity is not clear.

For SN~2003gs, \citet{Krisciunas09} used \emph{UBVRIJHK} photometry and Arnett's rule (\citealt{Arnett82}, but also see \citealt{Stritzinger05}) to estimate 0.25 M$_{\odot}$ of $^{56}$Ni was produced during the explosion. As for the optical spectra, SN~2003gs is similar to SN~2004eo (A.23) in that it is found to have absorption signatures that are consistent with its photometric characteristics; a FAINT$-$CL SN~Ia with a larger than normal $\mathcal{R}$(\ion{Si}{2}) and the presence of \ion{Ti}{2} features near 4000$-$4500~\AA. 

\subsection{SN~2003hv in NGC 1201}\label{ss:03hv}

    \citet{Leloudas09} studied SN~2003hv
out to very late phases (day $+$786). Notably, this seemingly spectroscopically normal SN~Ia has $\Delta$m$_{15}$(\emph{B}) = 1.61, while the late time light curves show a deficit in flux that follow the decay of radioactive $^{56}$Co, assuming
full and instantaneous positron trapping. \citet{Leloudas09} consider this as possibly due to a redistribution of flux for SN~2003hv (a.k.a. an infrared catastrophe, see \citealt{Axelrod80}) from a dense clumping of inner material, and would also explain the flat-topped nebular emission lines \citep{Motohara06,Gerardy07}. 

\citet{Mazzali11} also
studied the nebular spectrum of 2003hv and consider it as a
non-standard event. They note that its late time flux deficit, compared to normal SN~Ia, could be due to SN~2003hv having a lower mean density structure, possibly consistent with a sub-Chandrasekhar mass origin. 

\citet{Motohara06} presented NIR Subaru Telescope spectra of SN~2003du, 2003hv, and 2005W during their late phase evolution ($+$200 days post-maximum light). For both SN~2003du and 2003hv, they find a flat-topped [\ion{Fe}{2}] $\lambda$16440 emission feature that is blue-shifted by $\sim$ 2000 km~s$^{-1}$ from the SN rest frame (FWHM $\sim$ 4000 km~s$^{-1}$). \citet{Motohara06} further argue that the [\ion{Fe}{2}] emission would be rounded on top if the neutron-rich Fe-peak isotopes produced in the explosion were thoroughly mixed with the surrounding distribution of $^{56}$Ni; for SN~2003du and 2003hv they suggest that this is not the case. In fact, they find that SN~1991T and 2005W (see their Fig. 1), at least, may represent instances where the inner most regions have been thoroughly mixed. 

Similarly, \citet{Gerardy07} looked to address the nature of the thermonuclear burning front by utilizing late time ($+$135 days) mid-IR (5.2$-$15.2 $\mu$m) \emph{Spitzer\ Space\ Telescope} spectra of SN~2003hv and 2005df. In particular, \citet{Gerardy07} find direct evidence in SN~2005df for a small inner zone of nickel that is surrounded by $^{56}$Co and an asymmetric shell-like structure of Ar. While it is not clear \emph{why} a supposed initial deflagration phase of a DDT explosion mechanism produces little to no mixing for these two SN~Ia, the observations of \citet{Gerardy07} give strong support for a stratified abundance tomography like those seen in DDT-like models; the various species of material are restricted to radially confined zones, which is inconsistent with the large-scale mixing that is expected to occur in 3D deflagration models. This is also in agreement with X-ray observations of the Tycho supernova remnant \citep{Badenes06} in addition to optical and UV line resonance absorption imaging of SNR 1885 in M31 \citep{Fesen07}. 

\subsection{SN~2004dt in NGC 0799}\label{ss:04dt}

\citet{Wang06} and \citet{Altavilla07} studied the early spectral evolution
  of SN~2004dt from more than
a week before optical maximum, when line profiles show matter moving at
velocities as high as 25,000 km~s$^{-1}$. The variation of the
polarization across some \ion{Si}{2} lines approaches 2\%, making
SN~2004dt one of the most highly polarized SN~Ia observed and an outlier in the polarization-nebular velocity plane \citep{Maund10a}. In contrast with the polarization associated with \ion{Si}{2}, \citet{Wang06}
find that the strong 7400 \AA\ \ion{O}{1}$-$\ion{Mg}{2} absorption complex shows little or no
polarization signature. \citet{Wang06} conclude this is due to a spherical geometry of oxygen-rich material encompassing a lumpy distribution of IMEs.

\subsection{SN~2004eo in NGC 6928}\label{ss:04eo}

   \citet{Pastorello07a} presented optical and infrared observations
of the transitional normal, CL SN~2004eo \citep{Nakano04CBET}. The light curves and spectra appear normal (M$_{\emph{B}}$ = $-$19.08) while exhibiting low mean expansion velocities and a fast declining \emph{B}-band light curve ($\Delta$m$_{15}$(\emph{B}) = 1.46). The observed
properties of SN~2004eo signify it is intermediate between FAINT, LVG, and HVG SN~Ia. \citet{Mazzali08} also consider SN~2004eo as a spectroscopically normal SN~Ia that produced 0.43 $\pm$ 0.05 M$_{\odot}$ of $^{56}$Ni.

\subsection{SN~2005am in NGC 2811}\label{ss:05am}

 Between day $-$4 and day $+$69, \citet{Brown05} obtained UV, optical, and X-ray observations with the \emph{Swift} satellite of the SN~1992A-like 
   SN~2005am \citep{Kirshner93,Modjaz05CBET}. They place an upper limit
   on SN~2005am's X-ray luminosity (0.3$-$10~keV) of 6 x 10$^{39}$ erg s$^{-1}$.

\subsection{Under-luminous SN~2005bl in NGC 4070}\label{ss:05bl}

Both \citet{Taubenberger08} and \citet{Hachinger09} studied the sub-luminous SN
2005bl with observations made between
 day $-6$ and day $+$66, and carried out spectral analysis (``abundance tomography'') of SN
2005bl \citep{Morrell05CBET}. They find it to be one of incomplete burning similar to SN~1991bg and 1999by. Compared to SN~1991bg, a noteworthy difference of SN~1999by is the
likely presence of carbon \ion{C}{2} in pre-maximum spectra \citep{Taubenberger08}, whereas \ion{C}{1} $\lambda$10691 is also clearly detected in NIR spectra \citep{Hoflich02}. However, this is likely a biased comparison to SN~1991bg given that the earliest spectrum obtained was on day $-$1 (potentially too late to detect unburned material via \ion{C}{2} $\lambda$6580). To our knowledge no conspicuous \ion{C}{1} $\lambda$10691 absorption features have been documented for other SN~Ia. For example, \ion{C}{1} $\lambda$10691 is present in the pre-maximum spectra of SN~2011fe but it is not a conspicuous signature. Similarly, pre-maximum spectra of SN~2005bl show less conspicuous detections of \ion{C}{1} and \ion{C}{2} but still indicate low burning efficiency with a significant amount of
leftover unburned material \citep{Taubenberger08}. \citet{Hachinger09} suggest that a detonation at low pre-expanded densities is responsible for the abundance stratification of IMEs seen in the spectra of SN~2005bl. This would also explain the remaining carbon-rich material seen for some CL SN~Ia when caught early enough.

\subsection{SN~2005cf in MCG-01-39-003}\label{ss:05cf}

  \citet{WangX09} studied UV$-$optical$-$NIR observations of the normal SN~2005cf (see also \citealt{Pastorello07b}). During the early evolution of the spectrum, HVFs of \ion{Ca}{2} and \ion{Si}{2} are found to be present above 18,000 km~s$^{-1}$ (confirming observations of \citealt{Garavini07}). \citet{Gall12} studied the NIR spectra of
     SN~2005cf at epochs from day $-$10 to day $+$42, which show clear signatures of \ion{Co}{2} during post-maximum phases.
     In addition, they attribute fluorescence emission in making the underlying shape of the SED. 
     
\subsection{SN~2005cg in a low-luminosity, star forming host}\label{ss:05cg}

     \citet{Quimby06} presented and discussed
 the spectroscopic evolution and light curve of the SS SN~2005cg, which was discovered by ROTSE-IIIc. Pre-maximum spectra reveal HVFs of \ion{Ca}{2} and \ion{Si}{2} out to $\sim$ 24,000 km~s$^{-1}$ and \citet{Quimby06} find good consistency between observed and modeled \ion{Si}{2} profiles. They interpret the steep rise in the blue wing of the \ion{Si}{2} to be an indication of circumstellar interaction given that abundance estimates for HVFs suggest modest amounts of swept up material ($\sim$ 10$^{-4}$ $-$ 10$^{-3}$ M$_{\odot}$; see \citealt{Quimby06,Branch06}). 
 
\subsection{SN~2005hj}\label{ss:05hj}

 \citet{Quimby07} obtained optical spectra of the SS SN~2005hj during pre-maximum and post-maximum light phases. From a ROTSE-IIIb unfiltered light curve, SN~2005hj reached an over-luminous peak absolute magnitude of $-$19.6 (assuming z = 0.0574). Interestingly, the sharp and shallow 6100 \AA\ feature remains fairly stagnant at $\sim$ 10,600 km~s$^{-1}$ near and after maximum light, with a sudden decrease at later epochs. Similar to \citet{Quimby06}, \citet{Quimby07} find this also consistent with the interpretation that CSM is influencing spectral profiles of SN~1999aa-like SN~Ia (see also \citealt{Scalzo10,Scalzo12}). 
 
\subsection{SN~2006D in MCG-01-33-034}\label{ss:06D}

     \citet{Thomas07} obtained the spectra of the spectroscopically normal SN~2006D
from day $-$7 to day $+$13. The spectra show
one of the clearest signatures of carbon-rich material at photospheric velocities
observed in a \emph{normal} SN~Ia (below 10,000 km~s$^{-1}$). The 6300 \AA\ carbon feature
becomes weaker with time and disappears as the photosphere recedes and the SN reaches maximum brightness.
These observations$-$like all SN~Ia diversity studies$-$underscore the importance of obtaining spectra of SN~Ia during all phases.
If [\ion{O}{1}] and [\ion{C}{1}] lines are present in the spectra during post-maximum light phases
at velocities below 10,000 km~s$^{-1}$, this would indicate the presence of
unburned matter. These particular lines have not been detected, however the absence of said signatures does not imply a complete lack of C+O material at low velocities \citep{Baron03,Kozma05}.

\subsection{SN~2006X in M100}\label{ss:06X}

   \citet{WangX08b} presented \emph{UBVRI} and \emph{JK} light curves and optical
spectroscopy of the reddened BL SN~2006X \citep{Stockdale06CBET,Immler06CBET,Quimby06CBET} and find high mean expansion velocities during pre-maximum light phases ($\gtrsim$ 20,000 km~s$^{-1}$). \citet{WangX08b} suggest the observed properties of SN~2006X may be due to interaction with CSM. \citet{Yamanaka09b} also presented and discussed the early
 spectral evolution. They note that the $\mathcal{R}$(\ion{Si}{2}) ratio is unusually low for such a high-velocity gradient SN~Ia. However, 
 rather than this being an indication of low effective temperature, they suggest that the low $\mathcal{R}$(\ion{Si}{2}) value is due to line-blending, likely from a higher velocity component of \ion{Si}{2}. Both \citet{WangX08b} and \citet{Yamanaka09b} find the observed properties of SN~2006X to be consistent with characteristics of delayed detonation models. 

Equipped with high resolution spectra of narrow Na D signatures spanning $\sim$100 days post-maximum light, \citet{Patat07} infer the presence of intervening CSM and argue a mass loss history associated with SN~2006X in the decades prior to explosion. In fact, at least half of all SN~Ia with narrow rest frame, blue-shifted Na D absorption profiles are associated with high ejecta velocities \citep{Sternberg11,Foley12dust}. This indicates that CSM outflows are either present in some explosion scenarios \emph{or} associated with all progenitors scenarios at some point during the lead up to the explosion.

\citet{Patat09} later discussed the VLT spectropolarimetry of SN~2006X. In particular, they find that the presence of the high-velocity \ion{Ca}{2} is coincident with a relatively high polarization signature ($\sim$1.4\%) at day $-$10, that diminishes by only $\sim$15\% near maximum light, and is still present 41 days later. \citet{Patat09} note that this day $+$40 detection is not seen for SN~2001el \citep{Wang03a} or SN~2004du \citep{Leonard05}. As for the high-velocity \ion{Si}{2}, its polarization signature is seen to peak ($\sim$1.1\%) at day $-$6, drop by $\sim$30\% near maximum, and is undetected well into the post-maximum phase. While the findings of spectropolarimetry studies of SN~Ia are thought to be associated with, for example, ``deflagration phase plumes'' with time-dependent photospheric covering fractions, \citet{Patat09} are unable to conclude why SN~2006X exhibits a sizable post-maximum light re-polarization signature by day $+$39.

\subsection{SN~2006bt in CGCG 108-013}\label{ss:06bt}

\citet{Foley10b} obtained optical
light curves and spectra of transitional CN/CL SN~2006bt \citep{Lee06CBET}. The \emph{B}-band decline rate, $\Delta$m$_{15}$({\it B}) = 1.09, is within the range that is observed for normal SN~Ia, however SN~2006bt shows a larger than normal $\mathcal{R}$(\ion{Si}{2}), slightly lower mean expansion velocities, and a lack of a double peak in the \emph{I}-band; CL SN~1991bg-like properties. A tentative \ion{C}{2} 6300 feature is identified, however with a minimum at $\sim$ 6450 \AA. \citet{Foley10b} suggest this inferred lower projected Doppler velocity could be accounted for by a clump of carbon offset from the line of sight \emph{at} photospheric velocities. Because of an association within a halo population of its passive host galaxy, \citet{Foley10b}
conclude that the progenitor was also likely to be from an old population of stars.

\subsection{Over-luminous SN~2006gz in IC 1277}\label{ss:06gz}

\citet{Hicken07} studied SN~2006gz \citep{Prieto06CBET} and estimated a peak intrinsic \emph{V}-band brightness of $-$19.74 and $\Delta$m$_{15}$(\emph{B}) = 0.69, implying M($^{56}$Ni) $\sim$ 1.0$-$1.2 M$_{\odot}$ (assuming $R_{V}$ = 2.1$-$3.1; see also \citealt{Maeda09}). The spectroscopic signatures during early phases are relatively narrow on account of slightly lower mean expansion velocities. At two weeks before maximum light, \citet{Hicken07} attributed a relatively strong 6300 \AA\ feature to 
\ion{C}{2} $\lambda$6580 that diminishes in strength by day $-$10 \citep{Prieto06}. Compared to a 5 \AA\ equivalent width 6300 \AA\ absorption feature observed in the CN SN~1990N \citep{Leibundgut91, Jeffery92}, the absorption feature has an observed equivalent width of 25 \AA\ in the early spectra of SN~2006gz \citep{Hicken07}. So far, spectroscopic modeling that incorporate signatures of \ion{C}{2} $\lambda$6580 predict carbon mass fractions, X(C), that span an order of magnitude and are broadly consistent with both single- and double-degenerate scenarios. 

\citet{Maeda09} obtained Subaru and Keck observations of 2006gz 
at late-phases. Interestingly, SN~2006gz shows relatively weak pillars of iron emission that are usually seen in most SN~Ia subtypes.

\subsection{Extremely faint, SN~2007ax in NGC 2577}\label{ss:07ax}

SN~2007ax was a very faint, red, and peculiar SN~Ia. \citet{Kasliwal08} find that it shares similarities with a sub-luminous SN~2005ke \citep{Immler06,Hughes07,Patat12} and also shows clear excess UV emission $\sim$ 20 days post-maximum light. Based on the small amount of synthesized $^{56}$Ni that is inferred (0.05 $-$ 0.09 M$_{\odot}$), along with SN~Ia-like expansion velocities near maximum light ($\sim$9000 km~s$^{-1}$), \citet{Kasliwal08} conclude
that SN~2007ax is not compatible with a number of theoretical models that have been proposed to explain FAINT$-$CL SN~Ia. 

\subsection{Over-luminous SN~2007if}\label{ss:07if}

    \citet{Scalzo10} find that SN~2007if qualifies as a SCC SN~Ia, i.e. it is over-luminous (M$_{\emph{V}}$ = $-$20.4), has a slow-rise to peak brightness (t$_{rise}$ = 24 days), the early spectra contain signatures of stronger than normal \ion{C}{2}, and SN~2007if resides in a
low-luminosity host (M$_{\emph{g}}$ = $-$14.10). Despite having a red \emph{B}~$-$~\emph{V} color ($+$0.16) at \emph{B}-band maximum, signs of host reddening via Na D lines appear negligible. Utilizing Keck observations of the young metal-poor host galaxy, \citet{Childress11} concluded that SN~2007if is likely to have originated from a young,
metal-poor progenitor. From the H$\alpha$ line of the host galaxy, \citet{Yuan10} derived a redshift
of 0.0736. 

 Based on the
bolometric light curve and the sluggish \ion{Si}{2} velocity evolution, \citet{Scalzo10} conclude that SN~2007if
was the death of a super-Chandrasekhar mass progenitor. They estimate the total mass
of the system to be 2.4 M$_{\odot}$, with 1.6 M$_{\odot}$
 of $^{56}$Ni, and 0.3 to 0.5 M$_{\odot}$ in the
form of a C$+$O envelope. Given the possibility that other over-luminous events could potentially stem from similar super-Chandrasekhar mass origins, \citet{Scalzo12} searched the SNFactory sample (based on a criterion of SN~1991T/2007if-like selections) and found four additional super-Chandrasekhar mass candidates.

\subsection{SN~2007on in NGC 1404}\label{ss:07on}

SN~2007on was found associated with the elliptical galaxy,
NGC 1404 \citep{Pollas07}.
\citet{Voss08} reported the discovery of the progenitor of SN~2007on based on a detected X-ray source in pre-supernova archival X-ray images, located 0.9'' $\pm$ 1.3'' (later 1.15'' $\pm$ 0.27''; \citealt{Roelofs08}) from the position of SN~2007on within its host galaxy. However, \citet{Roelofs08} later reevaluated the detection of the progenitor of SN~2007on and concluded that given the offset discrepancy between the X-ray source and the SN location, the probability for a connection is of order 1 percent. However, should SN~Ia progenitors reveal themselves to be producers of pre-explosive X-ray sources, \citet{Voss08} suggest this would be consistent with a merger model with an accretion disc, formed from the disrupted companion star rather than an explosion immediately upon or soon after the merger of the two stars. 

\subsection{SN~2008J $-$ heavily reddened SN~2002ic-like in MGC-02-07-033}\label{ss:08J}

   \citet{Taddia12} studied SN~2008J \citep{Thrasher08CBET}, which provides additional observational
   evidence
 for hydrogen-rich CSM around an otherwise SN~1991T-like SS SN~Ia. They obtained a NIR spectrum extending up to 2.2
 $\mu$m, and find that SN~2008J is affected by a visual extinction of 1.9 mag. 

 \subsection{Sub-luminous SN~2008ha in UGC 12682}\label{ss:08ha}

   \citet{Foley10a} studied the optical spectrum of SN~2008ha near maximum
brightness \citep{Puckett08CBET,Soderberg09CBET}. It is found to be a dim thermonuclear SN~Ia with uncommonly slow projected expansion velocities. Carbon features at maximum light indicate that carbon-rich
material is present to significant depths in the SN ejecta. Consequently, \citet{Foley10a} conclude that SN~2008ha was a failed deflagration since late time
imaging and spectroscopy also give support to this idea \citep{Kromer13}.

\subsection{SN~2009nr in UGC 8255}\label{ss:09nr}

  \citet{Khan11} discuss the photometric and spectroscopic observations
of the over-luminous (M$_{\emph{V}}$ = $-$19.6, $\Delta$m$_{15}$(\emph{B}) = 0.95) SS SN~2009nr \citep{Balanustsa09CBET}. Similarly, \citet{Tsvetkov11} made \emph{UBVRI} photometric observations of SN~2009nr. They estimate that 0.78 $-$ 1.07 M$_{\odot}$ of $^{56}$Ni was synthesized during the explosion. \citet{Khan11} also find SN~2009nr is at a projected distance of 13.0 kpc from the nucleus of its star-forming
host galaxy. In turn, this indicates that the progenitor of SN~2009nr was
\emph{not} associated with a young stellar population, i.e. SN~2009nr may not have originated from a ``prompt'' progenitor channel as is often assumed for SN~Ia of its subtype. 

\subsection{Peculiar, sub-luminous PTF09dav}\label{ss:ptf09dav}

    \citet{Sullivan11} studied the peculiar PTF09dav
discovered by the Palomar Transient Factory. \citet{Sullivan11} find it to be faint (M$_{\emph{B}}$ = $-$15.5) compared to SN~1991bg, and does not
satisfy the faint end of the WLR. \citet{Sullivan11}
find estimates for both the $^{56}$Ni mass (0.019 M$_{\odot}$)
 and ejecta mass (0.36 M$_{\odot}$) significantly low for thermonuclear supernovae. The spectra are also consistent with signatures of \ion{Sc}{2}, \ion{Mn}{1}, \ion{Ti}{2}, \ion{Sr}{2} and low velocities of $\sim$6000 km~s$^{-1}$. 
The host galaxy of PTF09dav is not clear, however it appears this transient is not
associated with massive, old stellar populations. 
 \citet{Sullivan11} conclude that the observed properties of PTF09dav cannot be
explained by the known models of sub-luminous SN~Ia.

Notably, \citet{Kasliwal12} recently presented late time spectra of PTF09dav (and other similar low luminosity transients). They confirm that this class of objects look nothing like SN~Ia at all on account of little to no late-time iron emission, but instead with prominent emission from calcium in the NIR \citep{Perets10}, confirming previous suspicions of \citet{Sullivan11}. 

\subsection{PTF10ops, another peculiar cross-type SN~Ia}\label{ss:ptf10ops}

\citet{Maguire11} presented optical photometric and spectroscopic observations of a somewhat peculiar and sub-luminous SN~Ia, PTF10ops ($-$17.77 mag). Spectroscopically, this object has been noted as belonging to the CL class of SN~Ia on account of the presence of conspicuous \ion{Ti}{2} absorption features blue ward of 5000 \AA, in addition to a larger than normal $\mathcal{R}$(\ion{Si}{2}) ratio (partially indicative of cooler effective temperatures). Photometrically, PTF10ops overlaps ``normal'' SN~Ia properties in $\Delta$m$_{15}$(\emph{B}) (1.12 $\pm$ 0.06 mag) and its rise-time to maximum light (19.6 days). \citet{Maguire11} estimate $\sim$0.17 M$_{\odot}$ of $^{56}$Ni was produced during the explosion, which is well below what is expected for LVG$-$CN SN~Ia.  

\citet{Maguire11} also note that either PTF10ops remains without a visible host galaxy, or it resides within the outskirts of a massive spiral galaxy located at least 148 kpc away, which would be consistent with a possible influence of low metallicities or an old progenitor population. \citet{Maguire11} suggest the progenitor could have been the merger of two compact objects \citep{Pakmor10}, however time series spectrum synthesis is needed to confirm. 

\subsection{SN~2010jn in NGC 2929}\label{ss:10jn}

The BL SN~2010jn was discovered by the
Palomar Transient Factory (PTF10ygu) 15 days before it reached maximum light. \citet{Hachinger13} performed spectroscopic analysis of the photospheric phase observations and find that the outer layers of SN~2010jn are rich in iron-group elements. At such high velocities ($>$16,000 km~s$^{-1}$), iron-group elements have been tentatively identified in the spectra of SN~Ia before \citep{Hatano99} and may also be a ubiquitous property of SN~Ia. However, more early epoch, time series observations are needed in order to test and confirm such claims. For SN~2010jn at least, \citet{Hachinger13} favor a Chandrasekhar-mass delayed detonation, where the presence of iron-group elements within the outermost layers may be a consequence of outward mixing via hydrodynamical instabilities prior to or during the explosion (see \citealt{Piro11,Piro12}). 

\subsection{SN~2011iv in NGC 1404}\label{ss:11iv}

  \citet{Foley12b} presented the first maximum-light UV through NIR
  spectrum of a SN~Ia (SN~2011iv; \citealt{Drescher11CBET}). Despite having a normal looking spectrum, SN~2011iv declined in brightness fairly quickly ($\Delta$m$_{15}$(\emph{B}) = 1.69). Since the UV region of a SN~Ia spectrum
  is extremely sensitive to the composition of the outer layers, they offer the potential for strong constraints as soon as observational UV spectroscopic diversity is better understood.
  
\subsection{SN~2012cg in NGC 4424}\label{ss:12cg}

\citet{Silverman12a} presented early epoch observations of the nearby spectroscopically normal SN~2012cg \citep{Kandrashoff12,Cenko12,Marion12CBET}, discovered immediately after the event ($\sim$1.5 days after). Compared to the width of other normal SN~Ia \emph{B}-band light curves, \citet{Silverman12a} find that SN~2012cg's light curve relatively narrow for its peak absolute brightness, with $t_{rise}$ = 17.3 days (coincident photometry was also presented by \citealt{Munari13}). Mean expansion velocities within 2.5 days of the event were found to be more than 14,000 km~s$^{-1}$, while the earliest observations show high-velocity components of both \ion{Si}{2} and \ion{Ca}{2}. The \ion{C}{2} $\lambda\lambda$6580, 7234 absorption features were also detected very early.
   
\citet{Johansson13} obtained upper limits on dust emission via far infrared \emph{Herschel Space Observatory} flux measurements in the vicinity of the recent and nearby SN~2011by, 2011fe, and 2012cg. From non-detections during post-maximum epochs at 70 $\mu$m and 160 $\mu$m band-passes and archival image measurements, \citet{Johansson13} exclude dust masses $\gtrsim$~7~x~10$^{-3}$~M$_{\odot}$ for SN~2011fe, and $\gtrsim$~10$^{-1}$~M$_{\odot}$ for SN~2011by and 2012cg for $\sim$ 500 K dust temperatures, $\sim$ 10$^{17}$ cm dust shell radii, and peak SN bolometric luminosities of $\sim$ 10$^{9}$ L$_{\odot}$.

\subsection{SN~2000cx-like, SN~2013bh}\label{ss:13bh}  
  
\citet{Silverman13SN2013bh} discussed recent observations of SN~2013bh and found it similar to SN~2000cx on all accounts, with slightly higher mean expansion velocities. \citet{Silverman13SN2013bh} note that both of these SN~Ia reside on the fringes of their spiral host galaxies. In addition, both SN~2000cx and 2013bh lack narrow Na D lines that would otherwise indicate an environment of CSM. Given the extreme similarities between SN~2000cx and 2013bh, \citet{Silverman13SN2013bh} suggest identical explosion scenarios for both events.

\end{appendices}


\bibliographystyle{spr-mp-nameyear-cnd}

%

\end{document}